\documentclass{jfm}

\usepackage{graphicx}
\usepackage{natbib}
\usepackage{epstopdf}

\ifCUPmtlplainloaded \else
  \checkfont{eurm10}
  \iffontfound
    \IfFileExists{upmath.sty}
      {\typeout{^^JFound AMS Euler Roman fonts on the system,
                   using the 'upmath' package.^^J}%
       \usepackage{upmath}}
      {\typeout{^^JFound AMS Euler Roman fonts on the system, but you
                   dont seem to have the}%
       \typeout{'upmath' package installed. JFM.cls can take advantage
                 of these fonts,^^Jif you use 'upmath' package.^^J}%
      }
  \else
  \fi
\fi

\ifCUPmtlplainloaded \else
  \checkfont{msam10}
  \iffontfound
    \IfFileExists{amssymb.sty}
      {\typeout{^^JFound AMS Symbol fonts on the system, using the
                'amssymb' package.^^J}%
       \usepackage{amssymb}%
       \let\le=\leqslant  
       \let\ge=\geqslant  
      }{}
  \fi
\fi

\ifCUPmtlplainloaded \else
  \IfFileExists{amsbsy.sty}
    {\typeout{^^JFound the 'amsbsy' package on the system, using it.^^J}%
     \usepackage{amsbsy}}
    {}
\fi

\usepackage{graphicx}% Include figure files
\usepackage{dcolumn}% Align table columns on decimal point
\usepackage{bm}% bold math
\usepackage{amsmath} 
\usepackage{subfigure}
\usepackage{cases}
\usepackage{epstopdf}

\usepackage{color}

\usepackage{pdfsync}

\renewcommand{\theenumi}{\roman{enumi}}

		\def\beq{\begin{equation}}
		\def\eeq{\end{equation}}

\newsavebox{\astrutbox}
\sbox{\astrutbox}{\rule[-5pt]{0pt}{20pt}}

\newcommand{\comment}[1]{}
\newcommand{\Ref}{\mathrm{{Re}}_{_F}}
\newcommand{\Rof}{\mathrm{{Ro}}_{_F}}
\newcommand{\Reu}{\mathrm{{Re}}_{_U}}
\newcommand{\Rou}{\mathrm{{Ro}}_{_U}}
\newcommand{\Red}{\mathrm{{Re}}_{_D}}
\newcommand{\Rod}{\mathrm{{Ro}}_{_D}}

\title[Rotating Taylor-Green ]{Rotating Taylor-Green Flow }

\author[Alexandros Alexakis]%
{
A. \ns A\ls L\ls E\ls X\ls A\ls K\ls I\ls S %
  \thanks{Email address for correspondence: alexakis@lps.ens.fr}\ns 
}

\affiliation{Laboratoire de Physique Statistique, CNRS UMR 8550, Ecole Normale Sup\'erieure, \\24 rue Lhomond, Paris, 75005, France} %\\[\affilskip]
\pubyear{}
\volume{}
\pagerange{}
\date{\today }
\begin{document}

\maketitle

\begin{abstract}

The steady state of a forced Taylor-Green flow is investigated in a rotating frame of reference. The investigation involves the results of 184 numerical simulations for different Reynolds number $\mathrm{Re}_{_{F}}$ and Rossby number $\mathrm{Ro}_{_F}$. The large number of examined runs allows a systematic study that enables  the mapping of the different behaviors observed to the parameter space ($\Ref,\Rof$), and the examination of different limiting procedures for approaching the large $\mathrm{Re}_{_{F}}$ small $\mathrm{Ro}_{_F}$ limit. Four distinctly different states were identified: {\it laminar, intermittent bursts, quasi-2D condensates, and weakly rotating turbulence}.  These four different states are separated by power-law boundaries $\mathrm{Ro}_{_F} \propto \mathrm{Re}_{_{F}}^{-\gamma}$ in the small $\mathrm{Ro}_{_F}$ limit. In this limit, the predictions of asymptotic expansions can be directly compared to the results of the direct numerical simulations. While the first order expansion is in good agreement with the results of the linear stability theory, it fails to reproduce the dynamical behavior of the quasi-2D part of the flow in the nonlinear regime, indicating that higher order terms in the expansion need to be taken in to account. The large number of simulations allows also to investigate the scaling that relates the amplitude of the fluctuations  with the energy dissipation rate and the control parameters of the system for the different states of the flow. Different scaling was observed for different states of the flow, that are discussed in detail. The present results clearly demonstrate that the limits small Rossby and large Reynolds do not commute and it is important to specify the order in which they are taken.

\end{abstract}

\begin{keywords}
\end{keywords}

\maketitle

%%%%%%%%%%%%%%%%%%%%%%%%%%%%%%%%%%%%%%%%%%%%%%%%%%%%%%%%%%%%%%%%%%%%%%%%%%%%%%%%%%%%%%%%%%%%%%%%%%%%%%%%%%%%%%%%%%%%%%%%%%%%%%%%%%%
%%%%%%%%%%%%%%%%%%%%%%%%%%%%%%%%%%%%%%%%%%%%%%%%%%%%%%%%%%%%%%%%%%%%%%%%%%%%%%%%%%%%%%%%%%%%%%%%%%%%%%%%%%%%%%%%%%%%%%%%%%%%%%%%%%%
%%%%%%%%%%%%%%%%%%%%%%%%%%%%%%%%%%%%%%%%%%%%%%%%%%%%%%%%%%%%%%%%%%%%%%%%%%%%%%%%%%%%%%%%%%%%%%%%%%%%%%%%%%%%%%%%%%%%%%%%%%%%%%%%%%%
%%%%%%%%%%%%%%%%%%%%%%%%%%%%%%%%%%%%%%%%%%%%%%%%%%%%%%%%%%%%%%%%%%%%%%%%%%%%%%%%%%%%%%%%%%%%%%%%%%%%%%%%%%%%%%%%%%%%%%%%%%%%%%%%%%%

\section{Introduction}

%%%%%%%%%%%%%%%%%%%%%%%%%%%%%%%%%%%%%%%%%%%%%%%%%%%%%%%%%%%%%%%%%%%%%%%%%%%%%%%%%%%%%%%%%%%%%%%%%%%%%%%%%%%%%%%%%%%%%%%%%%%%%%%%%%%
%%%%%%%%%%%%%%%%%%%%%%%%%%%%%%%%%%%%%%%%%%%%%%%%%%%%%%%%%%%%%%%%%%%%%%%%%%%%%%%%%%%%%%%%%%%%%%%%%%%%%%%%%%%%%%%%%%%%%%%%%%%%%%%%%%%
%%%%%%%%%%%%%%%%%%%%%%%%%%%%%%%%%%%%%%%%%%%%%%%%%%%%%%%%%%%%%%%%%%%%%%%%%%%%%%%%%%%%%%%%%%%%%%%%%%%%%%%%%%%%%%%%%%%%%%%%%%%%%%%%%%%
%%%%%%%%%%%%%%%%%%%%%%%%%%%%%%%%%%%%%%%%%%%%%%%%%%%%%%%%%%%%%%%%%%%%%%%%%%%%%%%%%%%%%%%%%%%%%%%%%%%%%%%%%%%%%%%%%%%%%%%%%%%%%%%%%%%

%(\cite{Aref1984,Ottino1989,Ott1993}). 
%

An incompressible flow under rotation will experience the effect of the Coriolis force
altering its dynamical behavior \citep{Greenspan1968}. 
At sufficiently high rotation rates it will suppress the velocity gradients along the direction of rotation bringing the 
flow in a quasi-2D state in which the flow varies only along two dimensions. 
This behavior is due to the Taylor-Proudman theorem obtained for flows which the eddy turn over time is much longer 
than the rotation period.
In addition to rendering the flow quasi-2D a fast rotating system supports inertial waves 
whose frequency increases linearly with the 
rotation rate. Their fast dispersive dynamics weaken the nonlinear interactions allowing for some analytical treatment in the 
framework of weak wave turbulence theory \citep{Nazarenko_book2011,Galtier2003}. 
The interplay of the two {\it phenomena} quasi-2D and 
wave-turbulence can lead to a plethora of phenomena, often observed in natural flows.  

Many different situations can be considered for rotating fluids that can lead to distinct results depending 
on the forcing mechanism, and the value of the involved control parameters.
For example, different results can be envisioned for forced turbulence if the forcing mechanism injects energy exclusively 
to the quasi-2D component of the flow or if the forcing injects energy solely to the inertial waves. Differences are also
expected if the domain size is increased and more dynamical wavenumbers are introduced in the system.
The presence of helicity has also been shown to affect the behavior of the forward cascade \citep{Teitelbaum2009}.
%Finally the order in which the limits of fast rotation small viscosity are taken could lead to different conclusions. 

The large number of different possibilities has led to the emergence of numerous experimental and numerical studies.
These studies of rotating turbulence although numerous, they have to face among other challenges this wide parameter space
and different experimental setups have been considered.
First experiments in rotating tanks date back to \cite{Hopfinger1982}.
 Since then, numerous experiments have been constructed expanding the range of parameter space covered  
\citep{Boubnov1986,Sugihara2005,Swinney2005,Morize2005,Morize2006,Davidson2006,Sreenivasan2007,Staplehurst2008,Bokhoven2009,Sharon2009,Lamriben2011,Sharon2013}.
%%%%
The increase of computational power has also allowed the study of rotating flows by simulations.
Most numerical investigations have focused in decaying turbulence 
\citep{Bardina1985,Mansour1992,Hossain1994,Bartello1994,Squires1994,Godeferd1999,Smith1999,Morinishi2001,Muller2007,Muller2009,Teitelbaum2010,Yoshimatsu2011}, 
with more recent investigations of forced rotating turbulence both at large scales \citep{Yeung1998,Mininni2009,Mininni2009a,Mininni2010,Mininni2012} 
and at small scales in order to observe a development of an inverse cascade \citep{Smith1999,Teitelbaum2009,Mininni2009a}.
Computational cost however did not allow for an exhaustive  coverage of the parameter space. 
Small viscosity fluids (large Reynolds numbers) and high rotation rates (small Rossby numbers) put strong restrictions to simulations. 
Thus typically either moderate Reynolds numbers and small Rossby numbers are reached or moderate Rossby numbers and large Reynolds.
Additionally, these runs have not reached a steady state that requires long integration times. 

The present work attempts to overcome some of these limitations by focusing in one particular forcing mechanism and varying systematically 
the rotation rate and the viscosity. The aim is to understand and map the parameter space of a forced rotating flow.
Thus the focus here is on a large number of simulations at moderate resolutions rather than a few high resolution runs.
The object of the study is the steady state of a flow in a triple periodic square box of side $2\pi L$
%In this work we will focus in the steady state of a flow in a triple periodic square box of side $2\pi L$
forced by a body force ${\bf F}=F_0 {\bf f}$ of forcing amplitude $F_0$ in the presence of rotation $\Omega$ in the ${\bf z}$-direction.
To the authors knowledge this is the first study of forced rotating flows in the steady state.
In this setup the Navier-Stokes equations %(see \cite{Greenspan1968})
for a unit density fluid with viscosity $\nu$ non-dimensionalized by the forcing amplitude $F_0$ and the box size $L$
are:
%%%%%%%%%%%%%%%%%%%%%%%%%%%%%%%%%%%%%%%%%%%%%%%%%%%%%%%%%%%%%%%%%%%%%%%%%%%%%%%%%%%%%%%%%%%%%%%%%%%%%%%%
\begin{equation}                                                                                      %%
\partial_t {\bf u  }     =                       \mathbb{P}\left[ {\bf u  \times w }    \right]       %%
                               +\Rof^{-1} \mathbb{P}\left[ {\bf u  \times  e_z}  \right]              %%
                               +\Ref^{-1} \Delta {\bf u}                                              %%
                               + {\bf f},                                                             %%
%\qquad \mathrm{with}\qquad \nabla \cdot {\bf u}=0.                                                   %% 
\label{NSR}                                                                                           %%
\end{equation}                                                                                        %%
%%%%%%%%%%%%%%%%%%%%%%%%%%%%%%%%%%%%%%%%%%%%%%%%%%%%%%%%%%%%%%%%%%%%%%%%%%%%%%%%%%%%%%%%%%%%%%%%%%%%%%%%
${\bf u}$ is the velocity field (measured in units of $\sqrt{F_0L}$) satisfying $\nabla\cdot {\bf u}=0$.
The vorticity field is given  by ${\bf  w=\nabla\times u}$. The unit vector in the $z$-direction is ${\bf e_z}$ and $\mathbb{P}$ is the projection 
operator to solenoidal fields that in the examined triple periodic domain can be written as
%%%%%%%%%%%%%%%%%%%%%%%%%%%%%%%%%%%%%%%%%%%%%%%%%%%%%%%%%%%%%%%%%%%%%%%%%%%%%%%%%%%%%%%%%%%%%%%%%%%%%%%%
\beq                                                                                                  %%
\mathbb{P}[{\bf g}] \equiv -\Delta^{-1}\nabla \times \nabla \times {\bf g}                            %%
                    =       {\bf g} - \nabla  \Delta^{-1} (\nabla \cdot {\bf g})      \label{project} %% 
\eeq                                                                                                  %%
%%%%%%%%%%%%%%%%%%%%%%%%%%%%%%%%%%%%%%%%%%%%%%%%%%%%%%%%%%%%%%%%%%%%%%%%%%%%%%%%%%%%%%%%%%%%%%%%%%%%%%%%
with $\Delta^{-1}$ being the inverse Laplace operator. The term $\nabla \Delta^{-1} (\nabla \cdot {\bf g})$ 
in \ref{project}
is equivalent to a pressure gradient term $\nabla P$ that guaranties incompressibility. The two control  
parameters in equation \ref{NSR} are given by $\Ref \equiv \sqrt{F_0L}/\nu$ a Reynolds number based on 
the forcing amplitude\footnote{The square of $\Ref$ is sometimes referred to as the Grashof number}, and $\Rof \equiv \sqrt{F_0}/2\Omega\sqrt{L}$ a Rossby number based on the forcing 
amplitude. A more common choice for non-dimensionalization is the space time averaged squared velocity 
$U=\langle \langle {\bf u\cdot u} \rangle_{_V}  \rangle_{_T}^{1/2}$, where here the angular brackets stand 
for space and time average.
%%%%%%%%%%%%%%%%%%%%%%%%%%%%%%%%%%%%%%%%%%%%%%%%%%%%%%%%%%%%%%%%%%%%%%%%%%%%%%%%%%%%%%%%%%%%%%%%%%%%%%%%%%%%%%%%%%
\beq                                                                                                            %%
\langle f\rangle_{_V} \equiv \frac{1}{(2\pi )^3} \int_0^{2\pi} \int_0^{2\pi}\int_0^{2\pi} f \,\, dxdydz, \qquad %%
\langle f\rangle_{_T} \equiv \lim_{T\to \infty} \frac{1}{T} \int_0^T f \,\, dt.                                 %%
\eeq                                                                                                            %%
%%%%%%%%%%%%%%%%%%%%%%%%%%%%%%%%%%%%%%%%%%%%%%%%%%%%%%%%%%%%%%%%%%%%%%%%%%%%%%%%%%%%%%%%%%%%%%%%%%%%%%%%%%%%%%%%%%
Using $U$ we can obtain the usual definitions of the Reynolds $\Reu$ and Rossby $\Rou$ 
number as $\Reu = \Ref U$ and $\Rou = \Rof U$. 
We note however that $\mathrm{Re_{_U}}$ and $\mathrm{Ro_{_U}}$ are not true control 
parameters of the examined dynamical system, since they are measured a posteriori.
They are connected to the true control parameters  $\Ref,\Rof$ by a map that needs however to be determined.
As it turns out the map from the pair ($\Ref,\Rof$) to 
($\Reu,\Rou$) is neither {\it unique} (for the same ($\Ref,\Rof$) two different states with different
($\Reu,\Rou$) exist) nor {\it onto} ( not all pairs of ($\Reu,\Rou$) can be obtained by a suitable choice of 
($\Ref,\Rof$).) 

A third choice for non-dimensionalization is the energy dissipation rate $\epsilon$ that is defined as 
%%%%%%%%%%%%%%%%%%%%%%%%%%%%%%%%%%%%%%%%%%%%%%%%%%%%%%%%%%%%%%%%%%%%%%%%%%%%%%%%%%%%%%%%%%%%%%%%%%%%%%%%
\beq                                                                                                  %%
\epsilon\equiv \Ref^{-1} \langle \langle {\bf w \cdot w}\rangle_{_V}\rangle_{_T} =                    %%
\langle \langle {\bf f \cdot u}\rangle_{_V}\rangle_{_T}.                                              %%
\eeq                                                                                                  %%
%%%%%%%%%%%%%%%%%%%%%%%%%%%%%%%%%%%%%%%%%%%%%%%%%%%%%%%%%%%%%%%%%%%%%%%%%%%%%%%%%%%%%%%%%%%%%%%%%%%%%%%%
The last equality is due to the energy conservation property of the nonlinear and the rotation term.
The new parameters that can be defined are
   $\mathrm{Re_{_D}}=\mathrm{Re_{_F}}\epsilon^{1/3}$ and
   $\mathrm{Ro_{_D}}=\mathrm{Ro_{_F}}\epsilon^{1/3}$.
This choice of control parameters is mostly met in theoretical investigations (like weak wave turbulence theory). 
As with the case of ($\Reu,\Rou$), the parameters based on the energy dissipation  ($\Red,\Rod$),
can only be determined a posteriori and are not true control parameters. They do provide however
a better measure of the strength of turbulence than the other two definitions.
The three choices of control parameter pairs %($\Ref,\Rof$), ($\Reu,\Rou$) and ($\Red,\Rod$) 
can be summarized as $Re=\mathcal{U}L/\nu$, $Ro=\mathcal{U}/(2\Omega L)$ where the dimensional velocity 
$\mathcal{U}$ corresponds to the choice 
$\mathcal{U}=\sqrt{F_0L}                    $ for  ($\Ref,\Rof$),
$\mathcal{U}=\langle \tilde{u}^2\rangle_{_V}^{1/2}$ for  ($\Reu,\Rou$) and
$\mathcal{U}=(\tilde{\epsilon} L)^{1/3}     $ for  ($\Red,\Rod$),
where  $\tilde{u}$ and $\tilde{\epsilon}$ the dimensional velocity and energy dissipation rate respectively. 

The flow in this study is forced by the Taylor-Green (TG) vortex at wavenumber $q$: 
%%%%%%%%%%%%%%%%%%%%%%%%%%%%%%%%%%%%%%%%%%%%%%%%%%%%%%%%%%%%%%%%%%%%%%%%%%%%%%%%%%%%%%%%%%%%%%%%%%%%%%%%
\begin{equation}                                                                                      %%
{\bf f} = 2 \left\{                                                                                   %%
\begin{array}{ll}                                                                                     %% 
\,\,&{\bf e_x} \,\, \sin( q x) \cos( q y) \sin( q z) \\                                               %%
-   &{\bf e_y} \,\, \cos( q x) \sin( q y) \sin( q z) \\                                               %%
\,\,&{\bf e_z} \,\, 0                                                                                 %%
\end{array}                                                                                           %%
\right. .                                                                                             %% 
\label{TG}                                                                                            %% 
\end{equation}                                                                                        %%
%%%%%%%%%%%%%%%%%%%%%%%%%%%%%%%%%%%%%%%%%%%%%%%%%%%%%%%%%%%%%%%%%%%%%%%%%%%%%%%%%%%%%%%%%%%%%%%%%%%%%%%%  
It is normalized so that $\langle{\bf f\cdot f}\rangle_{_V}^{1/2}=1$.
Taylor-Green is a archetypal example of a nonhelical flow that leads to a fast generation of small scale structures 
and has been one of the first examples to study as a candidate for finite time singularity of the Euler equations \citep{Brachet1992}.
Due to its simplicity it has served as a model of many laboratory flows \citep{Mauer1998,Monchaux2007,Monchaux2009,Cortet2010,Salort2010}.
It has also been used as forcing in simulations for the study of non-helical hydrodynamic turbulent flows \citep{Mininni2006} and rotating flows
\citep{Mininni2009,Mininni2009a,Teitelbaum2009}. Finally it has been studied for its magnetic dynamo properties due to its similarities with laboratory dynamo 
experiments \citep{Ponty2008}.

This study investigates the properties of a TG forced flow, varying both the rotation rate and the Reynolds
number, covering the two dimensional parameter space. 

%Contrary to most of the past studies, in the present work we
%focused on the steady state turbulence where both large and small scale quantities fluctuate around a mean value. Note that 
%in the presence of an inverse cascade such a state is reached at very long time scales, much longer than the dynamical 
%(turn over time) time scale.

%%%%%%%%%%%%%%%%%%%%%%%%%%%%%%%%%%%%%%%%%%%%%%%%%%%%%%%%%%%%%%%%%%%%%%%%%%%%%%%%%%%%%%%%%%%%%%%%%%%%%%%%%%%%%%%%%%%%%%%%%%%%%%%
%%%%%%%%%%%%%%%%%%%%%%%%%%%%%%%%%%%%%%%%%%%%%%%%%%%%%%%%%%%%%%%%%%%%%%%%%%%%%%%%%%%%%%%%%%%%%%%%%%%%%%%%%%%%%%%%%%%%%%%%%%%%%%% 
%%%%%%%%%%%%%%%%%%%%%%%%%%%%%%%%%%%%%%%%%%%%%%%%%%%%%%%%%%%%%%%%%%%%%%%%%%%%%%%%%%%%%%%%%%%%%%%%%%%%%%%%%%%%%%%%%%%%%%%%%%%%%%% 
%%%%%%%%%%%%%%%%%%%%%%%%%%%%%%%%%%%%%%%%%%%%%%%%%%%%%%%%%%%%%%%%%%%%%%%%%%%%%%%%%%%%%%%%%%%%%%%%%%%%%%%%%%%%%%%%%%%%%%%%%%%%%%% 
\section{Parameter space - Phase diagram}\label{sec:Param} %%%%%%%%%%%%%%%%%%%%%%%%%%%%%%%%%%%%%%%%%%%%%%%%%%%%%%%%%%%%%%%%%%%%
%%%%%%%%%%%%%%%%%%%%%%%%%%%%%%%%%%%%%%%%%%%%%%%%%%%%%%%%%%%%%%%%%%%%%%%%%%%%%%%%%%%%%%%%%%%%%%%%%%%%%%%%%%%%%%%%%%%%%%%%%%%%%%%
%%%%%%%%%%%%%%%%%%%%%%%%%%%%%%%%%%%%%%%%%%%%%%%%%%%%%%%%%%%%%%%%%%%%%%%%%%%%%%%%%%%%%%%%%%%%%%%%%%%%%%%%%%%%%%%%%%%%%%%%%%%%%%% 
%%%%%%%%%%%%%%%%%%%%%%%%%%%%%%%%%%%%%%%%%%%%%%%%%%%%%%%%%%%%%%%%%%%%%%%%%%%%%%%%%%%%%%%%%%%%%%%%%%%%%%%%%%%%%%%%%%%%%%%%%%%%%%% 
%%%%%%%%%%%%%%%%%%%%%%%%%%%%%%%%%%%%%%%%%%%%%%%%%%%%%%%%%%%%%%%%%%%%%%%%%%%%%%%%%%%%%%%%%%%%%%%%%%%%%%%%%%%%%%%%%%%%%%%%%%%%%%% 

The numerical part of our study consists of 184 direct numerical simulations of the Navier-Stokes equations \ref{NSR} varying the 
values of the  control parameters $\Ref$ and $\Rof$. For all runs the forcing wave-number was kept fixed
to $q=2$. All runs were performed using a standard pseudo-spectral code, where each component of $\bm u$ and 
$\bm b$ is represented as truncated Galerkin expansion in terms of the Fourier basis. The nonlinear terms are initially 
computed in physical space and then transformed to spectral space using fast-Fourier transforms. Aliasing errors are 
removed using the 2/3  de-aliasing rule. The temporal integration was performed using a fourth-order Runge-Kutta method. 
Further details on the code can be found in \citep{Minini_code2}. The grid size varied depending on the value of
$\Ref$ and $\Rof$ from to $64^3$ to $512^3$. A run was considered well resolved if the value of enstrophy spectrum 
at the cut-off wavenumber was more than an order of magnitude smaller than its value at its peak.
Each run started from random multi-mode initial conditions and was continued for sufficiently long time so that long time 
averages in the steady state were be obtained. 
%The only exceptions are the highest resolution runs ($512^3$) for which
%the output of smaller resolution runs was used as initial conditions and then they were continued until saturation.

%%%%%%%%%%%%%%%%%%%%%%%%%%%%%%%%%%%%%%%%%%%%%%%%%%%%%%%%%%%%%%%%%%%%%%%%%%%%%%%%%%%%%%%%%%%%%%%%%%%%%%%%%%%%%%%%%%%%%%%%%%
\begin{figure*}                                                                                                         %%
%\begin{minipage}{\linewidth}                                                                                           %%
\centerline{\includegraphics[width=8cm]{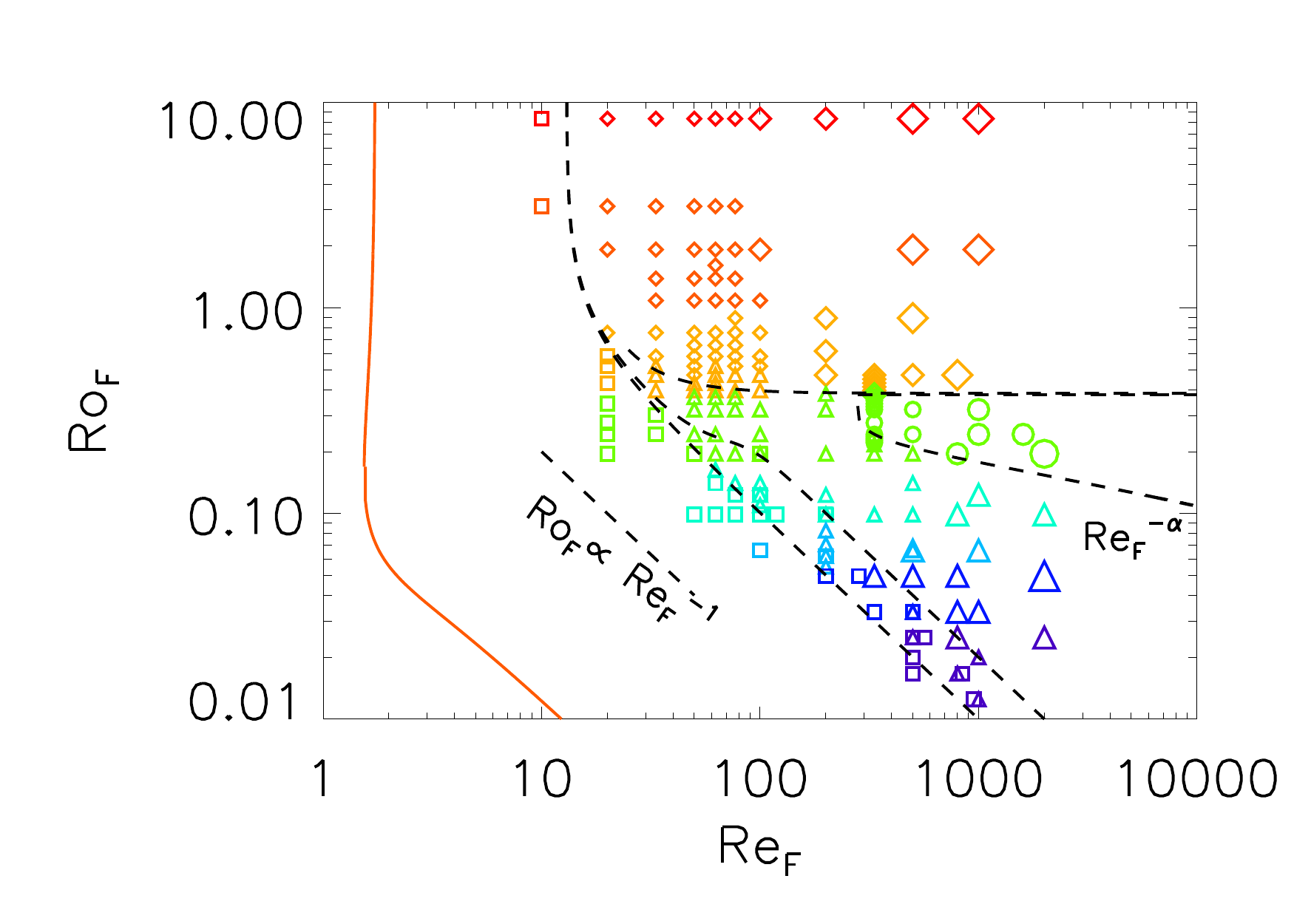} %}                                                         %%        
%\centerline{                                                                                                           %%
\includegraphics[width=7.2cm]{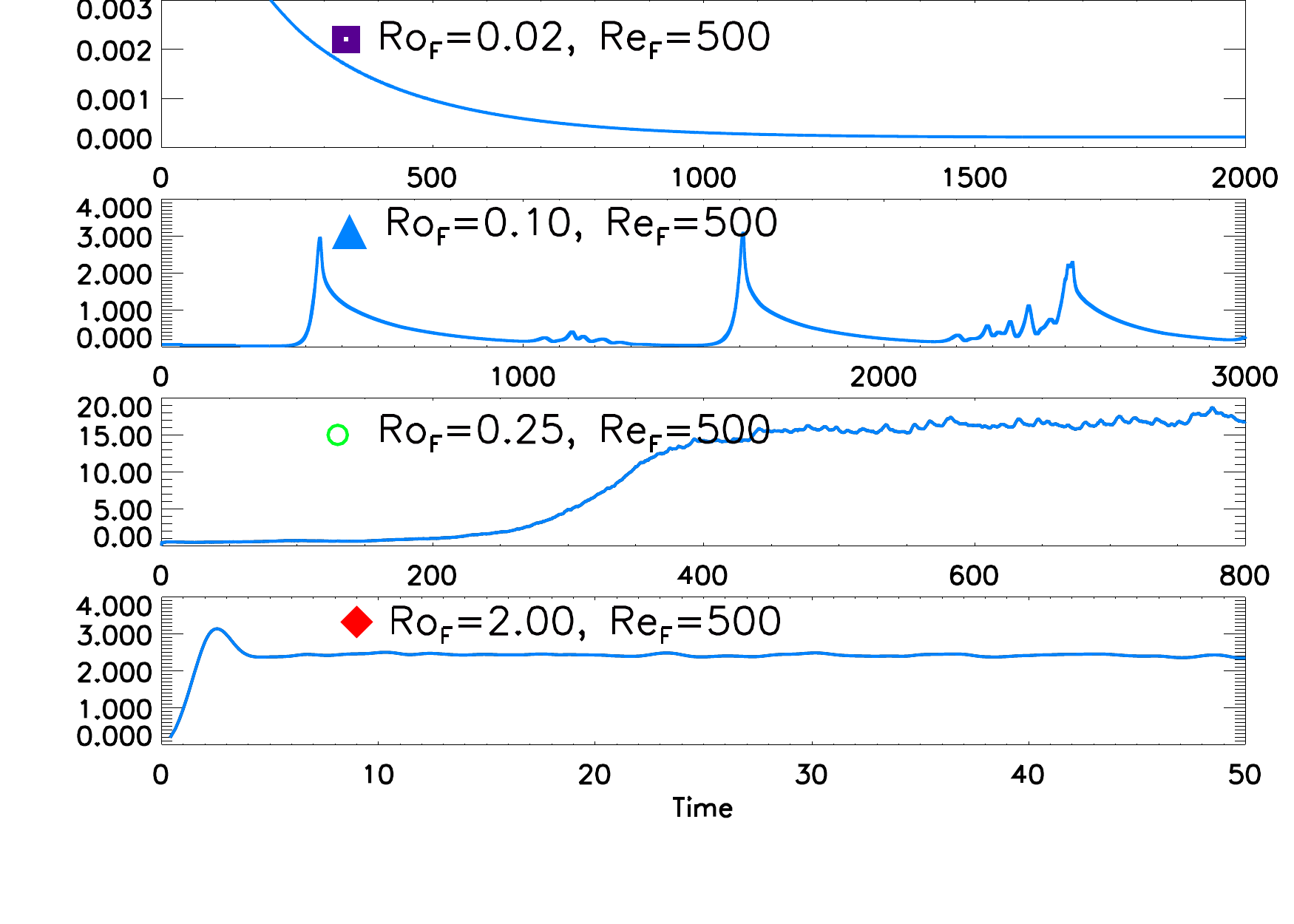}}                                                                    %%
%\end{minipage}                                                                                                         %%
\caption{Left panel: The parameter space $(\Rof,\Ref$). Each point  in this plane indicates a numerical simulation      %%
with this choice of parameters ($\Ref,\Rof$). Different symbols indicate different behavior:                            %%
{\it laminar} flows are indicated by squares, {\it intermittent bursts} are indicated by triangles,                     %% 
{\it quasi-2D condensates} are indicated by circles, {\it weakly rotating} flows are indicated by diamonds.             %%
The different shades of gray (colors online) used are indicative of the magnitude of $\Rof$.                            %%
Large symbols imply larger value of $\Ref$. Right panel: Energy evolution for four representative cases.        }       %%                  
\label{fig_1}                                                                                                           %%
\end{figure*}                                                                                                           %% 
%%%%%%%%%%%%%%%%%%%%%%%%%%%%%%%%%%%%%%%%%%%%%%%%%%%%%%%%%%%%%%%%%%%%%%%%%%%%%%%%%%%%%%%%%%%%%%%%%%%%%%%%%%%%%%%%%%%%%%%%%%

The location of all the performed runs in the ($\Ref,\Rof$) parameter space are shown in the first panel of figure 
\ref{fig_1}. The set of data with the largest value of $\Rof$ are in reality non-rotating runs ($\Rof=\infty$) that 
have been shifted to the finite value $\Rof=8$ so that they can appear in a logarithmic diagram.
The different shades of gray (colors online) used are indicative of the magnitude of $\Rof$ (thus light colors (red online) 
imply slow rotation while dark colors (violet online) imply fast rotation. Large symbols imply larger value of $\Ref$.
The same symbols, sizes and shades (colors online) are  used in all subsequent figures and thus the reader 
can always refer to figure \ref{fig_1} to estimate the value of $\Ref$ and $\Rof$.
The different symbols indicate the four different behaviors that were observed in the numerical runs.
Four representative cases of these behaviors are shown in the right panel of figure \ref{fig_1}. 
All cases have the same value of $\Ref$ but different values of $\Rof$. Note the differences in values in 
$y$-axis that indicate the different range of amplitude reached and the differences in values in $x$-axis that indicate the 
different time scales involved. The top sub-figure displays the evolution of energy from a run that displayed 
{\it laminar} behavior: after some transient period all energy is concentrated in the forcing scales and no fluctuations in 
the temporal behavior are observed. This behavior is observed in runs that occupy the lower and the left part of the shown 
parameter space and are indicated by squares $\Box$. The second from the top sub-figure displays a run for which the energy
evolution displayed {\it intermittent bursts}. It consists of sudden bursts of energy followed by long relaxation periods 
reaching very  small values of kinetic energy. In the left panel they are marked by triangles 
$\bigtriangleup$. They are found for low values of $\Rof$ and high $\Ref$. Note that triangles and squares sometimes overlap 
indicating a bimodal behavior of the flow. The third from the top sub-figure shows the energy evolution from a run that formed 
{\it quasi-2D condensates}.  The energy in these cases reaches very high values and it is concentrated in a few large scale 
2D-modes (ie $k_z=0$, $k_x\sim k_y\sim 1$). They a represented in the left panel by circles $\bigcirc$. They appear for 
intermediate values of $\Rof$ and for large $\Ref$. Finally the bottom sub-figure in  \ref{fig_1} displays the results from 
a {\it weakly rotating} flow. The behavior of these flows as the name suggests weakly deviates from the non-rotating ones: 
energy saturates at order one values and only weak anisotropy is observed. They are represented by diamonds $\Diamond$ in 
the left panel and occupy the higher and right part of the parameter space.

The dashed lines in the left panel of figure \ref{fig_1} separate the parameter space to the different phases that are
observed. The parameter space is then split in 5 different regions: {\it laminar, laminar and bursts, bursts, quasi 2D condensates}, and 
{\it weakly rotating flow}. For large values of $\Ref$ and $\Rof$ these boundaries are expected to take the form of power laws.
Indeed this assumption seems reasonable given the data. The scaling $\Rof\propto\Ref^{-1}$ seems to determine the boundary that
separates laminar region from the bursts and laminar bimodal behavior.
The region that exhibits quasi-2D condensates is bounded from below
by a weak power law $\Rof \propto \Ref^{-\alpha}$. The value of $\alpha$ however cannot be determined by the present data. 
The difficulty in resolving both high $\Ref$ and high $\Rof$ limits us only to a qualitative 
estimate of this lower boundary and prohibits us from measuring precisely $\alpha$. 
We will attempt however to argue the origin of these power laws in the following sections  using scaling arguments and asymptotic expansions.
From above the {\it quasi 2D condensates} appear to be bounded by the value $\Rof\simeq 0.4$ and appear only for $\Ref>240$.
For larger values of $\Rof$ {\it weakly rotating flows} are observed.

%%%%%%%%%%%%%%%%%%%%%%%%%%%%%%%%%%%%%%%%%%%%%%%%%%%%%%%%%%%%%%%%%%%%%%%%%%%%%%%%%%%%%%%%%%%%%%%%%%%%%%%%%%%%%%%%
\begin{figure*}                                                                                               %%
%\begin{minipage}{\linewidth}                                                                                 %%
\centerline{\includegraphics[width=8cm]{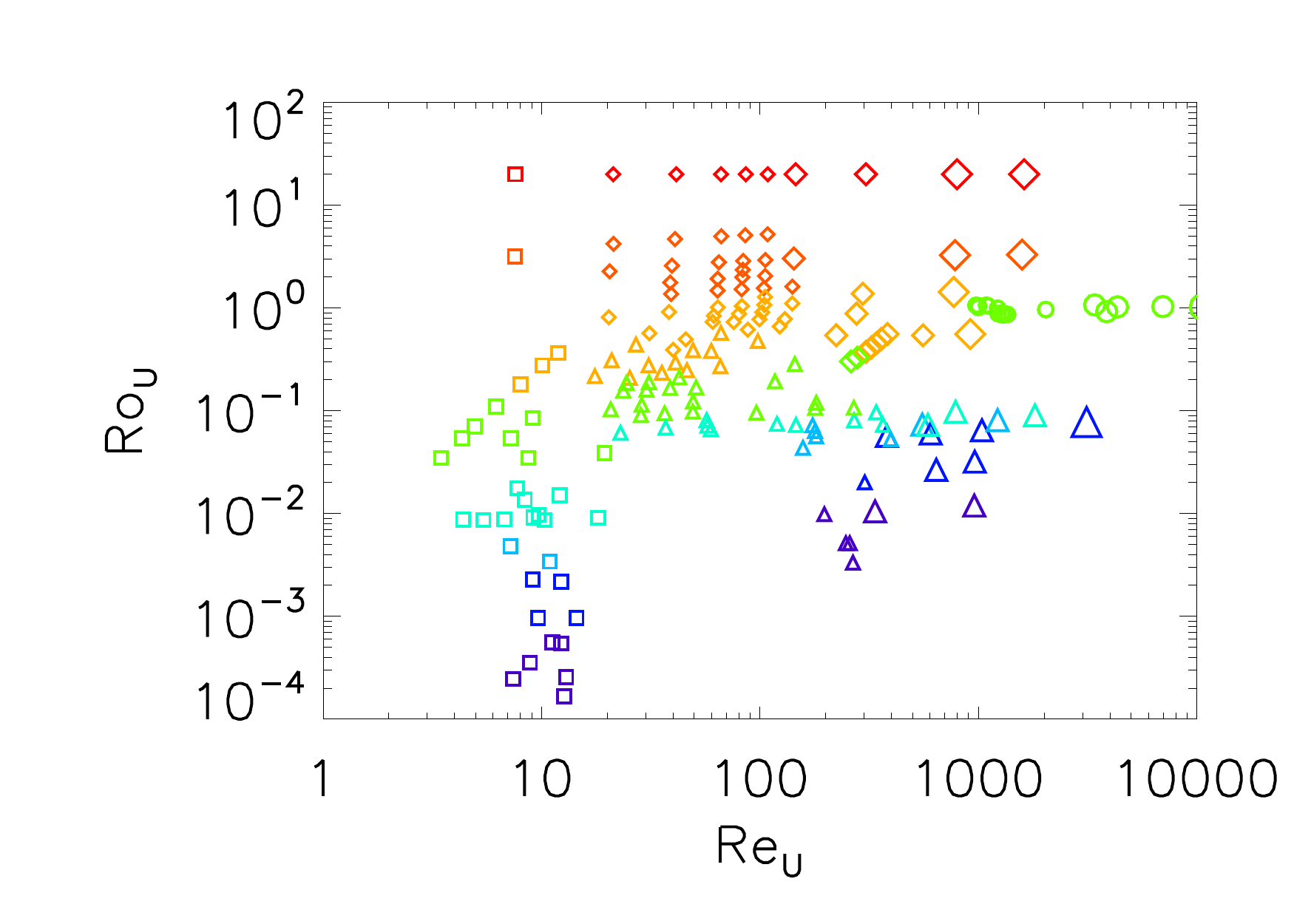}                                                  %%
%\centerline{                                                                                                 %%
\includegraphics[width=8cm]{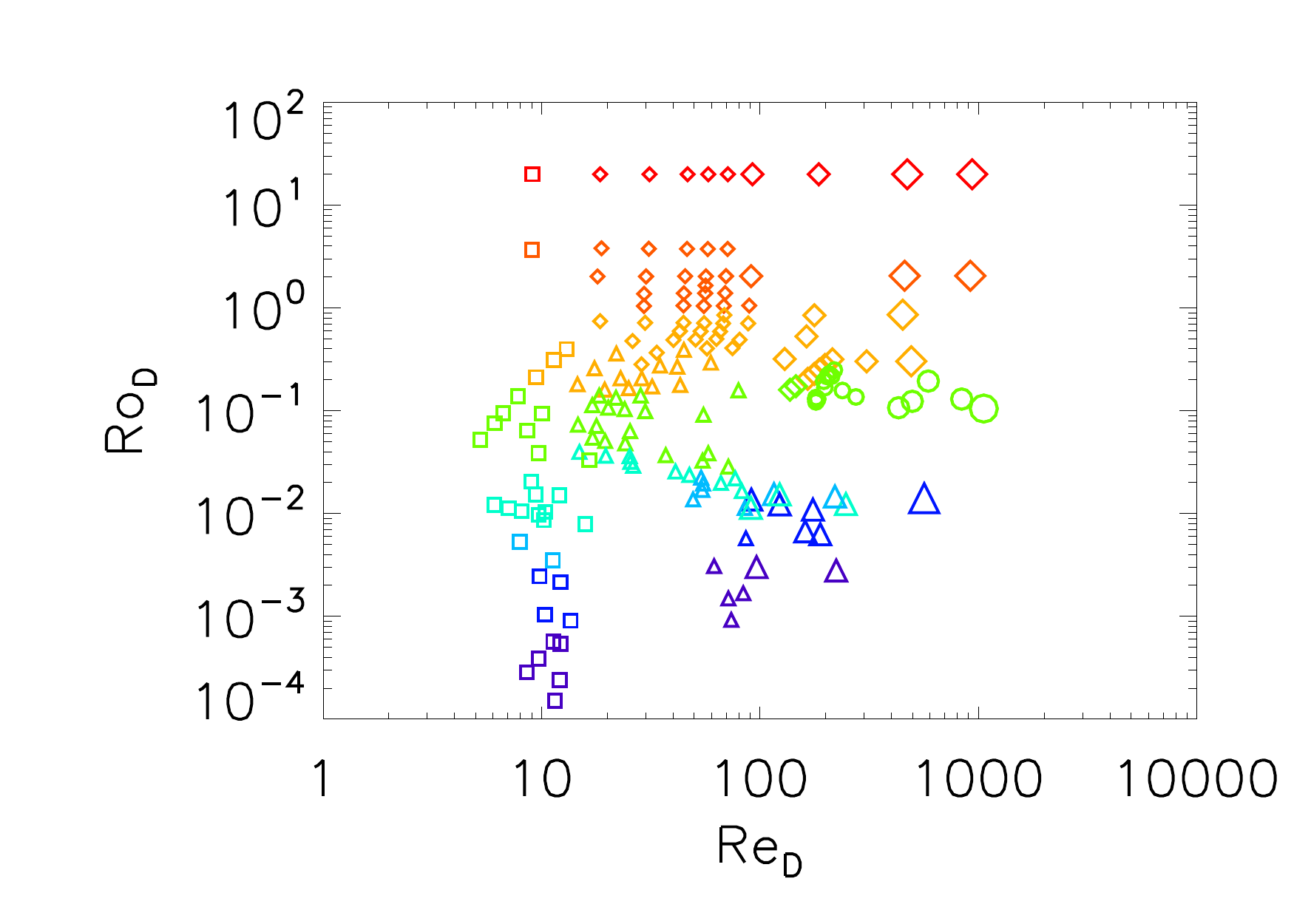} }                                                           %%
%\end{minipage}                                                                                               %%
\caption{The Parameter Space $(\Rou,\Reu$) in the left and $(\Rod,\Red$) in the right.                        %%
         Each point  in this plane indicates a numerical simulation that resulted in that value of            %%
         $(\Rou,\Reu)$ and $(\Rod,\Red$). Same symbols are used for the same runs as in figure                %%
         \ref{fig_1}. }                                                                                       %% 
\label{fig_2}                                                                                                 %%
\end{figure*}                                                                                                 %%
%%%%%%%%%%%%%%%%%%%%%%%%%%%%%%%%%%%%%%%%%%%%%%%%%%%%%%%%%%%%%%%%%%%%%%%%%%%%%%%%%%%%%%%%%%%%%%%%%%%%%%%%%%%%%%%%

As mentioned in the previous section %the velocity amplitude $\mathcal{U}$ used in 
the definition of the Reynolds and Rossby number 
is more than just a conventional formality in rotating turbulence. In figure \ref{fig_2} we show the data in the 
parameter space $(\Reu,\Rou)$ in the left panel and in the paramater space $(\Red,\Rod)$ in the right panel.
The presence of large scale condensates and the suppression of turbulence by rotation
drastically alters the range covered by the simulations for the different parameter choice. In the ($Re_{_U}$, $Ro_{_U}$)
parameter space the data appear more scattered. 
In particular the subcritical transition from laminar to intermittent bursts has lead to the 
to a region vacant of points around $\Rou=10^{-2},\Reu=100$ and $\Rod=10^{-2},\Red=30$.
The {\it quasi 2D condensates}  also have lead  
extremely large values of $Re_{_U}$ to be reached. On the other hand in the ($Re_{_D}$, $Ro_{_D}$),
the data seem to be much more concentrated and bounded on the right by $Re_{_D}\simeq 1000$. This only reflects 
the maximum grid size used $N=512^3$. Thus $Re_{_D}$ provides the best measure for the range of scales excited and
therefor on how turbulent is the flow is. In the following sections however 
the data will be presented based on the parameters ($Re_{_F}$, $Ro_{_F}$)
since these are the ones that are controlled in the numerical experiments.
 
%The parameter space thus seems to be very rich and can not be described by a single theory.
%Insight can be gained by looking in the low $Ro_{_F}$ limit that is examined in the following sections.

%%%%%%%%%%%%%%%%%%%%%%%%%%%%%%%%%%%%%%%%%%%%%%%%%%%%%%%%%%%%%%%%%%%%%%%%%%%%%%%%%%%%%%%%%%%%%%%%%%%%%%%%%%%%%%%%%%%%%%%%%%%%%%%
%%%%%%%%%%%%%%%%%%%%%%%%%%%%%%%%%%%%%%%%%%%%%%%%%%%%%%%%%%%%%%%%%%%%%%%%%%%%%%%%%%%%%%%%%%%%%%%%%%%%%%%%%%%%%%%%%%%%%%%%%%%%%%% 
%%%%%%%%%%%%%%%%%%%%%%%%%%%%%%%%%%%%%%%%%%%%%%%%%%%%%%%%%%%%%%%%%%%%%%%%%%%%%%%%%%%%%%%%%%%%%%%%%%%%%%%%%%%%%%%%%%%%%%%%%%%%%%% 
%%%%%%%%%%%%%%%%%%%%%%%%%%%%%%%%%%%%%%%%%%%%%%%%%%%%%%%%%%%%%%%%%%%%%%%%%%%%%%%%%%%%%%%%%%%%%%%%%%%%%%%%%%%%%%%%%%%%%%%%%%%%%%% 
\section{Stationary solutions in the $\Rof \to 0$ limit} \label{sec:Station} %%%%%%%%%%%%%%%%%%%%%%%%%%%%%%%%%%%%%%%%%%%%%%%%%%
%%%%%%%%%%%%%%%%%%%%%%%%%%%%%%%%%%%%%%%%%%%%%%%%%%%%%%%%%%%%%%%%%%%%%%%%%%%%%%%%%%%%%%%%%%%%%%%%%%%%%%%%%%%%%%%%%%%%%%%%%%%%%%%
%%%%%%%%%%%%%%%%%%%%%%%%%%%%%%%%%%%%%%%%%%%%%%%%%%%%%%%%%%%%%%%%%%%%%%%%%%%%%%%%%%%%%%%%%%%%%%%%%%%%%%%%%%%%%%%%%%%%%%%%%%%%%%% 
%%%%%%%%%%%%%%%%%%%%%%%%%%%%%%%%%%%%%%%%%%%%%%%%%%%%%%%%%%%%%%%%%%%%%%%%%%%%%%%%%%%%%%%%%%%%%%%%%%%%%%%%%%%%%%%%%%%%%%%%%%%%%%% 
%%%%%%%%%%%%%%%%%%%%%%%%%%%%%%%%%%%%%%%%%%%%%%%%%%%%%%%%%%%%%%%%%%%%%%%%%%%%%%%%%%%%%%%%%%%%%%%%%%%%%%%%%%%%%%%%%%%%%%%%%%%%%%%

 The complex phase space observed can be disentangled by
looking at the fast rotating limit in which the smallness of $\Rof$ can be used
to find asymptotic solutions. For $\Rof\ll 1$
the first order term in \ref{NSR} is linear and a solution can be obtained in terms of an expansion series
treating the nonlinearity in a perturbative manner.
We thus write  ${\bf u} =\Rof {\bf v}$ and expand ${\bf v} $ as
%We expand ${\bf u} $ as
%%%%%%%%%%%%%%%%%%%%%%%%%%%%%%%%%%%%%%%%%%%%%%%%%%%%%%%%%%%%%%%%%%%%%%%%%%%%%%%%%%%%%%%%%%%%%%%%%%%%%%%%
\beq                                                                                                  %% 
{\bf v} = {\bf v}^{(0)} + \Rof^2 \,\, {\bf v}^{(1)} + \Rof^4\,\,  {\bf v}^{(2)} +\dots  \label{expan} %% 
\eeq                                                                                                  %% 
%%%%%%%%%%%%%%%%%%%%%%%%%%%%%%%%%%%%%%%%%%%%%%%%%%%%%%%%%%%%%%%%%%%%%%%%%%%%%%%%%%%%%%%%%%%%%%%%%%%%%%%%
(The upper indices in parenthesis indicate order of the expansion and not powers.)
As a first step  we look for stationary solutions to the problem and thus assume no time dependence. 
It is further assumed the following scaling for the Reynolds number $\Ref^{-1}=\lambda_n \Rof^{2n-1}$ where now 
$\lambda_n$ is an order one number and $n$ an integer to be specified. It determines the order at which 
the viscous term will appear in the expansion and thus it will be referred to as the ordering index. 
After substitution equation \ref{NSR} becomes
%%%%%%%%%%%%%%%%%%%%%%%%%%%%%%%%%%%%%%%%%%%%%%%%%%%%%%%%%%%%%%%%%%%%%%%%%%%%%%%%%%%%%%%%%%%%%%%%%%%%%%%%
\begin{equation}                                                                                      %% 
\mathbb{L} \left[ {\bf v } \right] =                                                                  %%
{\bf f}  + \Rof^2 \,\,  \mathbb{P} \left[  {\bf v\times w}  \right]                                   %%
+ \lambda_n \Rof^{2n} \Delta {\bf u}.                                                                 %%
\label{exp1}                                                                                          %%
\end{equation}                                                                                        %%
%%%%%%%%%%%%%%%%%%%%%%%%%%%%%%%%%%%%%%%%%%%%%%%%%%%%%%%%%%%%%%%%%%%%%%%%%%%%%%%%%%%%%%%%%%%%%%%%%%%%%%%%
where now ${\bf w=\nabla\times v}$.
The linear operator $\mathbb{L}$ on the left hand side expresses the effect of rotation and is defined as
%%%%%%%%%%%%%%%%%%%%%%%%%%%%%%%%%%%%%%%%%%%%%%%%%%%%%%%%%%%%%%%%%%%%%%%%%%%%%%%%%%%%%%%%%%%%%%%%%%%%%%%%
\beq                                                                                                  %%
\mathbb{L} \left[ {\bf v } \right] \equiv                                                             %%
\mathbb{P} \left[ {\bf e_z} \times {\bf v } \right].                                                  %%
\eeq                                                                                                  %%
%%%%%%%%%%%%%%%%%%%%%%%%%%%%%%%%%%%%%%%%%%%%%%%%%%%%%%%%%%%%%%%%%%%%%%%%%%%%%%%%%%%%%%%%%%%%%%%%%%%%%%%%
In the triple periodic domain $\mathbb{L}$ can be written as:
%%%%%%%%%%%%%%%%%%%%%%%%%%%%%%%%%%%%%%%%%%%%%%%%%%%%%%%%%%%%%%%%%%%%%%%%%%%%%%%%%%%%%%%%%%%%%%%%%%%%%%%%
\beq                                                                                                  %%
\mathbb{L}    [{\bf f}]=  \Delta^{-1} \partial_z \nabla \times {\bf f}                                %%
\label{Lsmpl}                                                                                           %%
\eeq                                                                                                  %%
%%%%%%%%%%%%%%%%%%%%%%%%%%%%%%%%%%%%%%%%%%%%%%%%%%%%%%%%%%%%%%%%%%%%%%%%%%%%%%%%%%%%%%%%%%%%%%%%%%%%%%%%
Its kernel is composed of all solenoidal vector fields ${\bf g}$ such that $\partial_z {\bf g}=0$.
These vector fields are going to be referred to as two-dimensional-three-components (2D3C) fields
as they depend only on two coordinates but involve all three components.
The projection to the 2D3C-fields is just the vertical average that is denoted as
%%%%%%%%%%%%%%%%%%%%%%%%%%%%%%%%%%%%%%%%%%%%%%%%%%%%%%%%%%%%%%%%%%%%%%%%%%%%%%%%%%%%%%%%%%%%%%%%%%%%%%%%
\beq                                                                                                  %%
\overline{\bf g} \equiv \frac{1}{2\pi } \int_0^{2\pi} {\bf g} \,\,\, dz                               %%
\eeq                                                                                                  %%
%%%%%%%%%%%%%%%%%%%%%%%%%%%%%%%%%%%%%%%%%%%%%%%%%%%%%%%%%%%%%%%%%%%%%%%%%%%%%%%%%%%%%%%%%%%%%%%%%%%%%%%%
For any ${\bf f}$ that has zero projection in this set ($ie$ $\overline{\bf f}=0$) the 
general solution to the equation $\mathbb{L}{\bf g} = {\bf f}$ is:
%%%%%%%%%%%%%%%%%%%%%%%%%%%%%%%%%%%%%%%%%%%%%%%%%%%%%%%%%%%%%%%%%%%%%%%%%%%%%%%%%%%%%%%%%%%%%%%%%%%%%%%%
\beq                                                                                                  %% 
{\bf g}= \mathbb{L}^{-1}{\bf f}                + {\bf g}_{_{2D}}                                      %%
       =-\partial_z^{-1} \nabla \times {\bf f} + {\bf g}_{_{2D}}                                      %%
\eeq                                                                                                  %%
%%%%%%%%%%%%%%%%%%%%%%%%%%%%%%%%%%%%%%%%%%%%%%%%%%%%%%%%%%%%%%%%%%%%%%%%%%%%%%%%%%%%%%%%%%%%%%%%%%%%%%%% 
where ${\bf g}_{_{2D}}$ is an arbitrary (2D3C) field. If however $\overline{\bf f}\ne0$ no solution exists.

Equation \ref{exp1} can be treated perturbatively  and a stationary ${\bf v}$ can in principle be found
as a series expansion in $\Rof^2$. Since the Taylor-Green flow is not a solution of the Euler equations
at each order new wavenumbers will be excited. This procedure will terminate by viscosity that will introduce 
an exponential cutoff. It is not a surprise then that the ordering parameter $n$ that controls the relation
between $\Ref$ and $\Rof$ also controls the convergence of the expansion.
Thus the investigation begins by examining different possible values of the ordering parameter $n$.

%%%%%%%%%%%%%%%%%%%%%%%%%%%%%%%%%%%%%%%%%%%%%%%%%%%%%%%%%%%%%%%%%%%%%%%%%%%%%%%%%%%%%%%%%%%%%%%%%%%%%%%%%%%%%%%%%%
%%%%%%%%%%%%%%%%%%%%%%%%%%%%%%%%%%%%%%%%%%%%%%%%%%%%%%%%%%%%%%%%%%%%%%%%%%%%%%%%%%%%%%%%%%%%%%%%%%%%%%%%%%%%%%%%%%
\subsection{$n=0$, $\Ref^{-1}=\lambda_0 \Rof^{-1}$}      %%%%%%%%%%%%%%%%%%%%%%%%%%%%%%%%%%%%%%%%%%%%%%%%%%%%%%%%%%%
%%%%%%%%%%%%%%%%%%%%%%%%%%%%%%%%%%%%%%%%%%%%%%%%%%%%%%%%%%%%%%%%%%%%%%%%%%%%%%%%%%%%%%%%%%%%%%%%%%%%%%%%%%%%%%%%%%
%%%%%%%%%%%%%%%%%%%%%%%%%%%%%%%%%%%%%%%%%%%%%%%%%%%%%%%%%%%%%%%%%%%%%%%%%%%%%%%%%%%%%%%%%%%%%%%%%%%%%%%%%%%%%%%%%%
For $n=0$ the viscous term is of the same order with the rotation term and the operator
that needs to be inverted to obtain the first order term in ${\bf v}$ is $(\mathbb{L}-\lambda_0 \Delta)$.
This operator is positive definite and can always be inverted to obtain the zeroth order solution.
In detail for the Taylor-Green forcing eq. (\ref{TG}) it is obtained that 
%%%%%%%%%%%%%%%%%%%%%%%%%%%%%%%%%%%%%%%%%%%%%%%%%%%%%%%%%%%%%%%%%%%%%%%%%%%%%%%%%%%%%%%%%%%%%%%%%%%%%%%%%%%
\begin{equation}                                                                                         %%
{\bf v^{(0)}} =                                                                                          %%
\frac{ 2}{ 1 + 27 \lambda_0^2 q^4  }                                                                     %%
\left\{                                                                                                  %%
\begin{array}{lcl}                                                                                       %%
  {\bf e_x}(  - \cos(q x) \sin(q y) \sin(q z) &+ 9 \lambda_0 q^2 \sin(q x) \cos(q y) \sin(q z) &)\\      %%
  {\bf e_y}(  - \sin(q x) \cos(q y) \sin(q z) &- 9 \lambda_0 q^2 \cos(q x) \sin(q y) \sin(q z) &)\\      %%
  {\bf e_z}( \, 2 \,  \sin(q x) \sin(q y) \cos(q z) & &)                                                 %%
\end{array}                                                                                              %%
\right. .                                                                                                %%
\label{v00}                                                                                              %%
\end{equation}                                                                                           %%
%%%%%%%%%%%%%%%%%%%%%%%%%%%%%%%%%%%%%%%%%%%%%%%%%%%%%%%%%%%%%%%%%%%%%%%%%%%%%%%%%%%%%%%%%%%%%%%%%%%%%%%%%%%
The next order correction follows in a similar manner without a qualitative change the of the flow behavior.
We note that the stationary solution in this limit follows the scaling
$\langle {\bf u\cdot u}\rangle_{_V}= \Rof^2 \langle {\bf v\cdot v}\rangle_{_V} = 3\Rof^2 (1+27 \lambda_0^2 q^4)^{-1}$.
That implies that the relation between $(\Reu,\Rou)$ and $(\Ref,\Rof)$ for the laminar flow is given by
%%%%%%%%%%%%%%%%%%%%%%%%%%%%%%%%%%%%%%%%%%%%%%%%%%%%%%%%%%%%%%%%%%%%%%%%%%%%%%%%%%%%%%%%%%%%%%%%%%%%%%%%%%%
\beq                                                                                                     %%
\Reu = \frac{\sqrt{3} \, \Ref^2\Rof  }{ \sqrt{\Ref^2+27\Rof^2 q^4}},  \qquad                            %%
\Rou = \frac{\sqrt{3} \, \Rof^2\Ref  }{ \sqrt{\Ref^2+27\Rof^2 q^4}},                                    %%
\label{RR1}                                                                                              %%
\eeq                                                                                                     %%
%%%%%%%%%%%%%%%%%%%%%%%%%%%%%%%%%%%%%%%%%%%%%%%%%%%%%%%%%%%%%%%%%%%%%%%%%%%%%%%%%%%%%%%%%%%%%%%%%%%%%%%%%%%
Furthermore, $\langle {\bf u \cdot f}\rangle = 9 q^2 \Rof\lambda_0 (1+27\lambda_0^2 q^4)^{-1} $ and thus
%%%%%%%%%%%%%%%%%%%%%%%%%%%%%%%%%%%%%%%%%%%%%%%%%%%%%%%%%%%%%%%%%%%%%%%%%%%%%%%%%%%%%%%%%%%%%%%%%%%%%%%%%%%
\beq                                                                                                     %%
\Red = \frac{3^{2/3} \, \Ref^{4/3} \Rof^{1/3}  }{ (\Ref^2+27 \Rof^2 q^4)^{1/3}},  \qquad                %%
\Rod = \frac{3^{2/3} \, \Rof^{4/3} \Ref^{1/3}  }{ (\Ref^2+27 \Rof^2 q^4)^{1/3}},                        %%
\label{RR2}                                                                                              %%
\eeq                                                                                                     %%
%%%%%%%%%%%%%%%%%%%%%%%%%%%%%%%%%%%%%%%%%%%%%%%%%%%%%%%%%%%%%%%%%%%%%%%%%%%%%%%%%%%%%%%%%%%%%%%%%%%%%%%%%%%

These results \ref{RR1},\ref{RR2} provide the map  from the  $(\Reu,\Rou)$ space 
(left panel of figure \ref{fig_1}) to the $(\Reu,\Rou)$ and $(\Red,\Rod)$ space shown in figure \ref{fig_2} 
but only for the laminar solutions.

%%%%%%%%%%%%%%%%%%%%%%%%%%%%%%%%%%%%%%%%%%%%%%%%%%%%%%%%%%%%%%%%%%%%%%%%%%%%%%%%%%%%%%%%%%%%%%%%%%%%%%%%%%%%%%%%%%
%%%%%%%%%%%%%%%%%%%%%%%%%%%%%%%%%%%%%%%%%%%%%%%%%%%%%%%%%%%%%%%%%%%%%%%%%%%%%%%%%%%%%%%%%%%%%%%%%%%%%%%%%%%%%%%%%%
\subsection{$n=1$, $\Ref^{-1}=\lambda_1 \Rof^{1}$}           %%%%%%%%%%%%%%%%%%%%%%%%%%%%%%%%%%%%%%%%%%%%%%%%%%%%%
%%%%%%%%%%%%%%%%%%%%%%%%%%%%%%%%%%%%%%%%%%%%%%%%%%%%%%%%%%%%%%%%%%%%%%%%%%%%%%%%%%%%%%%%%%%%%%%%%%%%%%%%%%%%%%%%%%
%%%%%%%%%%%%%%%%%%%%%%%%%%%%%%%%%%%%%%%%%%%%%%%%%%%%%%%%%%%%%%%%%%%%%%%%%%%%%%%%%%%%%%%%%%%%%%%%%%%%%%%%%%%%%%%%%%
For this value of the ordering index, equation \ref{exp1} becomes 
%%%%%%%%%%%%%%%%%%%%%%%%%%%%%%%%%%%%%%%%%%%%%%%%%%%%%%%%%%%%%%%%%%%%%%%%%%%%%%%%%%%%%%%
\begin{equation}                                                                     %%
\mathbb{L} \left[ {\bf v } \right] = {\bf f}  + \Rof^2                                 %%
\,\, ( \mathbb{P}  \left[ {\bf v \times w}  \right] + \lambda_1 \Delta {\bf u} ).    %%
\end{equation}                                                                       %%
%%%%%%%%%%%%%%%%%%%%%%%%%%%%%%%%%%%%%%%%%%%%%%%%%%%%%%%%%%%%%%%%%%%%%%%%%%%%%%%%%%%%%%%
Since $\overline{\bf f}=0$ the zeroth order solution can be written as 
${\bf v}^{(0)} = {\bf v}^{(0)}_{\bf f} + {\bf v}_{_{2D}}^{(0)}$ where 
%%%%%%%%%%%%%%%%%%%%%%%%%%%%%%%%%%%%%%%%%%%%%%%%%%%%%%%%%%%%%%%%%%%%%%%%%%%%%%%%%%%%%%%
\beq {\bf v}^{(0)}_{\bf f} =                                                         %%
\mathbb{L}^{-1}{\bf f} =    2                                                        %%
\left\{                                                                              %%
\begin{array}{lcl}                                                                   %%
  -& {\bf e_x} &   \cos(q x) \sin(q y) \sin(q z)  \\                                 %%
  -& {\bf e_y} &   \sin(q x) \cos(q y) \sin(q z)  \\                                 %%
  2& {\bf e_z} &   \sin(q x) \sin(q y) \cos(q z)                                     %%
\end{array}                                                                          %%
\label{v01}                                                                          %%
\right., \eeq                                                                        %%
%%%%%%%%%%%%%%%%%%%%%%%%%%%%%%%%%%%%%%%%%%%%%%%%%%%%%%%%%%%%%%%%%%%%%%%%%%%%%%%%%%%%%%%
and ${\bf v}_{_{2D}}^{(0)}$ is an arbitrary 2D3C field. 
The velocity field in eq.\ref{v01} is the same as the one in \ref{v00} with $\lambda_0=0$.
The 2D3C field is determined at next order for which we have
%%%%%%%%%%%%%%%%%%%%%%%%%%%%%%%%%%%%%%%%%%%%%%%%%%%%%%%%%%%%%%%%%%%%%%%%%%%%%%%%%%%%%%%%%%%%%%%%%%%%%%%%%%%%
\begin{equation}                                                                                          %%
\mathbb{L}[{\bf v}^{(1)}]  = \mathbb{P} \left[   {\bf v}^{(0)} \times {\bf w}^{(0)}  \right]              %% 
                           + \lambda_1  \Delta   {\bf v}^{(0)}.                                           %%
\label{sord1}                                                                                             %%
\end{equation}                                                                                            %%
%%%%%%%%%%%%%%%%%%%%%%%%%%%%%%%%%%%%%%%%%%%%%%%%%%%%%%%%%%%%%%%%%%%%%%%%%%%%%%%%%%%%%%%%%%%%%%%%%%%%%%%%%%%%
Averaging over the $z$ direction we obtain by detailed calculation that:
%%%%%%%%%%%%%%%%%%%%%%%%%%%%%%%%%%%%%%%%%%%%%%%%%%%%%%%%%%%%%%%%%%%%%%%%%%%%%%%%%%%%%%%%%%%%%%%%%%%%%%%%%%%%
\beq                                                                                                      %%
     \overline{\mathbb{P} \left[   {\bf v}^{(0)}_f       \times {\bf w}_{_{2D}}^{(0)} \right] }=          %%
     \overline{\mathbb{P} \left[   {\bf v}_{_{2D}}^{(0)} \times {\bf w}_f^{(0)}  \right] }=               %% 
     \overline{\mathbb{P} \left[   {\bf v}^{(0)}_f       \times {\bf w}^{(0)}_f  \right] }=0              %%
\eeq                                                                                                      %%
%%%%%%%%%%%%%%%%%%%%%%%%%%%%%%%%%%%%%%%%%%%%%%%%%%%%%%%%%%%%%%%%%%%%%%%%%%%%%%%%%%%%%%%%%%%%%%%%%%%%%%%%%%%%
where ${\bf w}_{_{2D}}^{(0)}=\nabla\times {\bf v}_{_{2D}}^{(0)}$ and ${\bf w}^{(0)}_{\bf f} =\nabla\times {\bf v}^{(0)}_{\bf f}$.
We are thus left with:
%%%%%%%%%%%%%%%%%%%%%%%%%%%%%%%%%%%%%%%%%%%%%%%%%%%%%%%%%%%%%%%%%%%%%%%%%%%%%%%%%%%%%%%%%%%%%%%%%%%%%%%%%%%%%%%%%%%%%%%%%%%%%%%%
\beq                                                                                                                          %%
0 = \mathbb{P} \left[   {\bf v}_{_{2D}}^{(0)} \times {\bf w}^{(0)}_{_{2D}}  \right] + \lambda_1 \Delta {\bf v}_{_{2D}}^{(0)}. %%
\eeq                                                                                                                          %%
%%%%%%%%%%%%%%%%%%%%%%%%%%%%%%%%%%%%%%%%%%%%%%%%%%%%%%%%%%%%%%%%%%%%%%%%%%%%%%%%%%%%%%%%%%%%%%%%%%%%%%%%%%%%%%%%%%%%%%%%%%%%%%%%
Multiplying by ${\bf v}_{_{2D}}^{(0)}$ and horizontal averaging we get $\langle { (\nabla {\bf v}_{_{2D}}^{(0)})}^2\rangle_{_V} =0$
and thus the undetermined 2D3C field at first order becomes ${\bf v}_{_{2D}}^{(0)}=0$. 
%As a result a 2D3C component of the  velocity field will not appear at this order. 
A 2D3C flow can possibly appear at higher order but we will proceed no further than obtaining ${\bf v}^{(0)}$.

The relations given in \ref{RR1} simplify (by setting $\lambda_0=0$) to
\beq
\Reu=\sqrt{3}\, \Ref\Rof  \quad \mathrm{and} \quad \Rou=\sqrt{3}\, \Rof^2
\label{RefRouMap1}
\eeq
for the Reynolds numbers based on the root mean square velocity.
Similarly \ref{RR2} simplify to $\Red=3^{2/3}\,\Ref^{1/3}\Rof^{2/3}$ and  $\Rod=3^{2/3}\,\Ref^{-1/3}\Rof^{4/3}$
for the Reynolds numbers based on the energy injection rate.

%hline
%%%%%%%%%%%%%%%%%%%%%%%%%%%%%%%%%%%%%%%%%%%%%%%%%%%%%%%%%%%%%%%%%%%%%%%%%%%%%%%%%%%%%%%%%%%%%%%%%%%%%%%%%%%%%%%%
%%%%%%%%%%%%%%%%%%%%%%%%%%%%%%%%%%%%%%%%%%%%%%%%%%%%%%%%%%%%%%%%%%%%%%%%%%%%%%%%%%%%%%%%%%%%%%%%%%%%%%%%%%%%%%%%
 \subsection{$n=2$, $\Ref^{-1}=\lambda_2 \Rof^{3}$}       %%%%%%%%%%%%%%%%%%%%%%%%%%%%%%%%%%%%%%%%%%%%%%%%%%%%%%
%%%%%%%%%%%%%%%%%%%%%%%%%%%%%%%%%%%%%%%%%%%%%%%%%%%%%%%%%%%%%%%%%%%%%%%%%%%%%%%%%%%%%%%%%%%%%%%%%%%%%%%%%%%%%%%%
%%%%%%%%%%%%%%%%%%%%%%%%%%%%%%%%%%%%%%%%%%%%%%%%%%%%%%%%%%%%%%%%%%%%%%%%%%%%%%%%%%%%%%%%%%%%%%%%%%%%%%%%%%%%%%%%

For this ordering index the expansion follows as before but viscosity is not present at first order and
the solvability condition obtained in \ref{sord1} only restricts ${\bf v}_{_{2D}}$ at being 
a solution of the Euler equations $\mathbb{P} \left[   {\bf v}_{_{2D}}^{(0)} \times {\bf w}_{_{2D}}^{(0)}  \right]=0$. 
The first order correction is then 
%%%%%%%%%%%%%%%%%%%%%%%%%%%%%%%%%%%%%%%%%%%%%%%%%%%%%%%%%%%%%%%%%%%%%%%%%%%%%%%%%%%%%%%%%%%%%%%%%%%%%%%%%%%%%%%%%%
\beq                                                                                                            %%
{\bf v}^{(1)} = \mathbb{L}^{-1}[{\bf v^{(0)} \times w^{(0)}}]+{\bf v}_{_{2D}}^{(1)}.                            %%
\label{v11}                                                                                                     %%
\eeq                                                                                                            %%
%%%%%%%%%%%%%%%%%%%%%%%%%%%%%%%%%%%%%%%%%%%%%%%%%%%%%%%%%%%%%%%%%%%%%%%%%%%%%%%%%%%%%%%%%%%%%%%%%%%%%%%%%%%%%%%%%%
To proceed at second order and simplify the mathematical complexity that increases rapidly with the order of the expansion
we will assume at this point that ${\bf v}_{_{2D}}=0$ and verify a posteriori the validity of this assumption.  
To the next order the we obtain 
%%%%%%%%%%%%%%%%%%%%%%%%%%%%%%%%%%%%%%%%%%%%%%%%%%%%%%%%%%%%%%%%%%%%%%%%%%%%%%%%%%%%%%%%%%%%%%%%%%%%%%%%%%%%%%%%%%%%%%%%%%%
\beq                                                                                                                     %%
\mathbb{L}[{\bf v}^{(2)}]  = \mathbb{P} \left[  {\bf v}^{(1)}\times {\bf w}^{(0)}                                        %% 
                                              + {\bf v}^{(0)}\times {\bf w}^{(1)}\right] +\lambda_2\Delta {\bf v}^{(0)}. %%
\label{Lv2}                                                                                                              %%
\eeq                                                                                                                     %% 
%%%%%%%%%%%%%%%%%%%%%%%%%%%%%%%%%%%%%%%%%%%%%%%%%%%%%%%%%%%%%%%%%%%%%%%%%%%%%%%%%%%%%%%%%%%%%%%%%%%%%%%%%%%%%%%%%%%%%%%%%%%
Averaging over $z$ the nonlinear term drops to zero. This can be realized by noticing that the 0-th order components 
(${\bf v}^{(0)},{\bf w}^{(0)}$) only contain wavenumbers $\pm q$ in the $z$-direction while the first order component 
(${\bf v}^{(1)},{\bf w}^{(1)}$) 
has only $\pm 2q$ wavenumbers. Their product will thus have only $\pm q$ and $\pm 3q$ wave numbers and will thus average to zero.
%
%After averaging over $z$ the remaining terms lead to 
%\[
%   \overline{ \mathbb{P}
%   \Big{[}{\bf \mathbb{L}^{-1}[{\bf v^0_{_{2D}}\times w^0_f + u^0_f\times w^0_{_{2D}}} ] \times w^0_{_f} } \Big{]}   }+
%   \overline{ \mathbb{P}
%   \Big{[}{\bf v^0_{_{f}} \times (\nabla\times \mathbb{L}^{-1}[{\bf v^0_{_{2D}}\times w^0_f + v^0_f\times w^0_{_{2D}}} ])  }
%    \Big{]}} = - \nu \nabla^2 {\bf v^0_{_{2D}} } 
% \]
We obtain thus $\overline{\nabla^2 \bf v^{(0)}} =0 $ in agreement with our original assumption that 
${\bf v}_{_{2D}}=0$. 
%A detailed calculation with out this assumption indicates that 
%${\bf v}_{_{2D}}^{(0)}=0$ is indeed the only solution. 
%
Up to this scaling therefore the stationary solution obtained has zero projection to the
2D3C fields. 

%%%%%%%%%%%%%%%%%%%%%%%%%%%%%%%%%%%%%%%%%%%%%%%%%%%%%%%%%%%%%%%%%%%%%%%%%%%%%%%%%%%%%%%%%%%%%%%%%%%%%%%%%%%%%%%%%%
%%%%%%%%%%%%%%%%%%%%%%%%%%%%%%%%%%%%%%%%%%%%%%%%%%%%%%%%%%%%%%%%%%%%%%%%%%%%%%%%%%%%%%%%%%%%%%%%%%%%%%%%%%%%%%%%%%
 \subsection{$n=3$, $\Ref^{-1}=\lambda_3 \Rof^{5}$}    %%%%%%%%%%%%%%%%%%%%%%%%%%%%%%%%%%%%%%%%%%%%%%%%%%%%%%%%%%%
%%%%%%%%%%%%%%%%%%%%%%%%%%%%%%%%%%%%%%%%%%%%%%%%%%%%%%%%%%%%%%%%%%%%%%%%%%%%%%%%%%%%%%%%%%%%%%%%%%%%%%%%%%%%%%%%%%
%%%%%%%%%%%%%%%%%%%%%%%%%%%%%%%%%%%%%%%%%%%%%%%%%%%%%%%%%%%%%%%%%%%%%%%%%%%%%%%%%%%%%%%%%%%%%%%%%%%%%%%%%%%%%%%%%%
The assumption that ${\bf v}_{_{2D}}=0$ ceases to be true for this ordering.
If we assume that ${\bf v}^0_{_{2D}}=0$ then at third order where the dissipation term is present 
we will obtain
%%%%%%%%%%%%%%%%%%%%%%%%%%%%%%%%%%%%%%%%%%%%%%%%%%%%%%%%%%%%%%%%%%%%%%%%%%%%%%%%%%%%%%%%%%%%%%%%%%%%%%%%%%%%%%%%%%%%%
\beq                                                                                                               %%
\mathbb{L}[{\bf v}^{(3)}]  = \mathbb{P} \left[ {\bf v}^{(2)} \times {\bf w}^{(0)} +                                %%
                                               {\bf v}^{(0)} \times {\bf w}^{(2)} +                                %%
                                               {\bf v}^{(1)} \times {\bf w}^{(1)}    \right]                      %%
                                             + \lambda_3 \Delta {\bf v^{(0)} }.                                    %%
\label{Lv3}                                                                                                        %%
\eeq                                                                                                               %%
%%%%%%%%%%%%%%%%%%%%%%%%%%%%%%%%%%%%%%%%%%%%%%%%%%%%%%%%%%%%%%%%%%%%%%%%%%%%%%%%%%%%%%%%%%%%%%%%%%%%%%%%%%%%%%%%%%%%%
where ${\bf v}^{(0)},{\bf v}^{(1)}$ and ${\bf v}^{(2)}$ are obtained from eq. \ref{v01}, \ref{v11} and  \ref{Lv2} respectably
with $\lambda_2=0$.
Averaging over $z$ the nonlinear term will not become zero and thus will need to be balanced by the viscosity. 
%%%%%%%%%%%%%%%%%%%%%%%%%%%%%%%%%%%%%%%%%%%%%%%%%%%%%%%%%%%%%%%%%%%%%%%%%%%%%%%%%%%%%%%%%%%%%%%%%%%%%%%%%%%%%%%%%%%%%
\beq                                                                                                               %%
\lambda_3  \nabla^2 {\bf v_{_{2D}}} =                                                                              %%
\lambda_3  \nabla^2 \overline{\bf v^{(0)}} = - \overline{\mathbb{P} \left[   {\bf v}^{(2)} \times {\bf w}^{(0)} +  %%
                                                                             {\bf v}^{(0)} \times {\bf w}^{(2)} +  %%
                                                                             {\bf v}^{(1)} \times {\bf w}^{(1)}    %%
                                                                   \right]} \ne 0        \label{cntrd}             %%
\eeq                                                                                                               %%
%%%%%%%%%%%%%%%%%%%%%%%%%%%%%%%%%%%%%%%%%%%%%%%%%%%%%%%%%%%%%%%%%%%%%%%%%%%%%%%%%%%%%%%%%%%%%%%%%%%%%%%%%%%%%%%%%%%%%
Thus ${\bf v_{_{2D}}=0}$ is no longer an acceptable solution. Calculation of the nonlinear term \ref{cntrd}
leads to 
%%%%%%%%%%%%%%%%%%%%%%%%%%%%%%%%%%%%%%%%%%%%%%%%%%%%%%%%%%%%%%%%%%%%%%%%%%%%%%%%%%%%%%%%%%%%%%%%%%%%%%%%%%%%%%%%%%%%%
\beq                                                                                                               %%
                                               \overline{\mathbb{P} \left[   {\bf v}^{(2)} \times {\bf w}^{(0)} +  %%
                                                                             {\bf v}^{(0)} \times {\bf w}^{(2)} +  %%
                                                                             {\bf v}^{(1)} \times {\bf w}^{(1)}    %%
                                                                   \right]}                                        %%
%  \overline{\mathbb{P} \left[     {\bf v_2 \times w^0 +                                                           %%
%                                                  v^0 \times w^2 +                                                %%
%                                                  v^1 \times w^1 }\right]}                                        %%
=\left\{                                                                                                           %%
\begin{array}{r}                                                                                                   %%
 \partial_y \Psi \\                                                                                                %%
-\partial_x \Psi \\                                                                                                %%  
        0                                                                                                          %% 
\end{array}                                                                                                        %%
\right\}                                                                                                           %%
\eeq                                                                                                               %%
%%%%%%%%%%%%%%%%%%%%%%%%%%%%%%%%%%%%%%%%%%%%%%%%%%%%%%%%%%%%%%%%%%%%%%%%%%%%%%%%%%%%%%%%%%%%%%%%%%%%%%%%%%%%%%%%%%%%%
with 
%%%%%%%%%%%%%%%%%%%%%%%%%%%%%%%%%%%%%%%%%%%%%%%%%%%%%%%%%%%%%%%%%%%%%%%%%%%%%%%%%%%%%%%%%%%%%%%%%%%%%%%%%%%%%%%%%%%%%
\beq \Psi = \frac{9q}{2} [\cos(2qx)-\cos(2qy)]\sin(2qx)\sin(2qy) \eeq                                              %%
%%%%%%%%%%%%%%%%%%%%%%%%%%%%%%%%%%%%%%%%%%%%%%%%%%%%%%%%%%%%%%%%%%%%%%%%%%%%%%%%%%%%%%%%%%%%%%%%%%%%%%%%%%%%%%%%%%%%%
Thus even if the initial data have a zero projection to the 2D3C fields the nonlinearity 
at third order will force a 2D3C component to grow to zeroth order amplitude.

%%%%%%%%%%%%%%%%%%%%%%%%%%%%%%%%%%%%%%%%%%%%%%%%%%%%%%%%%%%%%%%%%%%%%%%%%%%%%%%%%%%%%%%%%%%%%%%%%%%%%%%%%%
%%%%%%%%%%%%%%%%%%%%%%%%%%%%%%%%%%%%%%%%%%%%%%%%%%%%%%%%%%%%%%%%%%%%%%%%%%%%%%%%%%%%%%%%%%%%%%%%%%%%%%%%%%
 \subsection{$n>4$,                           %%%%%%%%%%%%%%%%%%%%%%%%%%%%%%%%%%%%%%%%%%%%%%%%%%%%%%%%%%%%
 $\mathcal{O}(\Ref^{-1})<\mathcal{O}(\Rof^{5})$}  %%%%%%%%%%%%%%%%%%%%%%%%%%%%%%%%%%%%%%%%%%%%%%%%%%%%%%%%
%%%%%%%%%%%%%%%%%%%%%%%%%%%%%%%%%%%%%%%%%%%%%%%%%%%%%%%%%%%%%%%%%%%%%%%%%%%%%%%%%%%%%%%%%%%%%%%%%%%%%%%%%%
%%%%%%%%%%%%%%%%%%%%%%%%%%%%%%%%%%%%%%%%%%%%%%%%%%%%%%%%%%%%%%%%%%%%%%%%%%%%%%%%%%%%%%%%%%%%%%%%%%%%%%%%%%
At higher values of $n$ the nonlinearity at the 4th order term has non-zero
projection to the 2D3C fields and viscosity is not present to provide a balance.
Thus no solution can be obtained by the expansion. Physically this would imply that 
even if initial conditions were prepared carefully to avoid any instabilities 
the 2D3C component of the flow would grow  with time to values where 
the original scaling ${\bf u} =\Rof {\bf v}\ll1$ would not be valid and the expansion 
would fail. This also indicates that the {\it quasi}-2D condensates that were shown 
in \ref{fig_1} will show up at this order. This gives a lower bound on the unknown 
exponent $\alpha\ge \frac{1}{5}$ 
since at this order there is direct injection 
of energy in the 2D3C flow.

%%%%%%%%%%%%%%%%%%%%%%%%%%%%%%%%%%%%%%%%%%%%%%%%%%%%%%%%%%%%%%%%%%%%%%%%%%%%%%%%%%%%%%%%%%%%%%%%%%%%%%%%%%%%%%%%%%%%%%%%%%%%%%%
%%%%%%%%%%%%%%%%%%%%%%%%%%%%%%%%%%%%%%%%%%%%%%%%%%%%%%%%%%%%%%%%%%%%%%%%%%%%%%%%%%%%%%%%%%%%%%%%%%%%%%%%%%%%%%%%%%%%%%%%%%%%%%% 
%%%%%%%%%%%%%%%%%%%%%%%%%%%%%%%%%%%%%%%%%%%%%%%%%%%%%%%%%%%%%%%%%%%%%%%%%%%%%%%%%%%%%%%%%%%%%%%%%%%%%%%%%%%%%%%%%%%%%%%%%%%%%%% 
%%%%%%%%%%%%%%%%%%%%%%%%%%%%%%%%%%%%%%%%%%%%%%%%%%%%%%%%%%%%%%%%%%%%%%%%%%%%%%%%%%%%%%%%%%%%%%%%%%%%%%%%%%%%%%%%%%%%%%%%%%%%%%% 
\section{Time evolution and stability in the $Ro_{_F}\to 0$ limit} \label{sec:stab} %%%%%%%%%%%%%%%%%%%%%%%%%%%%%%%%%%%%%%%%%%%
%%%%%%%%%%%%%%%%%%%%%%%%%%%%%%%%%%%%%%%%%%%%%%%%%%%%%%%%%%%%%%%%%%%%%%%%%%%%%%%%%%%%%%%%%%%%%%%%%%%%%%%%%%%%%%%%%%%%%%%%%%%%%%%
%%%%%%%%%%%%%%%%%%%%%%%%%%%%%%%%%%%%%%%%%%%%%%%%%%%%%%%%%%%%%%%%%%%%%%%%%%%%%%%%%%%%%%%%%%%%%%%%%%%%%%%%%%%%%%%%%%%%%%%%%%%%%%% 
%%%%%%%%%%%%%%%%%%%%%%%%%%%%%%%%%%%%%%%%%%%%%%%%%%%%%%%%%%%%%%%%%%%%%%%%%%%%%%%%%%%%%%%%%%%%%%%%%%%%%%%%%%%%%%%%%%%%%%%%%%%%%%% 
%%%%%%%%%%%%%%%%%%%%%%%%%%%%%%%%%%%%%%%%%%%%%%%%%%%%%%%%%%%%%%%%%%%%%%%%%%%%%%%%%%%%%%%%%%%%%%%%%%%%%%%%%%%%%%%%%%%%%%%%%%%%%%%

The results in the previous section only imply the presence of stationary solutions.
Their realizability however is not guarantied since it can be unstable to infinitesimal 
or finite amplitude perturbations that can lead the system to different time dependent solutions.
It is thus also important to investigate the stability of the calculated solutions.
To follow the evolution of the flow we need to keep the time derivative term in eq. \ref{NSR}.
Since in rapidly rotating systems two distinct time scales exist one given by the rotation rate and 
one by the nonlinearity we introduce $\tau=\Rof t$ as the slow eddie turn over time and 
$t' =\Rof^{-1} t$ as the fast time scale. Equation \ref{NSR} then becomes
%%%%%%%%%%%%%%%%%%%%%%%%%%%%%%%%%%%%%%%%%%%%%%%%%%%%%%%%%%%%%%%%%%%%%%%%%%%%%%%%%%%%%%%%%%%%%%%%%%%%%%%%
\begin{equation}                                                                                      %% 
\partial_{t'}{\bf v} + \mathbb{L} \left[ {\bf v } \right] =                                           %%
{\bf f}  + \Rof^2\,\,  \mathbb{P} \left[  -\partial_\tau {\bf v + v\times w}  \right]                 %%
+ \lambda_n \Rof^{2n} \Delta   {\bf v}.               \label{ttNSR}                                   %%
\end{equation}                                                                                        %%
%%%%%%%%%%%%%%%%%%%%%%%%%%%%%%%%%%%%%%%%%%%%%%%%%%%%%%%%%%%%%%%%%%%%%%%%%%%%%%%%%%%%%%%%%%%%%%%%%%%%%%%%
%For reasons of clarity, from now on, we remove the tilde from the fast time scale unless otherwise stated. 
In principle this expansion remains valid only on the timescale $\tau$ and could fail at longer time scales \citep{Newell1969,Babin1996,Chen2005}.
In practical terms however such expansions are expected to capture the long time dynamics (eg saturation amplitude, dissipation rates ect) 
if slower time scale processes in the system do not become important. 
If such a third timescale exist the expansion should be carried out at 
the next order and a new timescale to be defined.
 
In addition to the stationary solution that was found in the previous section, the time dependence
allows for the existence of traveling inertial waves that propagate on the fast time scale $t'$ and
vary in amplitude on the slow dynamical time scale $\tau$. %Similarly 2D3C-flows also evolve
%on slow time scale $\tau$. % the dynamical time scale can co-exist with the inertial waves. 
The presence of the inertial waves (since they are not directly forced) depends on their stability properties.
Similar the 2D3C flows that are absent in the stationary solution for $\Ref\le \mathcal{O}(\Rof^{-3})$ can still 
be present as time dependent solutions evolving on slow time scale $\tau$ 
if an instability is present.
%in the presence  of an instability. 

%%%%%%%%%%%%%%%%%%%%%%%%%%%%%%%%%%%%%%%%%%%%%%%%%%%%%%%%%%%%%%%%%%%%%%%%%%%%%%%%%%%%%%%%%%%%%%%%%%%%%%%%
%%%%%%%%%%%%%%%%%%%%%%%%%%%%%%%%%%%%%%%%%%%%%%%%%%%%%%%%%%%%%%%%%%%%%%%%%%%%%%%%%%%%%%%%%%%%%%%%%%%%%%%%
\subsection{Energy stability } %%%%%%%%%%%%%%%%%%%%%%%%%%%%%%%%%%%%%%%%%%%%%%%%%%%%%%%%%%%%%%%%%%%%%%%%%
%%%%%%%%%%%%%%%%%%%%%%%%%%%%%%%%%%%%%%%%%%%%%%%%%%%%%%%%%%%%%%%%%%%%%%%%%%%%%%%%%%%%%%%%%%%%%%%%%%%%%%%%
%%%%%%%%%%%%%%%%%%%%%%%%%%%%%%%%%%%%%%%%%%%%%%%%%%%%%%%%%%%%%%%%%%%%%%%%%%%%%%%%%%%%%%%%%%%%%%%%%%%%%%%%

The energy stability method provides a sufficient criterion for stability of a stationary flow subject to
perturbations of arbitrary amplitude. It is based on the energy evolution equation of the perturbation.
Denoting the stationary solution as  ${\bf u}_f$  and without taking any asymptotic limit yet we can write 
the velocity field  ${\bf u}$ as the sum of the stationary solution  ${\bf u}_f$ and a 
fluctuating part ${\bf u}_p$. Multiplying  equation \ref{NSR} by the fluctuating part ${\bf u}_p$ and space 
averaging leads to the equation 
%%%%%%%%%%%%%%%%%%%%%%%%%%%%%%%%%%%%%%%%%%%%%%%%%%%%%%%%%%%%%%%%%%%%%%%%%%%%%%%%%%%%%%%%%%%%%%%%%%%%%%%%
\beq                                                                                                  %%
\frac{d}{dt} \left\langle \frac{1}{2} |{\bf u}_p|^2 \right\rangle_{_V} =                              %%
\left\langle {\bf u}_p \cdot \left( \nabla {\bf u}_f \right) {\bf u}_p \right\rangle_{_V}             %%
-\Ref^{-1} \left\langle |\nabla {\bf u}_p|^2 \right\rangle_{_V}                                     . %%
\label{pert_energy}                                                                                   %%
\eeq                                                                                                  %%  
%%%%%%%%%%%%%%%%%%%%%%%%%%%%%%%%%%%%%%%%%%%%%%%%%%%%%%%%%%%%%%%%%%%%%%%%%%%%%%%%%%%%%%%%%%%%%%%%%%%%%%%%
The stationary solution is then called energy stable if the right hand side is negative definite 
for all incompressible fields ${\bf u}_p$. Note that the Coriolis term is absent in eq.\ref{pert_energy} 
and the same equation is obtained if $\Rof=\infty$. However there is dependence on $\Rof$ through the 
functional form of ${\bf u}_f$. %whose first order approximation is given by $\Rof {\bf v}_f$ in \ref{v00}.
Stability of the stationary solution is then guarantied if 
%%%%%%%%%%%%%%%%%%%%%%%%%%%%%%%%%%%%%%%%%%%%%%%%%%%%%%%%%%%%%%%%%%%%%%%%%%%%%%%%%%%%%%%%%%%%%%%%%%%%%%%%%%%%%%%%%%%%%%%%%
\beq                                                                                                                   %% 
\mu(\Ref,\Rof)  =                                                                                                      %%
\sup_{{\bf u}_p} \frac{ \left\langle {\bf u}_p \cdot \left( \nabla {\bf u}_f \right) {\bf u}_p \right\rangle_{_V}      %%
-\Ref^{-1} \left\langle |\nabla {\bf u}_p|^2 \right\rangle_{_V}}{\left\langle |{\bf u}_p|^2 \right\rangle_{_V}}  \le 0 %%
\label{enstbgn}                                                                                                        %%
\eeq                                                                                                                   %% 
%%%%%%%%%%%%%%%%%%%%%%%%%%%%%%%%%%%%%%%%%%%%%%%%%%%%%%%%%%%%%%%%%%%%%%%%%%%%%%%%%%%%%%%%%%%%%%%%%%%%%%%%%%%%%%%%%%%%%%%%%
in which case the energy of the perturbation ${\bf u}_p$ is bounded by
%%%%%%%%%%%%%%%%%%%%%%%%%%%%%%%%%%%%%%%%%%%%%%%%%%%%%%%%%%%%%%%%%%%%%%%%%%%%%%%%%%%%%%%%%%%%%%%%%%%%%%%%%%%%%%%%%%%%%%
\beq                                                                                                                %%
\frac{1}{2}\langle |{\bf u}_p(t)|^2 \rangle_{_V}  \le \frac{1}{2}\langle |{\bf u}_p(0)|^2 \rangle_{_V} e^{2\mu t}   %%
\eeq                                                                                                                %%
%%%%%%%%%%%%%%%%%%%%%%%%%%%%%%%%%%%%%%%%%%%%%%%%%%%%%%%%%%%%%%%%%%%%%%%%%%%%%%%%%%%%%%%%%%%%%%%%%%%%%%%%%%%%%%%%%%%%%% 
by {\it Gronwall's} inequality. The energy stability analysis then reduces to solving the involved Euler-Lagrange
equations and determining $\mu(\Ref,\Rof)$. The energy stability boundary is then determined by the $\mu(\Ref,\Rof)=0$
curve in the parameter space. Here it is not attempted to solve the Euler-Lagrange equations that are beyond the point of this work 
but estimates are given based on simple inequalities and the previous approximations of steady state flow. 

We begin by noting that {\it Poincare's} inequality indicates that 
$\langle |\nabla {\bf u}_p|^2\rangle_{_V} \ge \langle |{\bf u}_p|^2\rangle_{_V}$ 
(where the fact that the flow is defined in a periodic box of non-dimensional size $2\pi$ has been used).
{\it H\"older's} inequality also indicates that 
%%%%%%%%%%%%%%%%%%%%%%%%%%%%%%%%%%%%%%%%%%%%%%%%%%%%%%%%%%%%%%%%%%%%%%%%%%%%%%%%%%%%%%%%%%%%%%%%%%%%%%%%
\beq                                                                                                  %%
\left\langle {\bf u}_p \cdot \left( \nabla {\bf u}_f \right) {\bf u}_p \right\rangle_{_V} \le         %%
\|\nabla {\bf u}_f\|_\infty \langle |{\bf u}_p|^2\rangle_{_V}                                         %%
\eeq                                                                                                  %%
%%%%%%%%%%%%%%%%%%%%%%%%%%%%%%%%%%%%%%%%%%%%%%%%%%%%%%%%%%%%%%%%%%%%%%%%%%%%%%%%%%%%%%%%%%%%%%%%%%%%%%%%
where by $\|\nabla {\bf u}_f\|_\infty$ we indicate the maximum of the modulus of the strain tensor 
$\nabla {\bf u}_f$ over space. These estimates together with the condition \ref{enstbgn} indicate that stability of the flow ${\bf u}_f$
is guarantied if $\|\nabla {\bf u}_f\|_\infty \le \Ref^{-1} $.

This criterion for stability can be calculated using the asymptotic solution we obtained in the previous section.
The expression \ref{v00} gives ${\bf u}_f$ up to an order $\Rof^{-2}$ correction. 
For $\lambda_0=0$ equation \ref{v00} reduces to the same zeroth-order solution as in the $n=1$ and $n=2$ case (see \ref{v01})
thus this choice is valid for all 
%$\lambda_0\ge0$ thus it also includes the
cases for which $\Ref^{-1} \le \mathcal{O}(\Rof^{3})$.
For higher ordering the zeroth order 2D3C field $\Rof^{-1}{\bf v}_{_{2D}}$ should also be included. Substituting thus \ref{v00} 
it is obtained
%%%%%%%%%%%%%%%%%%%%%%%%%%%%%%%%%%%%%%%%%%%%%%%%%%%%%%%%%%%%%%%%%%%%%%%%%%%%%%%%%%%%%%%%%%%%%%%%%%%%%%%
\beq                                                                                                 %%  
\|\nabla {\bf u}_f\|_\infty =   \Rof q \frac{ 4 + 18 \lambda_0 q^2 }{1 + 27 \lambda_0^2 q^4 }        %%
+  \mathcal{O}(\Rof^3)                                                                              .%%
\eeq                                                                                                 %%
%%%%%%%%%%%%%%%%%%%%%%%%%%%%%%%%%%%%%%%%%%%%%%%%%%%%%%%%%%%%%%%%%%%%%%%%%%%%%%%%%%%%%%%%%%%%%%%%%%%%%%%  
Substituting in $\|\nabla {\bf u}_f\|_\infty \le \Ref^{-1} $ the stability of the flow 
calculated for the approximate solution \ref{v00} is guarantied if
%%%%%%%%%%%%%%%%%%%%%%%%%%%%%%%%%%%%%%%%%%%%%%%%%%%%%%%%%%%%%%%%%%%%%%%%%%%%%%%%%%%%%%%%%%%%%%%%%%%%%%%%
\beq                                                                                                  %%
\Rof^{-2} - 4q\Ref \Rof^{-1} -(18 q^3- 27q^4\Ref^{-2} )  \ge 0                                        %%
\eeq                                                                                                  %%
%%%%%%%%%%%%%%%%%%%%%%%%%%%%%%%%%%%%%%%%%%%%%%%%%%%%%%%%%%%%%%%%%%%%%%%%%%%%%%%%%%%%%%%%%%%%%%%%%%%%%%%%
This leads to the following estimates that guarantee stability.
The flow is stable if
%%%%%%%%%%%%%%%%%%%%%%%%%%%%%%%%%%%%%%%%%%%%%%%%%%%%%%%%%%%%%%%%%%%%%%%%%%%%%%%%%%%%%%%%%%%%%%%%%%%%%%%%%%%%%
\beq                                                                                                       %%
\begin{array}{rl}                                                                                          %%
\Rof^{-1} \ge&  2q\Ref + q\sqrt{4\Ref^2 + 18 q - 27q^2\Ref^{-2}}   \qquad  \mathrm{or}          \\         %%
\Rof^{-1} \le&  2q\Ref - q\sqrt{4\Ref^2 + 18 q - 27q^2\Ref^{-2}}                                           %%
\end{array}   \label{enstb1}                                                                               %%
\eeq                                                                                                       %%
%%%%%%%%%%%%%%%%%%%%%%%%%%%%%%%%%%%%%%%%%%%%%%%%%%%%%%%%%%%%%%%%%%%%%%%%%%%%%%%%%%%%%%%%%%%%%%%%%%%%%%%%%%%%%
and for any value of $\Rof$ stability is guarantied if
%%%%%%%%%%%%%%%%%%%%%%%%%%%%%%%%%%%%%%%%%%%%%%%%%%%%%%%%%%%%%%%%%%%%%%%%%%%%%%%%%%%%%%%%%%%%%%%%%%%%%%%%%%%%%
\beq                                                                                                       %%
\Ref \le \frac{\sqrt{q}}{2} \sqrt{ (\sqrt{189}-9)  }\simeq 1.08\dots \, \sqrt{q}.           \label{enstb2} %%
\eeq                                                                                                       %%
%%%%%%%%%%%%%%%%%%%%%%%%%%%%%%%%%%%%%%%%%%%%%%%%%%%%%%%%%%%%%%%%%%%%%%%%%%%%%%%%%%%%%%%%%%%%%%%%%%%%%%%%%%%%%
%%%%%%%%%%%%%%%%%%%%%%%%%%%%%%%%%%%%%%%%%%%%%%%%%%%%%%%%%%%%%%%%%%%%%%%%%%%%%%%%%%%%%%%%%%%%%%%%%%%%%%%%%%%%%%
%\beq                                                                                                       %%
%\Ref^{-1} \le \sqrt{\frac{1}{4}(\sqrt{189}-9)q  }\simeq 1.08\dots \, \sqrt{q}              \label{enstb2}  %%
%\eeq                                                                                                       %%
%%%%%%%%%%%%%%%%%%%%%%%%%%%%%%%%%%%%%%%%%%%%%%%%%%%%%%%%%%%%%%%%%%%%%%%%%%%%%%%%%%%%%%%%%%%%%%%%%%%%%%%%%%%%%%
These conservatives estimates of the energy stability boundaries are shown in figure \ref{fig_1} with a solid grey
(orange online) line. 
For $\Rof \gg 1$ the estimate for stability is given by $\Ref \le \sqrt{3q/2}$ while the numerical simulations 
indicate a critical value for $\Ref=14.0$ which is considerable larger than the value obtained in eq. \ref{enstb2}.
For $\Rof\ll1$ equation \ref{enstb1} leads to the condition for stability $\Ref \le (4q\Rof)^{-1}$ 
while from the numerical simulations stability is shown in the same limit for q=2 if $\Ref \lesssim 10 \Rof^{-1}$.
%%%%%%
The analytical estimates for the stability boundaries \ref{enstb1} and \ref{enstb2} are thus very conservative
which is expected for the rough estimates used here. These estimates however do not aim at obtaining the exact values
but rather to give an understanding of the shape of the stability boundary that they clearly capture. 
They also prove  that for any value of $\Ref$, no mater how large, at sufficiently large $\Rof$ 
the flow will re-laminarize.

%%%%%%%%%%%%%%%%%%%%%%%%%%%%%%%%%%%%%%%%%%%%%%%%%%%%%%%%%%%%%%%%%%%%%%%%%%%%%%%%%%%%%%%%%%%%%%%%%%%%%%%%
%%%%%%%%%%%%%%%%%%%%%%%%%%%%%%%%%%%%%%%%%%%%%%%%%%%%%%%%%%%%%%%%%%%%%%%%%%%%%%%%%%%%%%%%%%%%%%%%%%%%%%%%
\subsection{Linear stability} %%%%%%%%%%%%%%%%%%%%%%%%%%%%%%%%%%%%%%%%%%%%%%%%%%%%%%%%%%%%%%%%%%%%%%%%%
%%%%%%%%%%%%%%%%%%%%%%%%%%%%%%%%%%%%%%%%%%%%%%%%%%%%%%%%%%%%%%%%%%%%%%%%%%%%%%%%%%%%%%%%%%%%%%%%%%%%%%%%
%%%%%%%%%%%%%%%%%%%%%%%%%%%%%%%%%%%%%%%%%%%%%%%%%%%%%%%%%%%%%%%%%%%%%%%%%%%%%%%%%%%%%%%%%%%%%%%%%%%%%%%%

%Since energy stability is established at order $\Ref=\mathcal{O}(\Rof^{-1})$ we examine the linear stability at a higher order. 
For $\Rof\to 0$ the energy stability indicates that stability is determined at $\Ref=\mathcal{O}(\Rof^{-1})$.
We thus begin the investigation of linear stability for this ordering.
Linear stability investigates the evolution of infinitesimal perturbations. 
Contrary to energy stability, linear stability does not guaranties stability %to perturbations of all amplitudes 
but linear instability guaranties instability. The two methods are thus complementary.
The linear equation for an infinitesimal fluctuation   
${\bf v}_p$ on the basic stationary state ${\bf v}_f$ (given at this order by eq. \ref{v01}) reads
%%%%%%%%%%%%%%%%%%%%%%%%%%%%%%%%%%%%%%%%%%%%%%%%%%%%%%%%%%%%%%%%%%%%%%%%%%%%%%%%%%%%%%%%%%%%%%%%%%%%%%%%
\begin{equation}                                                                                      %% 
\partial_{t'}{\bf v}_p + \mathbb{L} \left[ {\bf v}_p \right] =                                        %%
\Rof^2 \,\, \left( \mathbb{P} \left[  -                                                               %%
\partial_\tau {\bf v}_p +{\bf v}_p \times {\bf w}_f + {\bf v}_f \times {\bf w}_p   \right]            %%
+ \lambda_1  \Delta   {\bf v}_p.                                           \right)                    %%
\end{equation}                                                                                        %%
%%%%%%%%%%%%%%%%%%%%%%%%%%%%%%%%%%%%%%%%%%%%%%%%%%%%%%%%%%%%%%%%%%%%%%%%%%%%%%%%%%%%%%%%%%%%%%%%%%%%%%%%

Unlike in the energy stability method, in  linear stability the effect of rotation is not removed and
the smallness of $\Rof$ can be used to find the growth rate of the perturbations in an asymptotic way.
Like before we write ${\bf v}_p= {\bf v}_p^{(0)} + \Rof^2{\bf v}_p^{(1)} +\dots$. At zeroth order 
we obtain  
%%%%%%%%%%%%%%%%%%%%%%%%%%%%%%%%%%%%%%%%%%%%%%%%%%%%%%%%%%%%%%%%%%%%%%%%%%%%%%%%%%%%%%%%%%%%%%%%%%%%%%%%
\begin{equation}                                                                                      %%
\partial_{t'} {\bf v}_p^{(0)} + \mathbb{L}[{\bf v}_p^{(0)}] = 0                                       %%
\label{lnr}                                                                                           %%
\end{equation}                                                                                        %%
%%%%%%%%%%%%%%%%%%%%%%%%%%%%%%%%%%%%%%%%%%%%%%%%%%%%%%%%%%%%%%%%%%%%%%%%%%%%%%%%%%%%%%%%%%%%%%%%%%%%%%%%
The general solution of eq.\ref{lnr} is then given by
%%%%%%%%%%%%%%%%%%%%%%%%%%%%%%%%%%%%%%%%%%%%%%%%%%%%%%%%%%%%%%%%%%%%%%%%%%%%%%%%%%%%%%%%%%%%%%%%%%%%%%%%
\begin{equation}                                                                                      %% 
{\bf v}_p^{(0)} = {\bf v}_{_{2D}}^{(0)}(\tau,x,y) +{\bf v}_w^{(0)}(\tau,t',x,y,z)                     %%
\end{equation}                                                                                        %%
%%%%%%%%%%%%%%%%%%%%%%%%%%%%%%%%%%%%%%%%%%%%%%%%%%%%%%%%%%%%%%%%%%%%%%%%%%%%%%%%%%%%%%%%%%%%%%%%%%%%%%%%
${\bf v}_{_{2D}}$ is a 2D3C velocity field that depends only on the ($x,y$) coordinates and on the slow dynamical time $\tau$.
%\[ {\bf u}_{_{2D}}^0= [ \psi_y(x,y), -\psi_x(x,y),\phi(x,y)] \]
${\bf v}_w^{(0)}$ are inertial waves whose amplitudes also vary on the long time scale.
The general expression for the inertial waves in triple periodic boxes is given by:
\begin{eqnarray}                                                                                              %% 
{\bf v}_w^{(0)} = \sum_{s,{\bf k},k_z>0} &                                                                    %%
\left[ H_{\bf k}^{s}(\tau)   {\bf h}^{s}_{\bf k} e^{ (i{\bf k  x} + i\omega^s_{\bf k}t') }                    %%
%+      H_{\bf k}^-(\tau)   {\bf h}^-_{\bf k} e^{ (i{\bf k  x} + i\omega^-_{\bf k}t) }                        %%  
                                                             + c.c. \right]  \label{Wexpa}                    %%
\end{eqnarray}                                                                                                %%
%%%%%%%%%%%%%%%%%%%%%%%%%%%%%%%%%%%%%%%%%%%%%%%%%%%%%%%%%%%%%%%%%%%%%%%%%%%%%%%%%%%%%%%%%%%%%%%%%%%%%%%%%%%%%%%%
where $s=\pm1$, $c.c.$ stands for complex conjugate and represents the wave numbers with negative $k_z$.
${\bf h}^s_{\bf k}$ are the helical basis in Fourier space 
described in %\cite{Lesieur1972,Cambon1989,Waleffe1992,Waleffe1993}.
\cite{Cambon1989,Waleffe1992,Waleffe1993}. ${\bf h}^s_{\bf k}$ are eigenfunctions of the curl operator
with $i{\bf k} \times {\bf h}^s_{\bf k} = s|{\bf k}| {\bf h}^s_{\bf k}$ and thus the index $s$ determines the
sign of the helicity of the mode. According to eq. \ref{Lsmpl} then
\[
\mathbb{L}[{\bf h^{s}_k } e^{i{\bf kx}}]= -i \omega_{\bf k}^{s} {\bf h^{s}_k }e^{i{\bf kx}} \qquad \qquad \mathrm{ with} 
\qquad \qquad \omega_{\bf k}^s = s \frac{k_z}{|{\bf k}|  }.
\]
On the periodic box used here for any wavevector ${\bf k}=[k_x,k_y,k_z]$ with $k_\perp =\sqrt{k_x^2+k_y^2} \ne 0$,
${\bf h}^s_{\bf k}$ is defined as
\beq
h_{\bf k}^s = 
\frac{1}{\sqrt{2}\,|k|k_\perp }
\left[
\begin{array}{r}
 k_xk_z  \\
 k_yk_z  \\
 -k_\perp^2
\end{array}
\right]
+ 
\frac{is}{\sqrt{2}\,k_\perp}
\left[
\begin{array}{r}
-k_y  \\
\, k_x  \\
  0 \,\,
\end{array}
\right], \,\, \mathrm{while}\,\,
h_{\bf k}^s = 
\frac{1}{\sqrt{2}}
\left[
\begin{array}{r}
  1  \\
 is  \\
 0
\end{array}
\right]
\eeq
for the $k_\perp=0$ case.
% $h^s_{\bf k}$ becomes
%\beq
%h_{\bf k}^s = 
%\frac{1}{\sqrt{2}}
%\left[
%\begin{array}{r}
%  1  \\
% is  \\
% 0
%\end{array}
%\right]
%\eeq

On the fast time-scale no instability exists since rotation cannot transfer energy to the perturbation field.
The slow time scale evolution of the two fields ${\bf v}_{_{2D}}^{(0)}$ and ${\bf v}_w^{(0)}$ is 
obtained by a solvability condition at the next order
%%%%%%%%%%%%%%%%%%%%%%%%%%%%%%%%%%%%%%%%%%%%%%%%%%%%%%%%%%%%%%%%%%%%%%%%%%%%%%%%%%%%%%%%%%%%%%%%%%%%%%%%
\beq                                                                                                  %%
\partial_{t'} {\bf v}^{(1)}  + \mathbb{L} [{\bf v}^{(1)}]  =                                          %%
 -\partial_\tau {\bf v}_p^{(0)} +                                                                     %%
  \mathbb{P}[ {\bf v}_p^{(0)} \times {\bf w}_f + {\bf v}_f \times {\bf w}_p^{(0)}]                    %%
+ \lambda_1 \Delta {\bf v }_p^{(0)}                 .                                               %%
\label{ord1}                                                                                          %% 
\eeq                                                                                                  %%
%%%%%%%%%%%%%%%%%%%%%%%%%%%%%%%%%%%%%%%%%%%%%%%%%%%%%%%%%%%%%%%%%%%%%%%%%%%%%%%%%%%%%%%%%%%%%%%%%%%%%%%%
At this point we need to define the short time average $\langle g \rangle_\tau$ as:
%%%%%%%%%%%%%%%%%%%%%%%%%%%%%%%%%%%%%%%%%%%%%%%%%%%%%%%%%%%%%%%%%%%%%%%%%%%%%%%%%%%%%%%%%%%%%%%%%%%%%%%%%%%%%%%%%%%%%%%
%\begin{equation}                                                                                                    %% 
% \langle g(t) \rangle_\tau \, \equiv \, \frac{\Rof}{2} \int_{-1/\Rof}^{+1/\Rof} g\,\, dt.                           %%
%\end{equation}                                                                                                      %%
 \begin{equation}                                                                                                    %% 
  \langle g(t) \rangle_\tau \, \equiv \, \frac{\Rof^{\sigma}}{2} \int_{-1/\Rof^\sigma}^{+1/\Rof^\sigma} g\,\, dt.    %%
 \end{equation}                                                                                                      %%
%%%%%%%%%%%%%%%%%%%%%%%%%%%%%%%%%%%%%%%%%%%%%%%%%%%%%%%%%%%%%%%%%%%%%%%%%%%%%%%%%%%%%%%%%%%%%%%%%%%%%%%%%%%%%%%%%%%%%%%
with $-1\le\sigma<1$ (ie the time average is over a timescale much smaller than $\tau$ and much bigger than $t'$). 
With this definition the short time average of a fast oscillating function $g(t)=e^{i t'}=e^{i t/\Rof}$ becomes
$\langle  g \rangle_{\tau} = \mathcal{O}(\Rof^{1+\sigma})\ll 1$ and the short time average of a slow oscillating function
$g(t)=e^{i \tau}=e^{i t \Rof}$ becomes the identity $\langle  g \rangle_{\tau} = 1$. Note that the result is independent 
of the choice of $\sigma$.

To obtain then the evolution of ${\bf v}_{_{2D}}^{(0)}$ we perform a short time average and an average over the $z$-direction. 
The left hand side, and the advection term on the right averages to zero. The vorticity advection term also averages to zero 
because the vertical average will eliminate all Fourier modes of ${\bf v}_p$ with $k_z$ different from that of the forcing 
$k_z\ne q$. 
The remaining terms oscillate with frequency $\pm k_z/|{\bf k}|\ne 0$ and will be eliminated by the time average.
We are thus left with:
%%%%%%%%%%%%%%%%%%%%%%%%%%%%%%%%%%%%%%%%%%%%%%%%%%%%%%%%%%%%%%%%%%%%%%%%%%%%%%%%%%%%%%%%%%%%%%%%%%%%%%%
\beq                                                                                                 %%
\partial_\tau {\bf v}_{_{2D}}^{(0)} = \lambda_1 \Delta {\bf v}_{_{2D}}^{(0)}                       %%
\eeq                                                                                                 %%
%%%%%%%%%%%%%%%%%%%%%%%%%%%%%%%%%%%%%%%%%%%%%%%%%%%%%%%%%%%%%%%%%%%%%%%%%%%%%%%%%%%%%%%%%%%%%%%%%%%%%%%
which is a diffusion equation and thus any 2D3C perturbation will decay. Therefore at this order  
the stationary flow is linearly stable to two dimensional perturbations.

To obtain the evolution of ${\bf v}_w^{(0)}$ we multiply equation \ref{ord1} with 
              ${\bf h}^{-s_k}_{\bf k} e^{-i({\bf kx}+\omega_{\bf k}t')}$ 
space average and short time average over to obtain the evolution equation for complex amplitude $H_{\bf k}^{s_{ k}}$:
%%%%%%%%%%%%%%%%%%%%%%%%%%%%%%%%%%%%%%%%%%%%%%%%%%%%%%%%%%%%%%%%%%%%%%%%%%%%%%%%%%%%%%%%%%%%%%%%%%%%%%%%%%%%%%%%%%%%%%%%%%%
\beq                                                                                                                     %%
\partial_\tau H_{\bf k}^{s_k} (\tau) =  2                                                                                %% 
         \sum_{\tiny \begin{array}{c} {s_q,s_p,\bf q,p} \\                                                               %%
          {\bf q+p=k}\end{array} } \left\langle e^{ i (\omega_p-\omega_k) t'}                                            %% 
         \right\rangle_{\tau}                                                                                            %% 
         \mathcal{C}^{s_k,s_q,s_p}_{\bf k,q,p}  V_{\bf q}^{s_q} H_{\bf p}^{s_p} (\tau)                                   %%
                                   - \lambda_1 |{\bf k}|^2 H_{\bf k}^{s_k}(\tau)  .          \label{wteq}                %% 
\eeq                                                                                                                     %%
%%%%%%%%%%%%%%%%%%%%%%%%%%%%%%%%%%%%%%%%%%%%%%%%%%%%%%%%%%%%%%%%%%%%%%%%%%%%%%%%%%%%%%%%%%%%%%%%%%%%%%%%%%%%%%%%%%%%%%%%%%%
$ V_{\bf q}^{s_q} $ is the complex amplitude of the helical modes of
the stationary flow \ref{v01}. 
It is defined as $V_{\bf q}^{s_q}= \langle e^{-i{\bf q x}}  {\bf h}^{-s_q}_{\bf q} {\bf v}_f\rangle_{_V} $
where ${\bf q}=(\pm q,\pm q,\pm q)$ is one of the eight forcing wavenumbers.
The summation is over all wavenumber ${\bf p}\in \mathbb{Z}^3$ and ${\bf q}$
such that ${\bf k=q+p}$. 
The coupling tensor  $\mathcal{C}^{s_k,s_q,s_p}_{\bf k,p,q}$ is given by 
%%%%%%%%%%%%%%%%%%%%%%%%%%%%%%%%%%%%%%%%%%%%%%%%%%%%%%%%%%%%%%%%%%%%%%%%%%%%%%%%%%%%%%%%%%%%%%%%%%%%%%%%%%%%%%%%%%
\beq                                                                                                            %%
\mathcal{C}^{s_k,s_q,s_p}_{\bf k,q,p} =       \frac{1}{2}(s_q|{\bf q}|- s_p|{\bf p}| )                          %%
%\left\langle                                                                                                   %%
\bf  [{\bf h}_{\bf k}^{-s_k} \cdot({\bf h}_{\bf q}^{s_q} \times {\bf h}_{\bf p}^{s_p})].                        %%
%\right\rangle                                                                                                  %%
\eeq                                                                                                            %%
%%%%%%%%%%%%%%%%%%%%%%%%%%%%%%%%%%%%%%%%%%%%%%%%%%%%%%%%%%%%%%%%%%%%%%%%%%%%%%%%%%%%%%%%%%%%%%%%%%%%%%%%%%%%%%%%%%
The short time average leads to different possibilities that we discuss here in some detail.
First if $|\omega_p-\omega_k| = \mathcal{O}(1)$ then the averaged term $\langle e^{ i (\omega_p-\omega_k) t' }\rangle_{\tau}$,
becomes of smaller order and thus can be neglected. We refer to these terms as non-resonant terms.
If $|\omega_p-\omega_k| =0$ then the averaged term leads to an order one contribution $\langle e^{ i (\omega_p-\omega_k) t' }\rangle_{\tau}=1$ 
and needs to be taken into account. 
%%%%%%%%%%%%%%%%%%%%%%%%%%%%%%%%%%%%%%%

%%%%%%%%%%%%%%%%%%%%%%%%%%%%%%%%%%%%%%%
The third intermediate option is when the frequency difference is very small but non-zero. 
For $\Delta \omega =|\omega_p-\omega_k| = \mathcal{O}(\Rof^2)$ then the average 
$\langle e^{ i (\omega_p-\omega_k) t' }\rangle_{\tau}$
still remains an order one quantity. Here two conflicting arguments can be put forward for the role of these quasi resonances.
On the one hand since the value of $\Delta \omega$ for any two wavenumbers ${\bf p,q}$ is fixed and independent of $\Rof$ 
in the limit $\Rof\to 0$ the {\it quasi}-resonances can be neglected. On the other hand from the infinity of possible pairs ${(\bf k,p)}\in \mathbb{Z}^6$
one can always find a a pair of wavenumbers ${\bf k,p}$  such that $\Delta \omega= \mathcal{O}(\Rof^2)$  for any value of $\Rof$.
Thus there are always 
wave numbers for which quasi resonances are important. The resolution of course comes from the fact that not all wave numbers
are available. Any finite value of $\lambda_1$ will introduce a cut-off wavenumber $k_{_\lambda}$ such that all wavenumbers
${\bf k}$ with $|{\bf k}|>k_\lambda$ will be damped by viscosity. Thus pairs from
all remaining wavenumbers $|{\bf k}|\le k_\lambda$
will have finite {\it quasi}-resonance frequency $\Delta \omega$ and in the limit  $\Rof\to 0$ can be neglected.
Note the importance in this argument that $\lambda_1$ is finite. If the viscous term appeared at next order 
({\it ie} $\Ref=\mathcal{O}(\Rof)^{-3}$) this argument would break down and quasi resonances that appear for 
large enough $|{\bf k}|$ would need to be taken in to account. 

The cut-off wavenumber $k_{_\lambda}$ can be estimated by the balance of the shear of the basic flow 
$\langle (\nabla {\bf v}_f)^2\rangle_{_V}^{1/2} \propto q$  with the viscous damping $\lambda_1 k_{_\lambda}^2$ that leads to the estimate 
\beq
{k_{_\lambda}} \propto  \sqrt{\frac{q}{ \lambda_1 }} \propto \sqrt{\Ref\Rof q} . 
\eeq
Thus for small  values of $\Rof$ {\it quasi}-resonances can be neglected if 
$\Delta \omega \gg \mathcal{O}(\Rof^2)$ for all Fourier wavenumbers in a sphere of radius $k_{_\lambda}$.
In what follows we give estimates for $\Delta \omega$ and the density of the quasi resonances as well as solutions for the exact resonances
to estimate the validity of these assumptions in our simulations.
%%%%%%%%%%%%%%%%%%%%%%%%%%%%%%%%%%%%%%%
%%%%%%%%%%%%%%%%%%%%%%%%%%%%%%%%%%%%%%%

\begin{figure*}                                                                       %%
%\begin{center}                                                                       %%
\centerline{                                                                          %%
 \includegraphics[width=6cm]{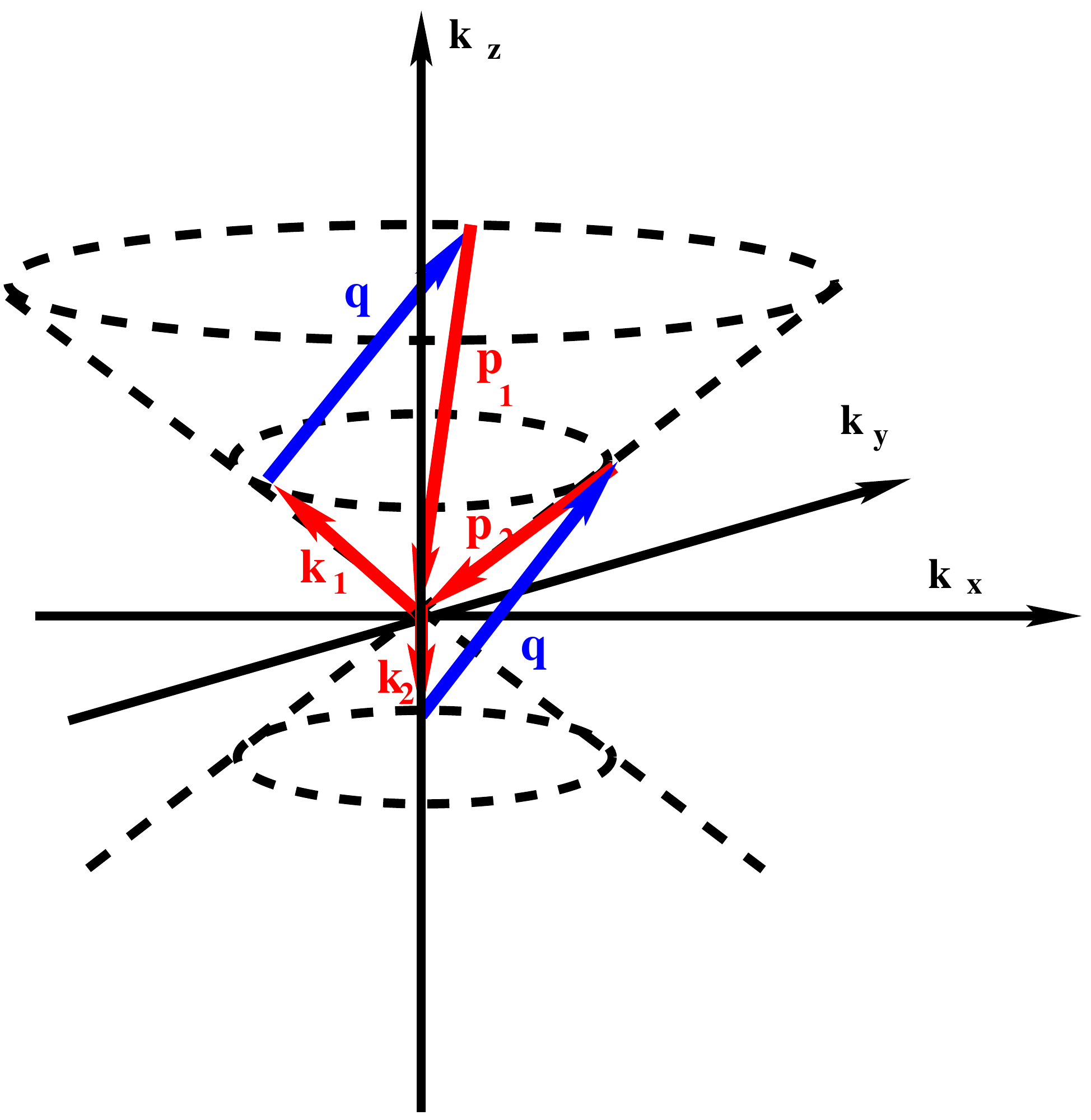}                                     %%
 \includegraphics[width=6cm]{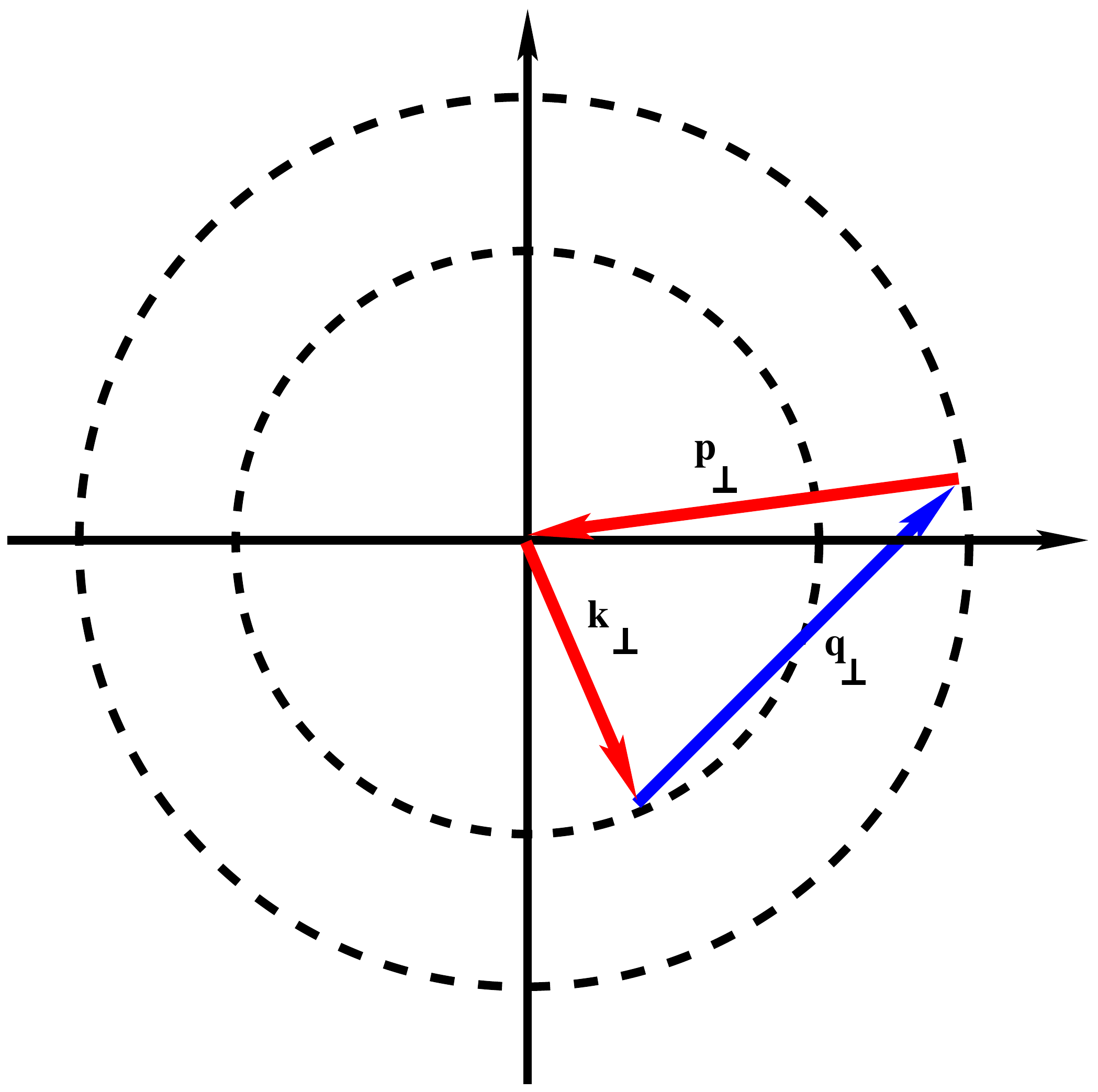}                                    %%
}                                                                                     %%
\caption{Left panel shows two possible couplings of the wave vectors                  %%
        ${\bf k,p}$ (red online) with the forcing wave number ${\bf q}$ (blue online).%%
         Right panel shows the $(k_x,k_y)$-plane where two circles of radius          %%  
         $|{\bf p_\perp}|$ and $|{\bf q_\perp}|$ have been drawn.                     %% 
         In order for the forcing wave vector ${\bf q}_\perp=(q,q,0)$ to be able to   %% 
         form a triangle with two wave numbers ${\bf p_\perp}$ and ${\bf k_\perp}$    %%
         its two ends need to be placed at the edges of the two circles as shown by   %%
         the blue       arrow.  Thus ${\bf q_\perp}$ needs to be longer than the      %%
         distance                                                                     %%
         $|p_\perp-k_\perp|$ and shorter than the distance $|p_\perp+k_\perp|$ }      %% 
\label{fig_3}                                                                         %%
\end{figure*}

For exact resonances of eq.\ref{wteq} the following resonance conditions need to be satisfied for the wave numbers 
$ {\bf k,p,q}$: 
%%%%%%%%%%%%%%%%%%%%%%%%%%%%%%%%%%%%%%%%%%%%%%%%%%%%%%%%%%%%%%%%%%%%%%%%%%%%%%%%%%%%%%%%%%%%%%%%%%%%%%%
\beq                                                                                                 %%
(a)  \quad {\bf k = p+q } \qquad \mathrm{and} \quad                                                  %%
(b) \qquad \frac{k_z}{|\bf k|}= s\frac{p_z}{|\bf p|}.                                                %%
\label{exact_reslin}                                                                                    %%
\eeq                                                                                                 %%
%%%%%%%%%%%%%%%%%%%%%%%%%%%%%%%%%%%%%%%%%%%%%%%%%%%%%%%%%%%%%%%%%%%%%%%%%%%%%%%%%%%%%%%%%%%%%%%%%%%%%%%
where $\bf q$ is one of the 8 forcing wave vectors $(\pm q,\pm q, \pm q)$
that are located at the corners of a $2q$-length cube centered at the origin.
The sign $s=s_ps_k=\pm1$ in \ref{exact_reslin}b indicates whether the coupling is between waves traveling in the same 
direction ($s=+1$) or opposite direction ($s=-1$). Waves traveling in the same direction implies that the 
two wavenumbers lie in the same cone ${\bf k,p}$ of opening angle $\theta$ such that 
$\tan(\theta)=k_z/|k|=p_z/|p|=\omega$ while for opposite traveling waves the coupling is between 
wavenumbers ${\bf k,p}$ that are on opposite cones.  The left panel of figure \ref{fig_3} indicates two such couplings. 
%%%%%%%%%%%%%%%%%%%%%%%%%%%%%%%%%%%%%%%%%%%%%%%%%%%%%%%%%%%%%%%%%%%%%%%%%%%%%%%%%%%%%%%%
                                                                         %%
%%%%%%%%%%%%%%%%%%%%%%%%%%%%%%%%%%%%%%%%%%%%%%%%%%%%%%%%%%%%%%%%%%%%%%%%%%%%%%%%%%%%%%%%
In the right panel of figure \ref{fig_3}
the projection of the three vectors ${\bf k,q,p}$ in the plane perpendicular to
the rotation axis is shown.  It is obvious that the three projected vectors 
${\bf k_\perp,q_\perp,p_\perp}$ can satisfy eq.\ref{exact_reslin}$a$ if the following triangle inequalities hold 
${\bf \big{|}  |p_\perp|-|k_\perp| \big{|} \le |q_\perp| \le|k_\perp|+|p_\perp|} $. 
The extreme cases being when ${\bf k_\perp,p_\perp}$
are parallel or anti-parallel to  ${\bf q_\perp}$. 
Using  eq.\ref{exact_reslin}b we then end up with the following bounds for the allowed wavenumbers lying on the same cone:
%%%%%%%%%%%%%%%%%%%%%%%%%%%%%%%%%%%%%%%%%%%%%%%%%%%%%%%%%%%%%%%%%%%%%%%%%%%%%%%%%%%%%%%%%%%%%%%%%%%%%%%
\beq                                                                                                 %%  
  \frac{\sqrt{2} q|k_z|}{ q + 2 |k_z|} \le k_\perp  \le {\sqrt{2}} |k_z|,                            %%
\label{bound_res1}                                                                                   %%
\eeq                                                                                                 %%
%%%%%%%%%%%%%%%%%%%%%%%%%%%%%%%%%%%%%%%%%%%%%%%%%%%%%%%%%%%%%%%%%%%%%%%%%%%%%%%%%%%%%%%%%%%%%%%%%%%%%%%
while for opposite cones the bounds are:
%%%%%%%%%%%%%%%%%%%%%%%%%%%%%%%%%%%%%%%%%%%%%%%%%%%%%%%%%%%%%%%%%%%%%%%%%%%%%%%%%%%%%%%%%%%%%%%%%%%%%%%
\beq                                                                                                 %%
 \sqrt{2}|k_z| \le   k_\perp   \quad \mathrm{and} \quad                                              %%
\frac{ qk_\perp}{ 2k_\perp  + \sqrt{2} q} \le |k_z|  \le \frac{ qk_\perp}{ 2k_\perp  - \sqrt{2} q}   %%
\label{bound_res2}                                                                                   %%
\eeq                                                                                                 %%
%%%%%%%%%%%%%%%%%%%%%%%%%%%%%%%%%%%%%%%%%%%%%%%%%%%%%%%%%%%%%%%%%%%%%%%%%%%%%%%%%%%%%%%%%%%%%%%%%%%%%%%

\begin{figure*}                                                                                                             %%
%\begin{center}                                                                                                             %%
\centerline{                                                                                                                %%
 \includegraphics[width=8cm]{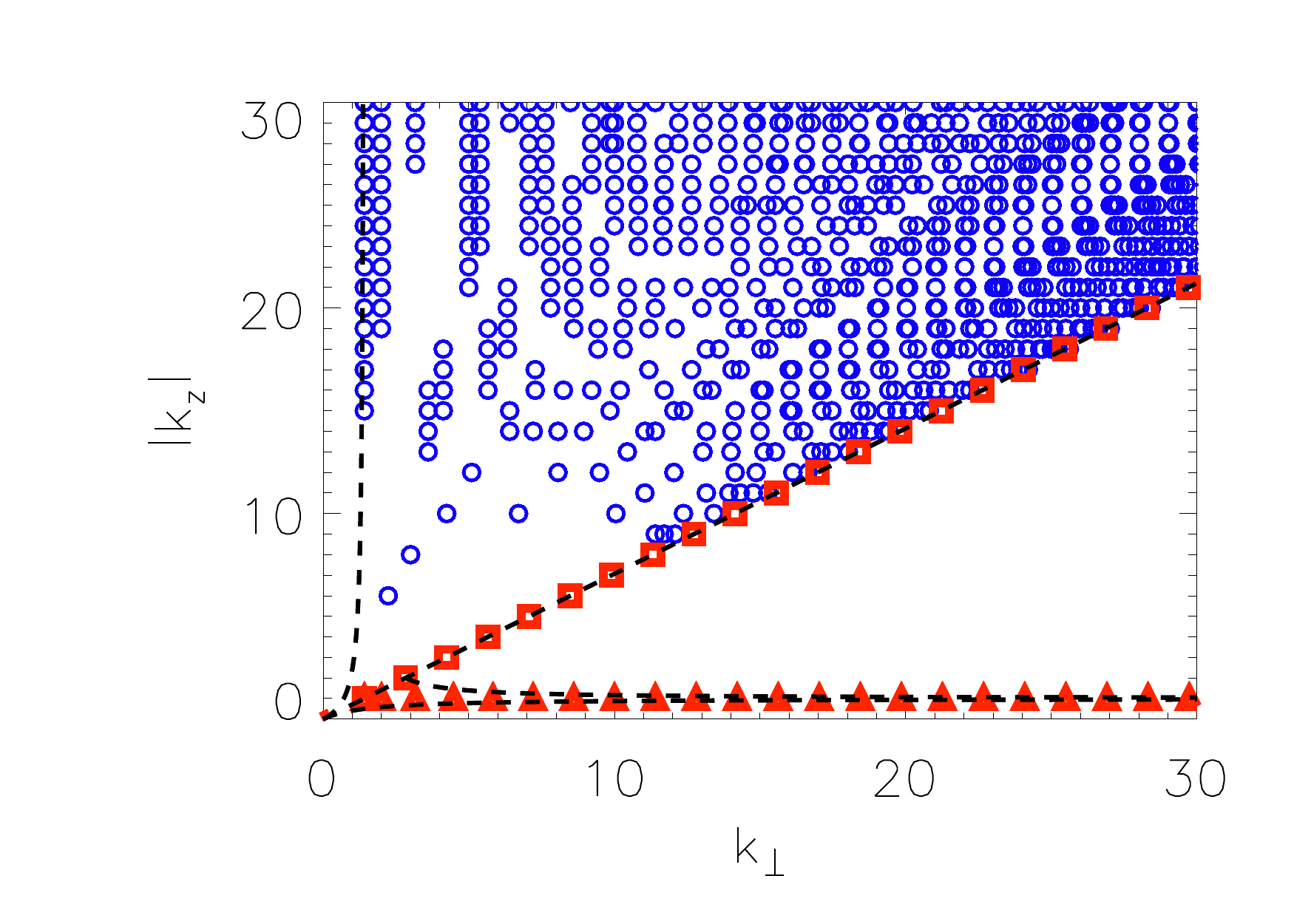}                                                                           %%
 \includegraphics[width=8cm]{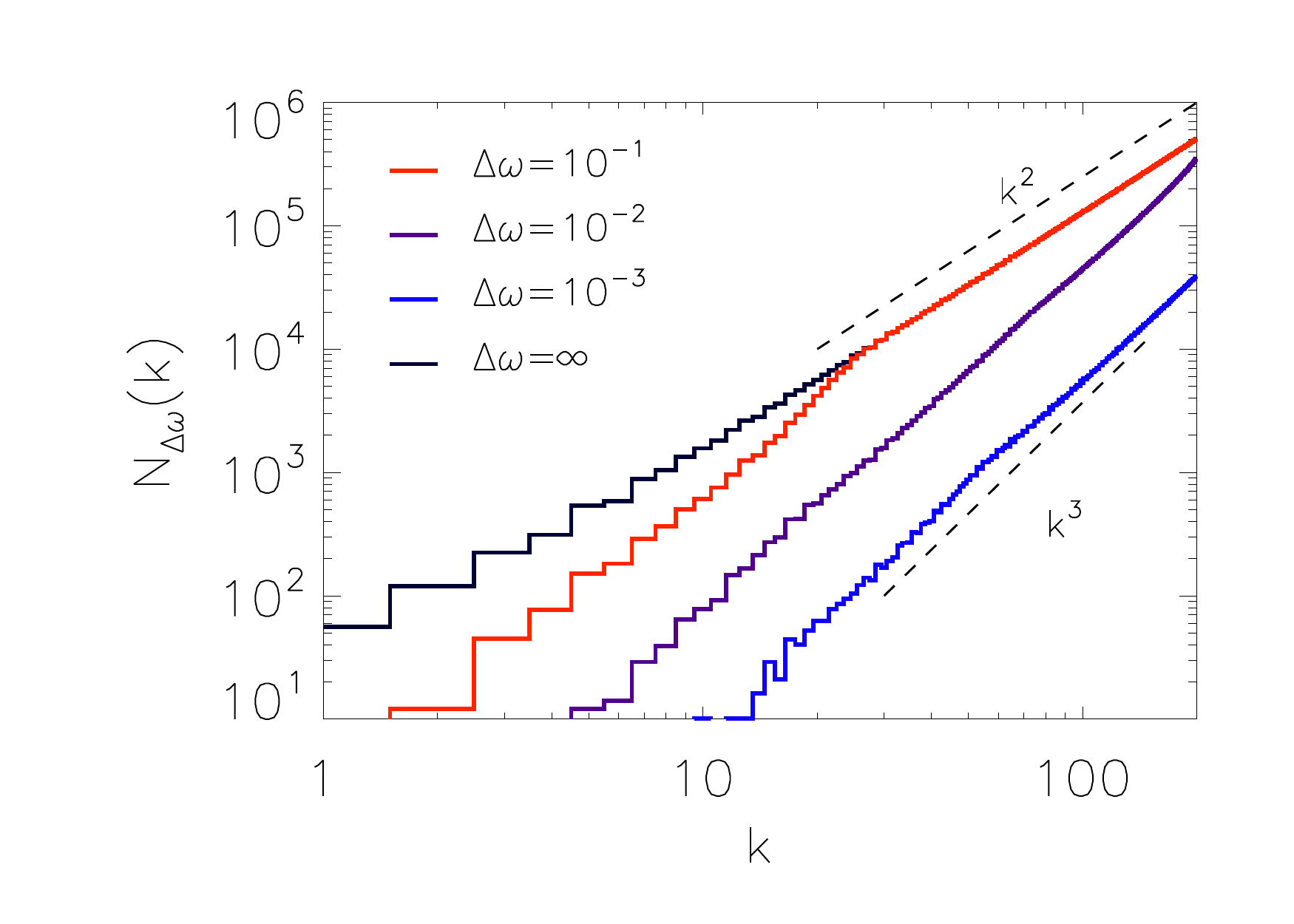}                                                                          %%
}                                                                                                                           %%
\caption{Left panel: T he location of the exact resonances ${\bf k} = m {\bf q}/q$ by red squares, the exact resonances     %%
         (eq.\ref{exact2}) by red triangles, {\it quasi}-resonances by blue circles for a frequency tolerance               %%
         $\Delta \omega=10^{-3}$ in the plane $k_z,k_\perp=\sqrt{k_x^2+k_y^2}$. The dashed lines indicate the bounds        %%
         \ref{bound_res1} and  \ref{bound_res2}. Right panel: Number of modes $N_{\Delta \omega}(k)$ that satisfy the       %%
         quasi-resonance conditions up to a a frequency ambiguity $\Delta \omega$ on the spherical shell                    %% 
         $k <  |{\bf k}| \le  k+1$. }                                                                                       %%
%\end{center}                                                                                                               %%
\label{fig_4}                                                                                                               %%
\end{figure*}

The triangle inequalities however do not guarantee the existence of resonances because the wave numbers ${\bf,k,q,p}$ are discrete
and the equations \ref{exact_reslin} need to be solved on a discrete lattice. General solutions of such problems 
however are nontrivial and have been solved but for very few cases \citep{Bustamante2013}. 
However there are two simple solutions that can be found by simple inspection. 
The first is simply when ${\bf p} = m {\bf q}/q$ for any integer $m\in \mathbb{Z}$.
Then ${\bf k }$ lies in the same cone as ${\bf p }$ and is given by $ {\bf k }= (m\pm1) {\bf q}/q$ and the resonance condition 
\ref{exact_reslin}b is satisfied exactly. This condition however leads to zero nonlinearity as it corresponds to coupling of 
parallel shear layers that are exact solutions of the Euler equations. The second type of exact resonances is obtained for 
 %%%%%%%%%%%%%%%%%%%%%%%%%%%%%%%%%%%%%%%%%%%%%%%%%%%%%%%%%%%%%%%%%%%%%%%%%%%%%%%%%%%%%%%%%%%%%%%%%%%%%%%
\beq                                                                                                  %%
{\bf k}=\left[                                                                                        %%
\begin{array}{c}                                                                                      %%
 -s_x \,  m    \, q   \\                                                                              %%
 s_y \,(m+1) \, q   \\                                                                                %%
 s_z \, q/2                                                                                           %%
\end{array}                                                                                           %%
\right], \,\,                                                                                         %%
\mathrm{and}                                                                                          %%
\,\,                                                                                                  %%
{\bf p}=\left[                                                                                        %%
\begin{array}{c}                                                                                      %%
-s_x \, (m+1)   \,q   \\                                                                               %%
s_y \,  m      \, q   \\                                                                              %%
        - s_z q/2                                                                                     %%
\end{array}                                                                                           %%
\right], \,\,                                                                                         %%
\mathrm{for}                                                                                          %%
\,\,                                                                                                  %%
{\bf q}=\left[                                                                                        %%
\begin{array}{c}                                                                                      %%
  s_x q   \\                                                                                          %%
  s_y q   \\                                                                                          %%
  s_z q                                                                                               %%
\end{array}                                                                                           %%
\right],                                                                                              %%
\label{exact2}                                                                                        %%
\eeq                                                                                                  %%
%%%%%%%%%%%%%%%%%%%%%%%%%%%%%%%%%%%%%%%%%%%%%%%%%%%%%%%%%%%%%%%%%%%%%%%%%%%%%%%%%%%%%%%%%%%%%%%%%%%%%%%%
true for every $s_x,s_y,s_z=\pm1$ and $m\in \mathbb{Z}$ . This solution couples modes in opposite cones and results in non zero coupling.
Note that these modes only exist when $q$ is even since $q/2$ that appears in \ref{exact2} has to be an integer.
For $q=2$, that is examined here, these two solutions were the only exact solutions found by numerically solving \ref{exact_reslin}.
For larger values of $q$ more exact solutions were found but not a simple analytical formula for them.
The left panel of figure \ref{fig_4} shows in the plane $k_z,k_\perp=\sqrt{k_x^2+k_y^2}$ and for $q=2$ 
the location of the exact resonances of 
the first type ${\bf k} = m {\bf q}/q$ by red squares and of the second type (eq.\ref{exact2}) by red triangles.
The dashed lines indicate the bounds \ref{bound_res1} and  \ref{bound_res2}.

The modes with the smallest value of $|{\bf k}|$ and $|{\bf p}|$ that fall in the set \ref{exact2} for $q=2$ are 
obtained for $m=0$ and $m=-1$. For example, a simple pair of coupled modes is given by
${\bf k}=(2 ,0,1)$ and ${\bf p}=(0,-2,-1)$ for ${\bf q}=(2,2,2)$. 
Keeping only these modes we can write a minimal model for their linear evolution: 
%(valid in the large $\Rof$ limit but for small $\lambda_1$ to justify ignoring the rest of the modes) :
%%%%%%%%%%%%%%%%%%%%%%%%%%%%%%%%%%%%%%%%%%%%%%%%%%%%%%%%%%%%%%%%%%%%%%%%%%%%%%%%%%%%%%%%%%%%%%%%%%%%%%%%%%%%%%%%%%%%%%%%%%%%
\begin{equation}                                                                                                          %%
\begin{array}{lccc}                                                                                                       %% 
\partial_\tau H_{\bf k}^{+}  & = &                                                                                        %%
%        \left[\mathcal{C}^{+,+,-}_{\bf k,q,p} V_{\bf q}^{+} +\mathcal{C}^{+,-,-}_{\bf k,q,p} V_{\bf q}^{-} \right]       %%
         \mathcal{M}_k                                                                                                    %% 
                                                                                                         H_{\bf p}^{-} &  %% 
         -  |{\bf p}|^2 \lambda_1 H_{\bf k}^{+}                                                               \\          %% 
\partial_\tau H_{\bf p}^{ -}   &= &                                                                                       %%
%        \left[\mathcal{C}^{-,+,+}_{\bf p,q,k} V_{\bf q}^{+} +\mathcal{C}^{-,-,+}_{\bf p,q,k} V_{\bf q}^{-} \right]       %%
         \mathcal{M}_p                                                                                                    %%
                                                                                                          H_{\bf k}^{+} & %%
         - |{\bf k}|^2\lambda_1 H_{\bf p}^{+}                                                                             %%
\end{array}  \label{2mode}                                                                                                %%
\end{equation}                                                                                                            %%
%%%%%%%%%%%%%%%%%%%%%%%%%%%%%%%%%%%%%%%%%%%%%%%%%%%%%%%%%%%%%%%%%%%%%%%%%%%%%%%%%%%%%%%%%%%%%%%%%%%%%%%%%%%%%%%%%%%%%%%%%%%%
where $\mathcal{M}_k=\mathcal{C}^{+,+,-}_{\bf k,q,p} V_{\bf q}^{+} +\mathcal{C}^{+,-,-}_{\bf k,q,p} V_{\bf q}^{-}$ and
      $\mathcal{M}_p=\mathcal{C}^{-,+,+}_{\bf p,q,k} V_{\bf q}^{+} +\mathcal{C}^{-,-,+}_{\bf p,q,k} V_{\bf q}^{-}$.
It predicts instability when 
%$\lambda_1 \le |\mathcal{C}^{1,1,1}_{\bf p,q,k} V_{\bf q}^{1})|/|{\bf k}|^2 \simeq 88888$. 
$\lambda_1$ is smaller than a critical value $\lambda_{1c}= 3/20\sqrt{5}\simeq 0.065\dots $ 
implying instability for $\Ref \ge 14.9\dots \Rof^{-1}$. 
The stability from the numerical was estimated closer to $\Ref \ge 20 \Rof^{-1}$ indicating
that ignoring the higher exact resonances underestimated the stability boundary and/or that
sufficiently large $\Rof$ (so that quasi-resonances can be neglected) has not be reached yet in our simulations. 
%The value of this critical $\lambda_{1c}$ measured by the simulations is $\lambda_{1c}\simeq 0.045$. 
%Despite its simplicity the model gives a good estimate of $\lambda_{1c}$.

As discussed before in the $\Rof\to 0$ limit only the exact resonances need to be retained.
However since numerical simulations always operate on finite values of $\Rof$
beside the exact resonances we also need to consider quasi resonances for which \ref{exact_reslin}b is satisfied
up to an $\Delta \omega =\mathcal{O}(\Rof^2)$ accuracy. 
The resonance conditions \ref{exact_reslin} define a surface embedded in the 
$\mathbb{R}^3$ space of ${\bf k}$ where they are satisfied.
Allowing for a  $\Delta \omega$ ambiguity implies that  modes within a distance $\Delta k$ 
normal to the resonance surface should also be considered. Here  $\Delta k$ is estimated by Taylor expansion  
$\Delta \omega \simeq (\Delta {\bf k}) \cdot \nabla_{\bf k}(\omega_{\bf k}-\omega_{\bf k+q}) \propto (\Delta k)q/k^2 $, 
where the last relation is valid in the $q\ll|{\bf k}|$ limit.
%The number of modes that satisfy the quasi-resonance conditions
The volume $V_k$ in Fourier space occupied by these allowed wavenumbers inside a sphere of radius $k$ is then 
proportional the area of the resonance surface times the thickness $\Delta k$ thus $V_k \propto k^2 \Delta k$.
The total number of quasi-resonances on a spherical shell of radius $k$ and width $\delta k =1$ will scale as 
$N_{\Delta \omega}(k) = n_k \partial_k V_k$ thus:
%%%%%%%%%%%%%%%%%%%%%%%%%%%%%%%%%%%%%%%%%%%%%%%%%%%%%%%%%%%%%%%%%%%%%%%%%%%%%%%%%%%%%%%%%%%%%%%%%%%
\beq N_{\Delta \omega}(k) \propto k \Delta k \propto k^3 \Delta \omega. \label{Nomega} \eeq      %%
%%%%%%%%%%%%%%%%%%%%%%%%%%%%%%%%%%%%%%%%%%%%%%%%%%%%%%%%%%%%%%%%%%%%%%%%%%%%%%%%%%%%%%%%%%%%%%%%%%%
This scaling is valid 
provided that $\Delta k\simeq \Delta \omega k^2/q$ remains small. For large values of $k$ however, $\Delta k$ becomes 
so large that all the modes inside the spherical shell satisfy the resonance conditions \ref{exact_reslin}. The 
scaling will thus transition to $N_{\Delta \omega}(k) = 4\pi k^2$. 
Quasi-resonances were sought for numerically and the results are shown in figure \ref{fig_4}.
%
%%%%%%%%%%%%%%%%%%%%%%%%%%%%%%%%%%%%%%%%%%%%%%%%%%%%%%%%%%%%%%%%%%%%%%%%%%%%%%%%%%%%%%%%%%%%%%%%%%%%%%%%%%%%%%%%%%%%%%%%%%%%%%
                                                                                                               %%
%%%%%%%%%%%%%%%%%%%%%%%%%%%%%%%%%%%%%%%%%%%%%%%%%%%%%%%%%%%%%%%%%%%%%%%%%%%%%%%%%%%%%%%%%%%%%%%%%%%%%%%%%%%%%%%%%%%%%%%%%%%%%%
% 
The left panel of figure \ref{fig_4} shows  the location of the {\it quasi}-resonances by blue circles 
for a frequency tolerance $\Delta \omega=10^{-3}$. 
The right panel of the same figure indicates number of modes that satisfy the quasi-resonance conditions
up to a a frequency ambiguity $\Delta \omega$ on the spherical shell $k <  |{\bf k}| \le  k+1$.
The notation $\Delta \omega=\infty$ implies all pairs of modes in this spherical shell are counted independent of resonances.
The scalings $k^2$ and $k^3$ predicted in the previous paragraph are shown as a reference.
It is clear that {\it quasi}-resonances increase very rapidly and unless the viscous cut-off wavenumber
$k_{_\lambda}$ is sufficiently small, very high rotation rates will be required to justify
their neglection.

\section{Nonlinear behavior                      } \label{Sec:nonlin}%%%%%%%%%%%%%%%%%%%%%%%%%%%%%%%%%%%%%%%%%%%%%%%%%%%%%%%%%%
%%%%%%%%%%%%%%%%%%%%%%%%%%%%%%%%%%%%%%%%%%%%%%%%%%%%%%%%%%%%%%%%%%%%%%%%%%%%%%%%%%%%%%%%%%%%%%%%%%%%%%%%%%%%%%%%%%%%%%%%%%%%%%%
%%%%%%%%%%%%%%%%%%%%%%%%%%%%%%%%%%%%%%%%%%%%%%%%%%%%%%%%%%%%%%%%%%%%%%%%%%%%%%%%%%%%%%%%%%%%%%%%%%%%%%%%%%%%%%%%%%%%%%%%%%%%%%% 
%%%%%%%%%%%%%%%%%%%%%%%%%%%%%%%%%%%%%%%%%%%%%%%%%%%%%%%%%%%%%%%%%%%%%%%%%%%%%%%%%%%%%%%%%%%%%%%%%%%%%%%%%%%%%%%%%%%%%%%%%%%%%%% 
%%%%%%%%%%%%%%%%%%%%%%%%%%%%%%%%%%%%%%%%%%%%%%%%%%%%%%%%%%%%%%%%%%%%%%%%%%%%%%%%%%%%%%%%%%%%%%%%%%%%%%%%%%%%%%%%%%%%%%%%%%%%%%%

\subsection{Reduced equations}

The previous section determined the values of the parameters for which a dynamical behavior will be observed in the fast rotating limit.
The properties of this dynamical behavior however can only be determined when nonlinearities are taken in to account.
%The nonlinear behavior as was shown in section \ref{sec:Param} is very rich.
To understand the nonlinear behavior the asymptotic expansion in the fast rotating limit is examined including the nonlinear terms.
Following the same steps as we did in the previous section we begin from eq.\ref{ttNSR} with $n$=1 and 
expand ${\bf v}= {\bf v}^{(0)} + \Rof^2{\bf v}^{(1)} +\dots$. At zeroth order 
we obtain the same linear equation for ${\bf v}$ as we did for ${\bf v}_p$ in \ref{lnr}.
The general solution of which is then given by
%%%%%%%%%%%%%%%%%%%%%%%%%%%%%%%%%%%%%%%%%%%%%%%%%%%%%%%%%%%%%%%%%%%%%%%%%%%%%%%%%%%%%%%%%%%%%%%%%%%%%%%%%%
\begin{equation}                                                                                        %% 
{\bf v}^{(0)} ={\bf v}_f^{(0)}(x,y,z) + {\bf v}_{_{2D}}^{(0)}(\tau,x,y) +{\bf v}_w^{(0)}(\tau,t,x,y,z). %%
\end{equation}                                                                                          %%
%%%%%%%%%%%%%%%%%%%%%%%%%%%%%%%%%%%%%%%%%%%%%%%%%%%%%%%%%%%%%%%%%%%%%%%%%%%%%%%%%%%%%%%%%%%%%%%%%%%%%%%%%%
As before
${\bf v}_f      ^{(0)}$ is the zeroth order stationary solution \ref{v01},
${\bf v}_{_{2D}}^{(0)}$ is a velocity field that depends only on the ($x,y$) coordinates and 
                        on the slow dynamical time $\tau$, 
${\bf v}_w      ^{(0)}$ are inertial waves whose amplitude also varies on the long time scale
and their exact form is given by \ref{Wexpa}.

To next order we then have
%%%%%%%%%%%%%%%%%%%%%%%%%%%%%%%%%%%%%%%%%%%%%%%%%%%%%%%%%%%%%%%%%%%%%%%%%%%%%%%%%%%%%%%%%%%%%%%%%%%%%%%%
\beq                                                                                                  %%
\partial_{t} {\bf v}^{(1)}  + \mathbb{L} [{\bf v}^{(1)}]  =                                           %%
 -\partial_\tau {\bf v}^{(0)} +                                                                       %%
  \mathbb{P}[ {\bf v}^{(0)} \times {\bf w}^{(0)} ]                                                    %%
+ \lambda_1 \Delta   {\bf v }^{(0)}                                                                   %%
\label{nord1}                                                                                         %% 
\eeq                                                                                                  %%
%%%%%%%%%%%%%%%%%%%%%%%%%%%%%%%%%%%%%%%%%%%%%%%%%%%%%%%%%%%%%%%%%%%%%%%%%%%%%%%%%%%%%%%%%%%%%%%%%%%%%%%%
To obtain then the evolution of ${\bf v}_{_{2D}}^{(0)}$ we perform a short time average and an average over the $z$-direction. 
The left hand side averages to zero and we are left with:
%%%%%%%%%%%%%%%%%%%%%%%%%%%%%%%%%%%%%%%%%%%%%%%%%%%%%%%%%%%%%%%%%%%%%%%%%%%%%%%%%%%%%%%%%%%%%%%%%%%%%%%%%%%%%%%%%%%%%%%%%%%%%%%%%%%%%%%%%%%%%%%%%%%%%%%%%%%%%%%%
\beq                                                                                                                                                          %%
\begin{array}{cll}                                                                                                                                            %%
\partial_\tau {\bf v}_{_{2D}}^{(0)} &=& \left\langle   \overline{ \mathbb{P}[ {\bf v}^{(0)} \times {\bf w}^{(0)} ]   }                   \right\rangle_\tau   %%
                                     +  \lambda_1 \Delta   {\bf v}_{_{2D}}^{(0)}                                                                              %%
                                    \\                                                                                                                        %%
                                    &=&  \left\langle  \overline{ \mathbb{P}[ {\bf v}_f      ^{(0)} \times {\bf w}_f      ^{(0)} ]   }   \right\rangle_\tau   %%
                                     +   \left\langle  \overline{ \mathbb{P}[ {\bf v}_f      ^{(0)} \times {\bf w}_w      ^{(0)} ]   }   \right\rangle_\tau   %%
                                     +   \left\langle  \overline{ \mathbb{P}[ {\bf v}_f      ^{(0)} \times {\bf w}_{_{2D}}^{(0)} ]   }   \right\rangle_\tau   %%
                                    \\                                                                                                                        %%
                                    &+&  \left\langle  \overline{ \mathbb{P}[ {\bf v}_w      ^{(0)} \times {\bf w}_f      ^{(0)} ]   }   \right\rangle_\tau   %%
                                     +   \left\langle  \overline{ \mathbb{P}[ {\bf v}_w      ^{(0)} \times {\bf w}_w      ^{(0)} ]   }   \right\rangle_\tau   %%
                                     +   \left\langle  \overline{ \mathbb{P}[ {\bf v}_w      ^{(0)} \times {\bf w}_{_{2D}}^{(0)} ]   }   \right\rangle_\tau   %%
                                    \\                                                                                                                        %%
                                    &+&  \left\langle  \overline{ \mathbb{P}[ {\bf v}_{_{2D}}^{(0)} \times {\bf w}_f      ^{(0)} ]   }   \right\rangle_\tau   %%
                                     +   \left\langle  \overline{ \mathbb{P}[ {\bf v}_{_{2D}}^{(0)} \times {\bf w}_w      ^{(0)} ]   }   \right\rangle_\tau   %%
                                     +   \left\langle  \overline{ \mathbb{P}[ {\bf v}_{_{2D}}^{(0)} \times {\bf w}_{_{2D}}^{(0)} ]   }   \right\rangle_\tau   %%
                                    \\                                                                                                                        %%
                                    &+&  \lambda_1 \Delta   {\bf v}_{_{2D}}^{(0)}                       \\                                                    %%
\end{array}                                                                                                                                                   %%
\eeq                                                                                                                                                          %%
%%%%%%%%%%%%%%%%%%%%%%%%%%%%%%%%%%%%%%%%%%%%%%%%%%%%%%%%%%%%%%%%%%%%%%%%%%%%%%%%%%%%%%%%%%%%%%%%%%%%%%%%%%%%%%%%%%%%%%%%%%%%%%%%%%%%%%%%%%%%%%%%%%%%%%%%%%%%%%%%
The only quadratic terms that remain non-zero after the vertical average are 
\[
\left\langle  \overline{ \mathbb{P}[ {\bf v}_f      ^{(0)} \times {\bf w}_w      ^{(0)} ]   }   \right\rangle_\tau, 
\left\langle  \overline{ \mathbb{P}[ {\bf v}_w      ^{(0)} \times {\bf w}_f      ^{(0)} ]   }   \right\rangle_\tau,
\left\langle  \overline{ \mathbb{P}[ {\bf v}_w      ^{(0)} \times {\bf w}_w      ^{(0)} ]   }   \right\rangle_\tau \,\, \mathrm{and} \,\,
\left\langle  \overline{ \mathbb{P}[ {\bf v}_{_{2D}}^{(0)} \times {\bf w}_{_{2D}}^{(0)} ]   }   \right\rangle_\tau .
 \]
The first two correspond to the terms that appear in linear theory and we have already showed that lead to zero contribution 
after the short time average. The third term can be written using the helical mode expansion (eq. \ref{Wexpa}) as
%%%%%%%%%%%%%%%%%%%%%%%%%%%%%%%%%%%%%%%%%%%%%%%%%%%%%%%%%%%%%%%%%%%%%%%%%%%%%%%%%%%%%%%%%%%%%%%%%%%%%%%%%%%%%%%%%%%%%%%%%%%%%%%%%%%%%%%%%%%%%%%%%%%%%%%%%%%%%%%%%
\beq                                                                                                                                                           %%
\left\langle  \overline{ \mathbb{P}[ {\bf v}_w      ^{(0)} \times {\bf w}_w      ^{(0)} ]   }   \right\rangle_\tau =                                           %% 
         \sum_{\tiny \begin{array}{c} {s_k,s_p,\bf k,p} \\ {k_z+p_z=0}\end{array} }                                                                                 %%
         \left(                                                                                                                                                %%
         \left\langle e^{ i (\omega_k+\omega_p) t }  \right\rangle_{\tau}  e^{i({\bm k} +\bm p )\cdot \bm x }                                                  %% 
         \mathcal{D}^{s_k,s_p}_{\bf k,p}  H_{\bf k}^{s_k} H_{\bf q}^{s_q}                                                                                      %%
         \right).                                                                                                                                              %%
\label{wave2Dterm}                                                                                                                                             %%
\eeq                                                                                                                                                           %%
%%%%%%%%%%%%%%%%%%%%%%%%%%%%%%%%%%%%%%%%%%%%%%%%%%%%%%%%%%%%%%%%%%%%%%%%%%%%%%%%%%%%%%%%%%%%%%%%%%%%%%%%%%%%%%%%%%%%%%%%%%%%%%%%%%%%%%%%%%%%%%%%%%%%%%%%%%%%%%%%%
The vector $\mathcal{D}^{s_k,s_p}_{\bf k,p}$ is given by
%%%%%%%%%%%%%%%%%%%%%%%%%%%%%%%%%%%%%%%%%%%%%%%%%%%%%%%%%%%%%%%%%%%%%%%%%%%%%%%%%%%%%%%%%%%%%%%%%%%%%%%%%%%%%%%%%%
\beq                                                                                                            %%
\mathcal{D}^{s_k,s_p}_{\bf k,p} =       \frac{1}{2}(s_k|{\bf k}|- s_p|{\bf p}| )                                %%
\mathbb{P}_{\bf r} \bf  [{\bf h}_{\bf k}^{s_k} \times {\bf h}_{\bf p}^{s_p}]                                    %%
=   \mathcal{C}^{s_{r},s_k,s_p}_{\bf r,k,p} h^{s_{r}}_{\bf r}                                                   %%
\label{Ddef}                                                                                                    %%
\eeq                                                                                                            %%
%%%%%%%%%%%%%%%%%%%%%%%%%%%%%%%%%%%%%%%%%%%%%%%%%%%%%%%%%%%%%%%%%%%%%%%%%%%%%%%%%%%%%%%%%%%%%%%%%%%%%%%%%%%%%%%%%%
(where ${\bf r =k+p}$ and $\mathbb{P}_{\bf k}[{\bf g_k}]={\bf k \times k\times g_k}/|{\bf k}^2|$ is the projection operator $ \mathbb{P}$ in Fourier space).
A remarkable result of \cite{Waleffe1993} was that for exact resonances $\omega_k+\omega_p=0$ this term \ref{wave2Dterm} 
also averages to zero.  This can be realized by noting that $k_z+p_z=0$ and $\omega_k+\omega_p=0$ implies that $|{\bf k}|=|{\bf p}|$ and 
$s_p=s_k$ and therefore the vector $\mathcal{D}^{s_k,s_p}_{\bf k,p}$ in \ref{Ddef} becomes zero and does not effect the 2D3C flow. 
At this order then the only nonlinear term left for the evolution  of the 2D3C flow is the coupling with itself, that leads to
%%%%%%%%%%%%%%%%%%%%%%%%%%%%%%%%%%%%%%%%%%%%%%%%%%%%%%%%%%%%%%%%%%%%%%%%%%%%%%%%%%%%%%%%%%%%%%%%%%%%%%%%%%%%%%%%%%%%%%%%%%%%%%%%%%%%%%%%%%%%%%%%%%%%%%%%%%
\beq                                                                                                                                                    %% 
\partial_\tau {\bf v}_{_{2D}}^{(0)} = \mathbb{P}[ {\bf v}_{_{2D}}^{(0)} \times {\bf w}_{_{2D}}^{(0)} ] + \lambda_1 \nabla^2 {\bf v}_{_{2D}}^{(0)}       %%
\label{2DNS}                                                                                                                                            %%
\eeq                                                                                                                                                    %%
%%%%%%%%%%%%%%%%%%%%%%%%%%%%%%%%%%%%%%%%%%%%%%%%%%%%%%%%%%%%%%%%%%%%%%%%%%%%%%%%%%%%%%%%%%%%%%%%%%%%%%%%%%%%%%%%%%%%%%%%%%%%%%%%%%%%%%%%%%%%%%%%%%%%%%%%%%
which is the unforced 2D Navier-Stokes Equation. As result at this order the nonlinear evolution equation for the 2D3C flow completely decouples from the rest 
of the flow, and since there is no forcing term in \ref{2DNS}, ${\bf v}_{_{2D}}^{(0)}$ will decay to zero at long times. 

In the simulations however for all dynamical regimes a 2D3C flow was present. In order to explain the presence of a 2D3C flow
in the simulation we have to evoke either quasi-resonances and higher order terms in the expansion or a break-down of 
the fast rotating limit. 
%%%%%%%%%%%%%%%%%%%%%%%%%%%%%%%%%%%%%%%%%%%%%%%%%%%%%%%%%%%%%%%%%%%%%%%%%%%%%%%%%%%%%%%%%%%%%%%%%%%%%%%%
\begin{figure*}                                                                                       %%
\centerline{                                                                                          %%
 \includegraphics[width=8cm]{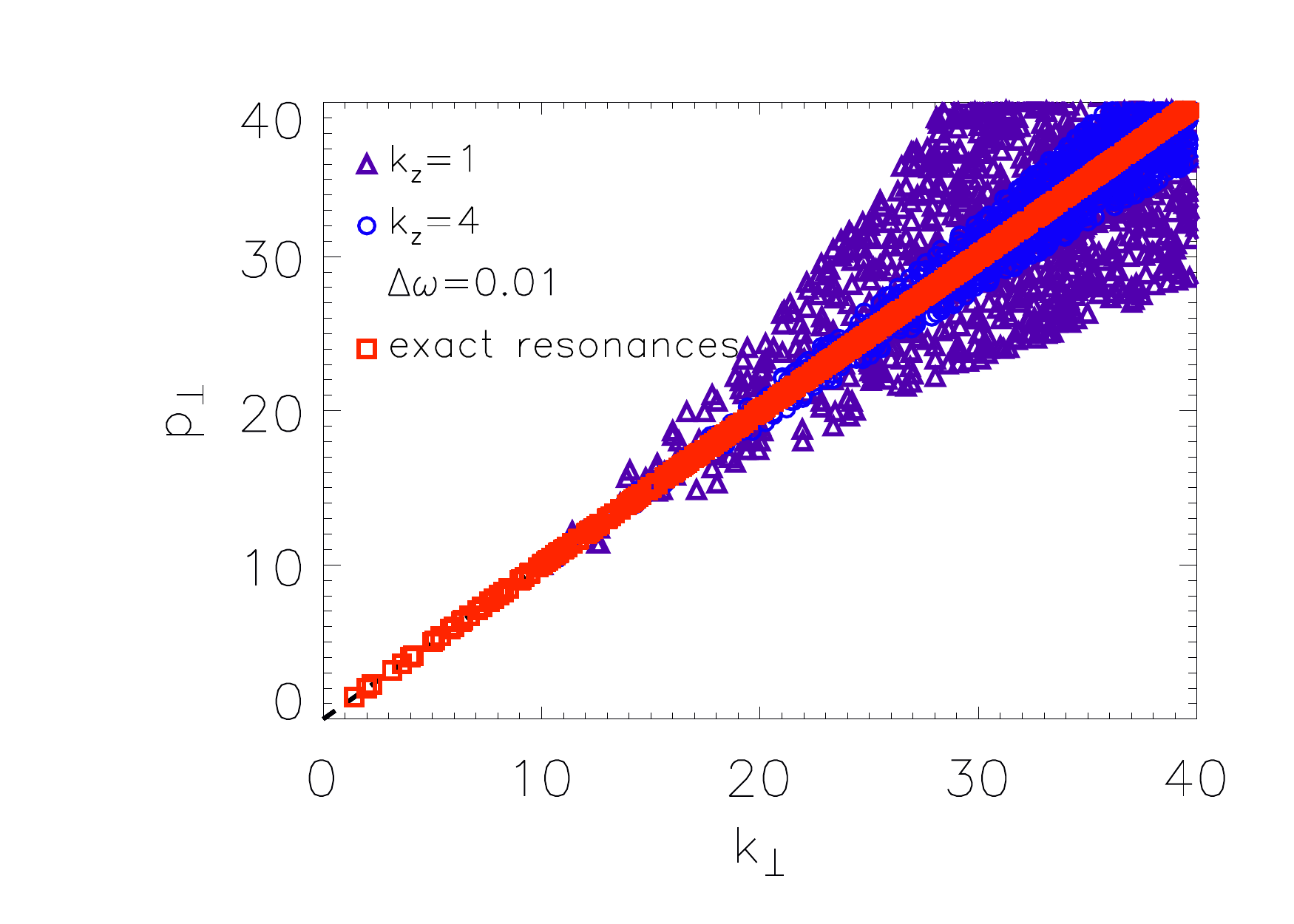}                                                     %%
 \includegraphics[width=8cm]{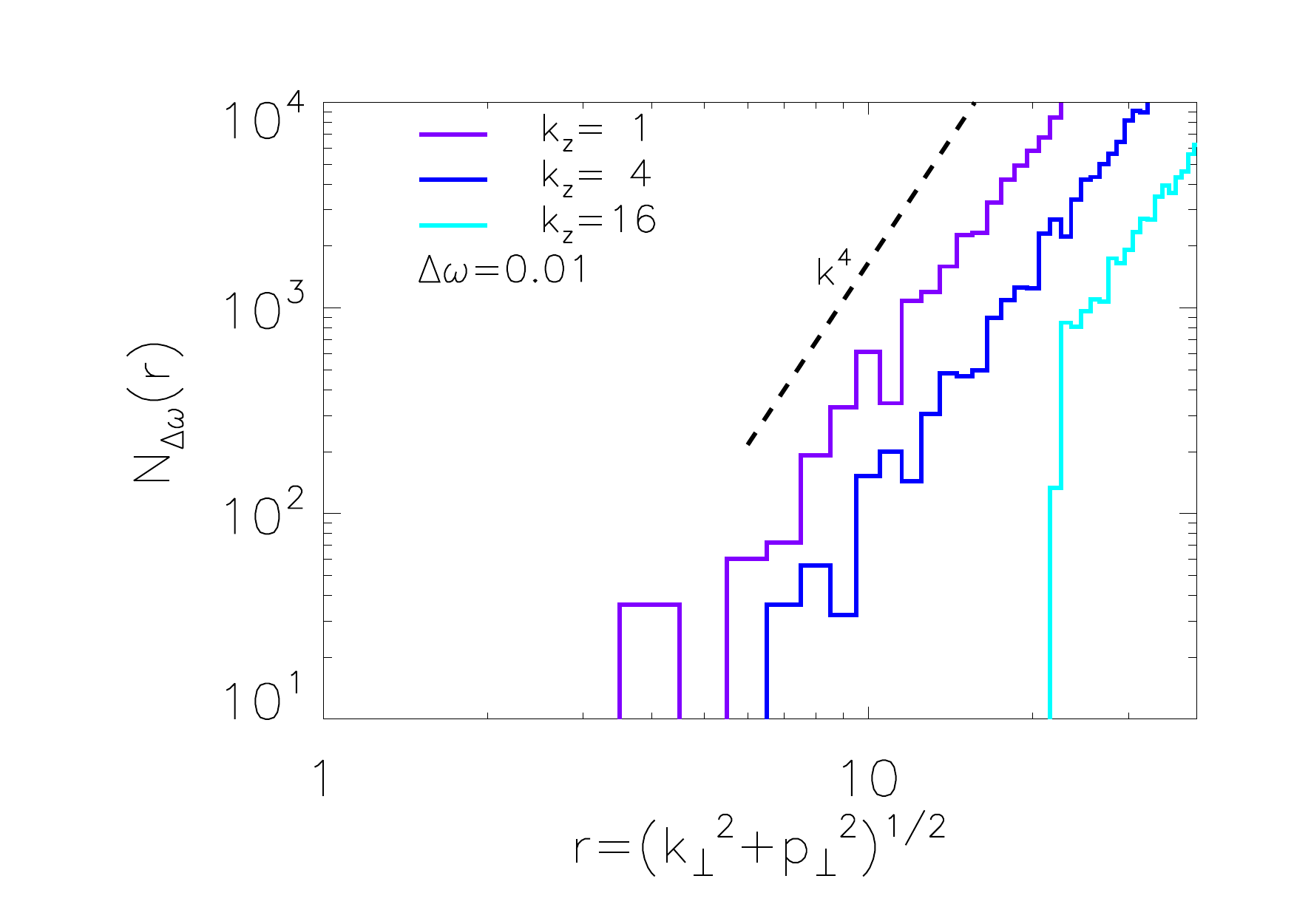}                                                    %%
}                                                                                                     %%
\caption{Left panel: The location of resonances for the modes ${\bf k,p}$.                            %% 
        Exact resonances are shown by red squares, and  the                                           %%
        {\it quasi}-resonances by blue circles ($k_z=-p_z=4$) and purple triangles ($k_z=-p_z=1$)     %%
        for a frequency tolerance $\Delta \omega=10^{-2}$.                                            %%
        Right panel: Number of modes $N_{\Delta \omega}(r)$                                           %%
        that satisfy the quasi-resonance conditions up to a a frequency ambiguity $\Delta \omega$     %%
        on the spherical shell  $r <  \sqrt{ k_\perp^2+p_\perp^2}| \le  r+1$.}                        %%
\label{fig_5}                                                                                         %%
\end{figure*}                                                                                         %%
%%%%%%%%%%%%%%%%%%%%%%%%%%%%%%%%%%%%%%%%%%%%%%%%%%%%%%%%%%%%%%%%%%%%%%%%%%%%%%%%%%%%%%%%%%%%%%%%%%%%%%%%
We consider then quasi-resonances that could lead to the generation of a quasi-2D flow
and allow for a frequency tolerance $\Delta \omega \ll 1$ 
over which a violation of the resonance condition is allowed.
For fixed $k_z=-p_z$ the coupling triads reside in the four dimensional space defined by $(k_x,k_y,p_x,p_y)$.
In this space a 4-spherical shell of unit thickness and of radius $r=(k_\perp^2+p_\perp^2)^{1/2}$, has 
$N\propto r^3$ triads out of which  
the quasi-resonance condition confines the allowed triads to reside in a subspace of thickness 
$\Delta k_\perp$ that limits their number to $N_{\Delta\omega}\propto r^2\Delta k$. The thickness $\Delta k$
is obtained from the dispersion relation
$\Delta \omega \simeq \Delta k_\perp \partial_{k_\perp} \omega= - \Delta k_\perp k_zk_\perp/k^3$.
Thus $ \Delta k_\perp \propto \Delta \omega k^3/k_zk_\perp \simeq \Delta \omega r^2/k_z$ for $k_\perp\gg k_z$.
The number of these allowed modes then will scale as $N_{\Delta\omega}(r) \propto r^2 \Delta k \propto r^4/k_z \Delta \omega$
(where locality between the modes $k_\perp\sim p_\perp\sim r$ has been assumed). The number of quasi-resonances therefore
grows with $k$ even faster than the quasi resonances of the linear term in the previous section. 
An other point that worths notice is that the smallest values of $k_z$ have the largest number quasi-resonances ($k_z=1$ for our case).

At figure \ref{fig_5} we plot on the left panel the exact resonances, and the quasi resonances for $k_z=1,4$ and $\Delta \omega=0.01$ on the 
$k_\perp,p_\perp$ plane. Due to their large number only one in 100 points has been plotted.
The right panel of the same figure shows the density $N_{\Delta\omega}(r)$ for for different values of $k_z$.
These modes however will contribute to the 2D3C dynamics with coupling coefficients of the order 
$\mathcal{D}^{s_k,s_p}_{\bf k,p}=\mathcal{O}(\Delta k)=\mathcal{O}(\Delta \omega)$.
Their contribution to the ${\bf v}_{_{2D}}^{(0)}$ evolution will be of the order 
$N_{\Delta\omega}\mathcal{D}^{s_k,s_p}_{\bf k,p}=\mathcal{O}( \Delta \omega^2)$,
and thus the same order as the next order term in the expansion.

In order thus to capture this contribution the expansion will need to be carried at next order introducing a second slow time scale
and increasing $\Ref$ to the scaling $\Ref=\lambda_2\Rof^3$. The process that ``lives" on this second time scale
although slower, on the long time limit that we investigate here could be dominant and explain the 2D3C dynamics observed in 
the simulations. 
%This however is not attempted here as it will require a long derivation and it is beyond the purpose of
%present work.

To obtain the evolution equation for the amplitude $H_{\bf k}^{s_k}$of the inertial waves we multiply equation \ref{nord1} with ${\bf h_k}^{-s_k}e^{-i({\bf k \cdot x}+\omega_kt)}$ 
and space time average. The right hand side will then drop out and we are going to be left with the evolution equation for the complex amplitude:
%%%%%%%%%%%%%%%%%%%%%%%%%%%%%%%%%%%%%%%%%%%%%%%%%%%%%%%%%%%%%%%%%%%%%%%%%%%%%%%%%%%%%%%%%%%%%%%%%%%%%%%%%%%%%%%%%%%%%%%%%%%%%%%%%%%%%%%%%%%%%%%%%%%%%%%%%%%%%%%%%%%%%%%%%%%%%%%%%%%%%%% 
\begin{equation}                                                                                                                                                                     %%
\begin{array}{ccl}                                                                                                                                                                   %%
\partial_\tau H_{\bf k}^{s_k} (\tau)                                                                                                                                                 %%
  &=& \left\langle \left\langle  {\bf h_k}^{-s_k}e^{-i({\bf k \cdot x}+\omega_kt)} { \mathbb{P}[ {\bf v}^{(0)} \times {\bf w}^{(0)} ]   }  \right\rangle_{_V} \right\rangle_\tau     %%
                                       -  \lambda_1 |{\bf k}|^2 H_{\bf k}^{s_k}(\tau)                                                                                              \\%%
  &=& \left\langle \left\langle  {\bf h_k}^{-s_k}e^{-i({\bf k \cdot x}+\omega_kt)}                                                                                                   %%
                        { \mathbb{P}[  {\bf v}_f^{(0)} \times {\bf w}_w^{(0)}    + {\bf v}_w^{(0)} \times {\bf w}_{f}^{(0)} ]  }                                                     %%
      \right\rangle_{_V} \right\rangle_\tau \\                                                                                                                                       %%
  &+& \left\langle \left\langle  {\bf h_k}^{-s_k}e^{-i({\bf k \cdot x}+\omega_kt)}                                                                                                   %%
                        { \mathbb{P}[  {\bf v}_{_{2D}}^{(0)} \times {\bf w}_w^{(0)}    + {\bf v}_w^{(0)} \times {\bf w}_{_{2D}}^{(0)} ]  }                                           %%
      \right\rangle_{_V} \right\rangle_\tau \\                                                                                                                                       %%
  &+& \left\langle \left\langle  {\bf h_k}^{-s_k}e^{-i({\bf k \cdot x}+\omega_kt)}                                                                                                   %%
                        { \mathbb{P}[  {\bf v}_{w}^{(0)} \times {\bf w}_w^{(0)}  ]  }                                                                                                %%
      \right\rangle_{_V} \right\rangle_\tau \\                                                                                                                                       %%
  &-&  \lambda_1 |{\bf k}|^2 H_{\bf k}^{s_k}(\tau)  .                                                                                                                                %% 
\end{array}                                                                                                                                                                          %%
\end{equation}                                                                                                                                                                       %%
The first term couples the inertial waves to the stationary flow. It is the same term that was studied in the 
linear stability section and is the term that injects energy to inertial waves. The second term corresponds to the 
term that couples the 2D3C flow to the inertial waves. This coupling term does not exchange energy with the 2D3C part 
of the flow as was discussed before. It can however redistribute the energy among the inertial wave modes modes ${\bf k,p}$ with
$k_z=p_z$ and $k_\perp=p_\perp$
(see \cite{Waleffe1993}). The third term couples all the inertial waves with each other. 
In the fast rotating limit only exact resonances with will survive at this order after the short time average.
For the wave numbers $ {\bf k_1,k_2,k_3}$ the following resonance conditions hold:
%%%%%%%%%%%%%%%%%%%%%%%%%%%%%%%%%%%%%%%%%%%%%%%%%%%%%%%%%%%%%%%%%%%%%%%%%%%%%%%%%%%%%%%%%%%%%%%%%%%%%%%
\beq                                                                                                 %%
(a)  \quad {\bf k_3 = k_1+k_2 } \qquad \mathrm{and} \quad                                            %%
(b) \qquad s_3\frac{k_{3z}}{|\bf k_3|}= s_1\frac{k_{1z}}{|\bf k_1|}+s_2\frac{k_{2z}}{|\bf k_2|}.     %%
\label{exact_res}                                                                                    %%
\eeq                                                                                                 %%
%%%%%%%%%%%%%%%%%%%%%%%%%%%%%%%%%%%%%%%%%%%%%%%%%%%%%%%%%%%%%%%%%%%%%%%%%%%%%%%%%%%%%%%%%%%%%%%%%%%%%%%
Analyzing the statistical properties of the flow through these exact resonances is the subject of weak wave turbulence theory
that has been discussed in \cite{Galtier2003} and leads to the prediction for the energy spectrum $E(k_\perp)\propto k_\perp^{-5/2}$.

%%%%%%%%%%%%%%%%%%%%%%%%%%%%%%%%%%%%%%%%%%%%%%%%%%%%%%%%%%%%%%%%%%%%%%%%%%%%%%%%%%%%%%%%%%%%%%%%%%%%%%%%%%%%%%%%%%%%%%%%%%%%%%%
%%%%%%%%%%%%%%%%%%%%%%%%%%%%%%%%%%%%%%%%%%%%%%%%%%%%%%%%%%%%%%%%%%%%%%%%%%%%%%%%%%%%%%%%%%%%%%%%%%%%%%%%%%%%%%%%%%%%%%%%%%%%%%%
\subsection{Intermittent bursts     }%%%%%%%%%%%%%%%%%%%%%%%%%%%%%%%%%%%%%%%%%%%%%%%%%%%%%%%%%%%%%%%%%%%%%%%%%%%%%%%%%%%%%%%%%%
%%%%%%%%%%%%%%%%%%%%%%%%%%%%%%%%%%%%%%%%%%%%%%%%%%%%%%%%%%%%%%%%%%%%%%%%%%%%%%%%%%%%%%%%%%%%%%%%%%%%%%%%%%%%%%%%%%%%%%%%%%%%%%%
%%%%%%%%%%%%%%%%%%%%%%%%%%%%%%%%%%%%%%%%%%%%%%%%%%%%%%%%%%%%%%%%%%%%%%%%%%%%%%%%%%%%%%%%%%%%%%%%%%%%%%%%%%%%%%%%%%%%%%%%%%%%%%%

\begin{figure*}                                                                       %%
%\begin{center}                                                                       %%
\centerline{                                                                          %%
 \includegraphics[width=8cm]{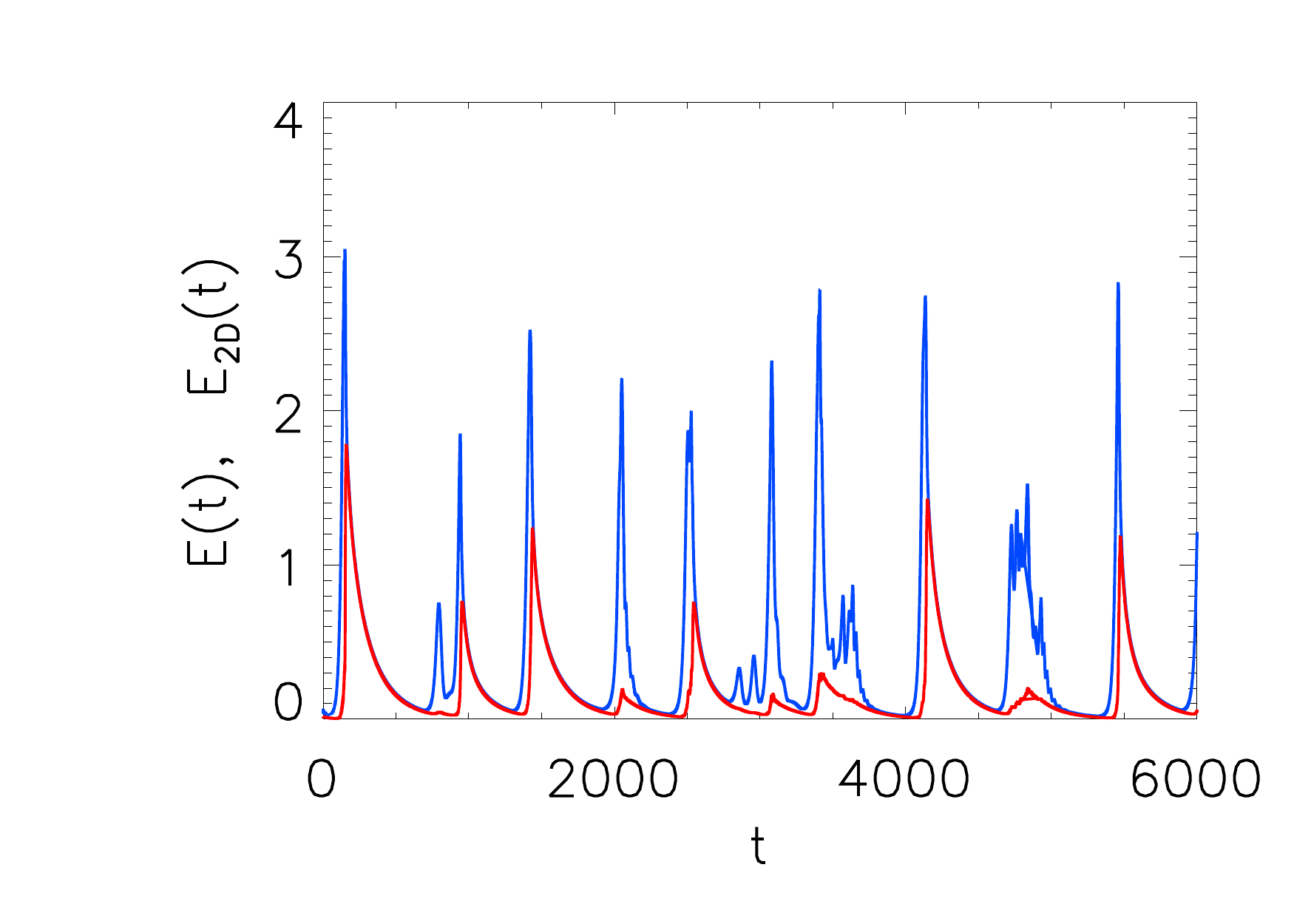}                                     %%
 \includegraphics[width=8cm]{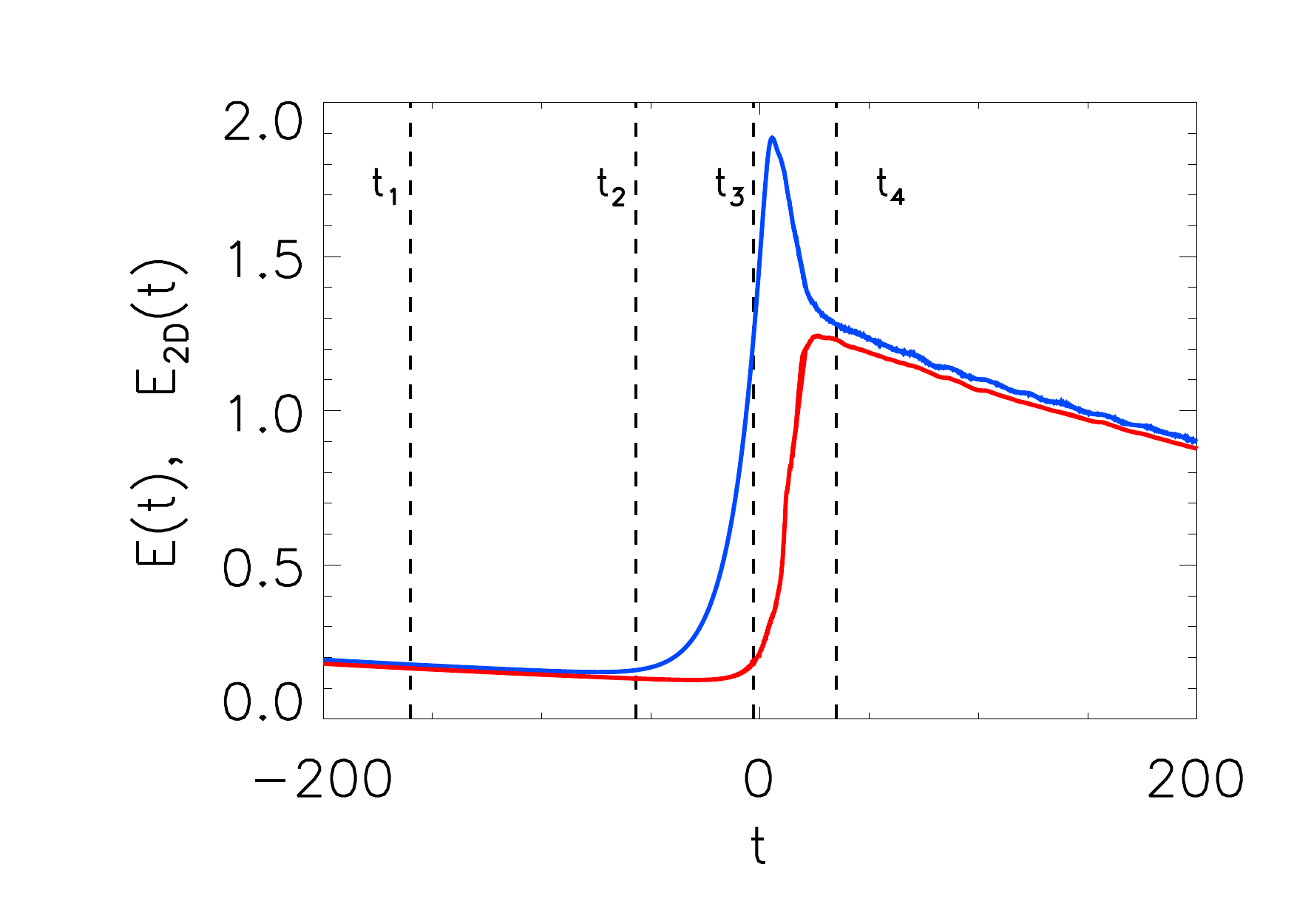}                                    %%
}                                                                                     %% 
\caption{Left panel: A series of bursts for ($\Ref=1000,\Rof=0.125$). The dark        %%
        (blue online) curve shows the evolution of the total energy and the           %%
        light gray (red online) line shows the evolution of the energy of the 2D3C    %%
        component of the flow. Right panel: The same signal focused on a single       %%
        burst that has been shifted to t=0. }                                         %%
\label{fig_6}                                                                         %%
\end{figure*}

Given the theoretical background discussed in the previous section we can try to interpret the observed results in the simulations.
As discussed in section \ref{sec:Param} the striking feature of the flow close to the linear stability boundary is bursts of
energy followed by a slow decay. The flow during this slow decay phase is dominated by a large 
2D3C component. The top left panel of figure \ref{fig_6} shows a series of such burst for a typical run in this range ($\Ref=1000,\Rof=0.125$).
The dark (blue online) curve shows the evolution of the total energy and the light gray (red online) line shows
the evolution of the energy of the 2D3C component of the flow.    
The right panel of figure \ref{fig_6} shows the same signal focused on a single burst that has been shifted to t=0.
It can be seen that at the birth of the burst the inertial waves (the non-2D3C component of the flow)
become unstable and increase to  a high amplitude. The inertial waves then drive the increase of the 2D3C component
that follows after. After the burst the 2D3C flow dominates and slowly decays exponentially 
by a rate proportional to the viscosity. As a result for large value of $\Ref$ the decay is slow. 
%and the frequency of appearance of a burst decreases. 
Thus in order to get well converged averages, runs of long durations are needed.

%%%%%%%%%%%%%%%%%%%%%%%%%%%%%%%%%%%%%%%%%%%%%%%%%%%%%%%%%%%%%%%%%%%%%%%%%%%%%%%%%%%%%%%%
                                                                         %%
%%%%%%%%%%%%%%%%%%%%%%%%%%%%%%%%%%%%%%%%%%%%%%%%%%%%%%%%%%%%%%%%%%%%%%%%%%%%%%%%%%%%%%%%
%%%%%%%%%%%%%%%%%%%%%%%%%%%%%%%%%%%%%%%%%%%%%%%%%%%%%%%%%%%%%%%%%%%%%%%%%%%%%%%%%%%%%%%%
\begin{figure*}                                                                       %%
\centerline{                                                                          %%
 \includegraphics[width=8cm]{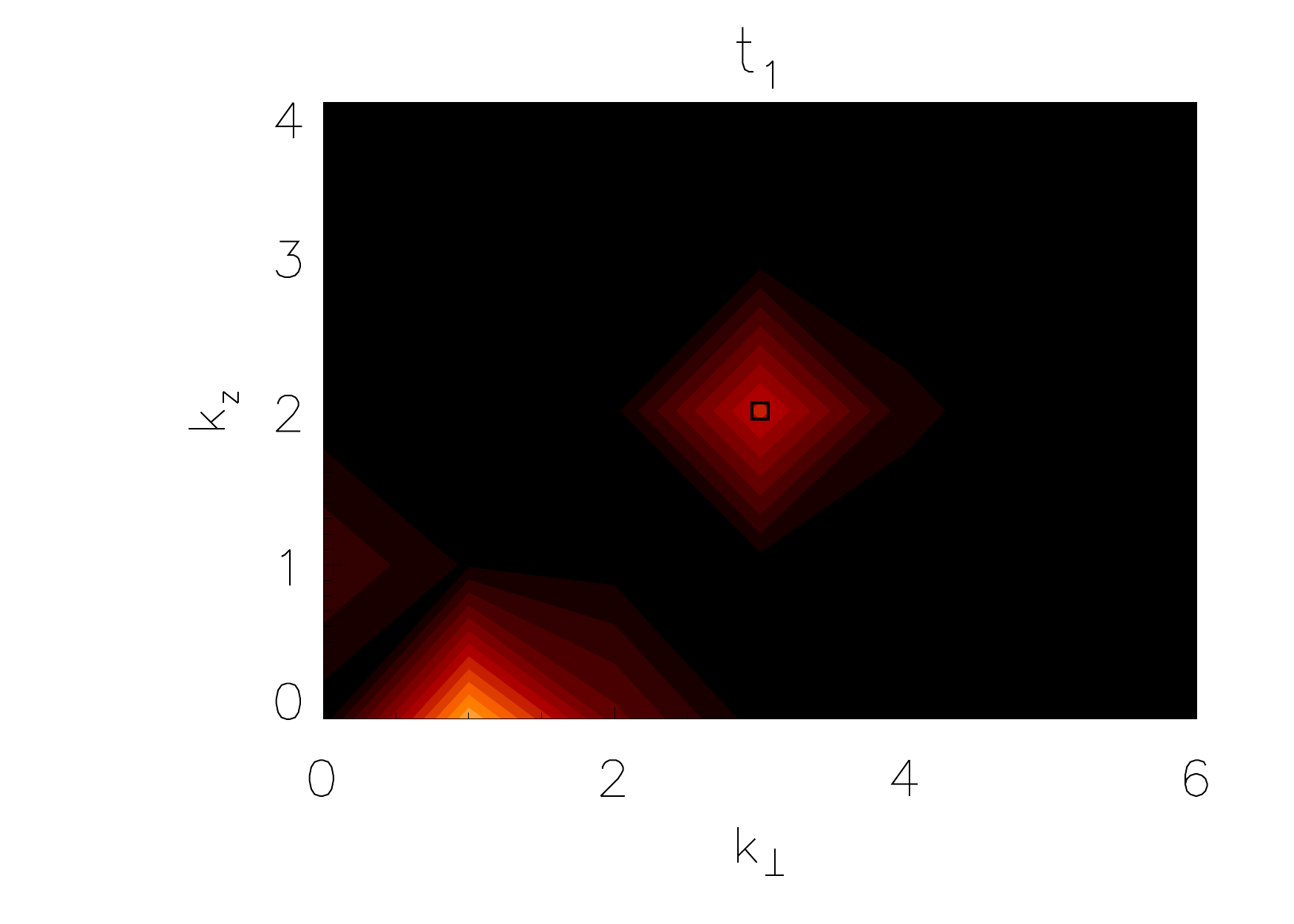}                                 %%
 \includegraphics[width=8cm]{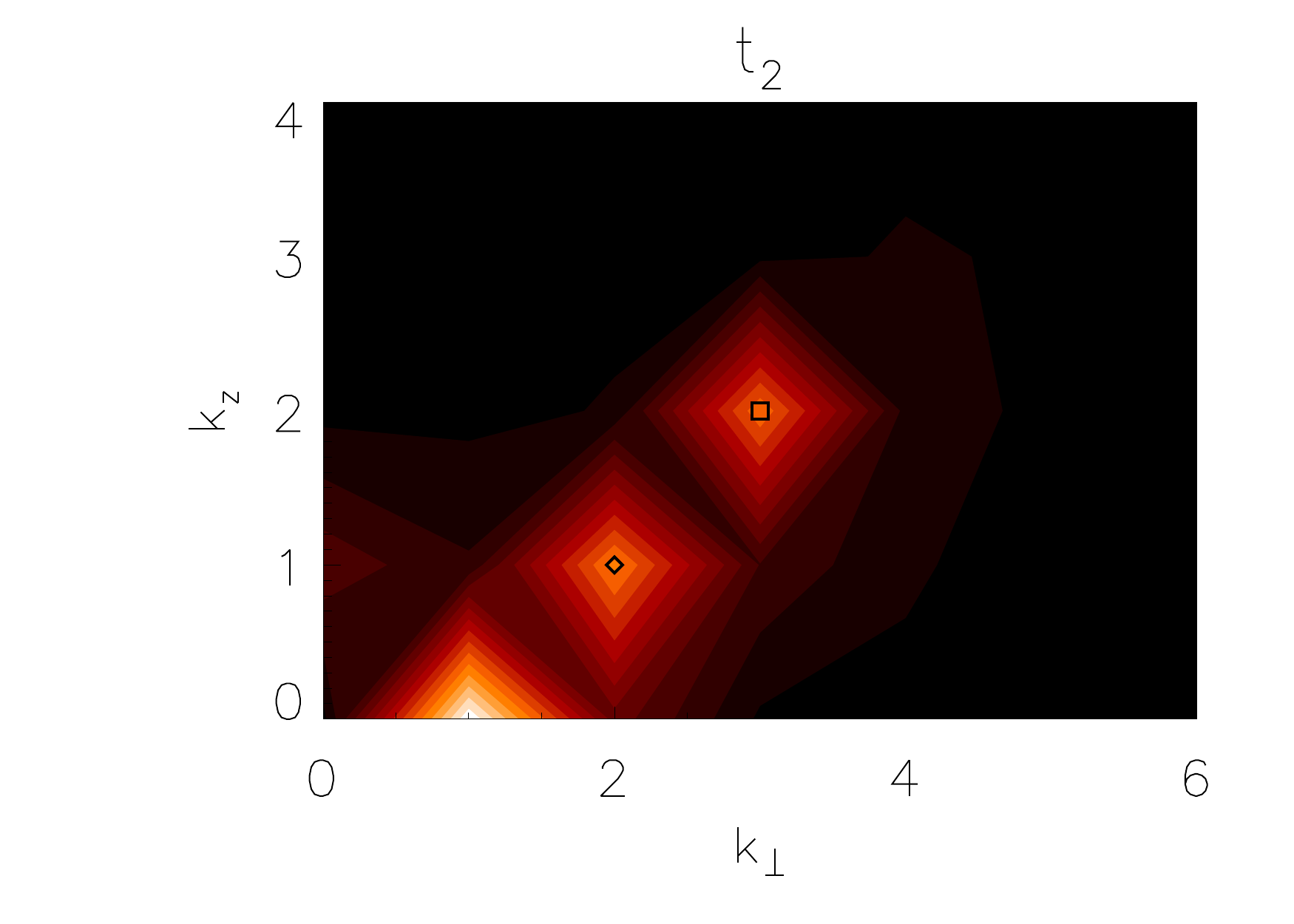}                                %%
}                                                                                     %%
\centerline{                                                                          %%
 \includegraphics[width=8cm]{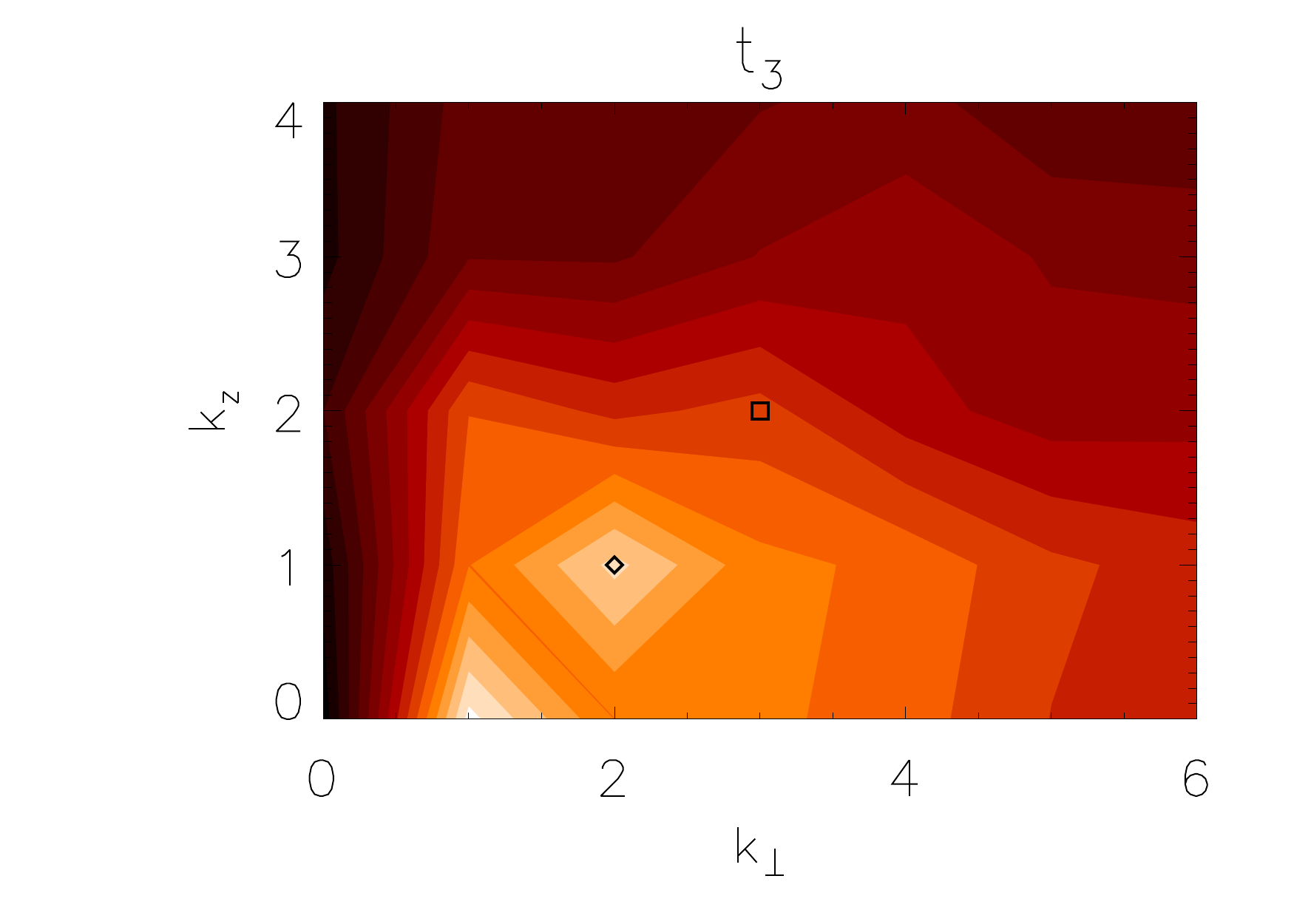}                              %%
 \includegraphics[width=8cm]{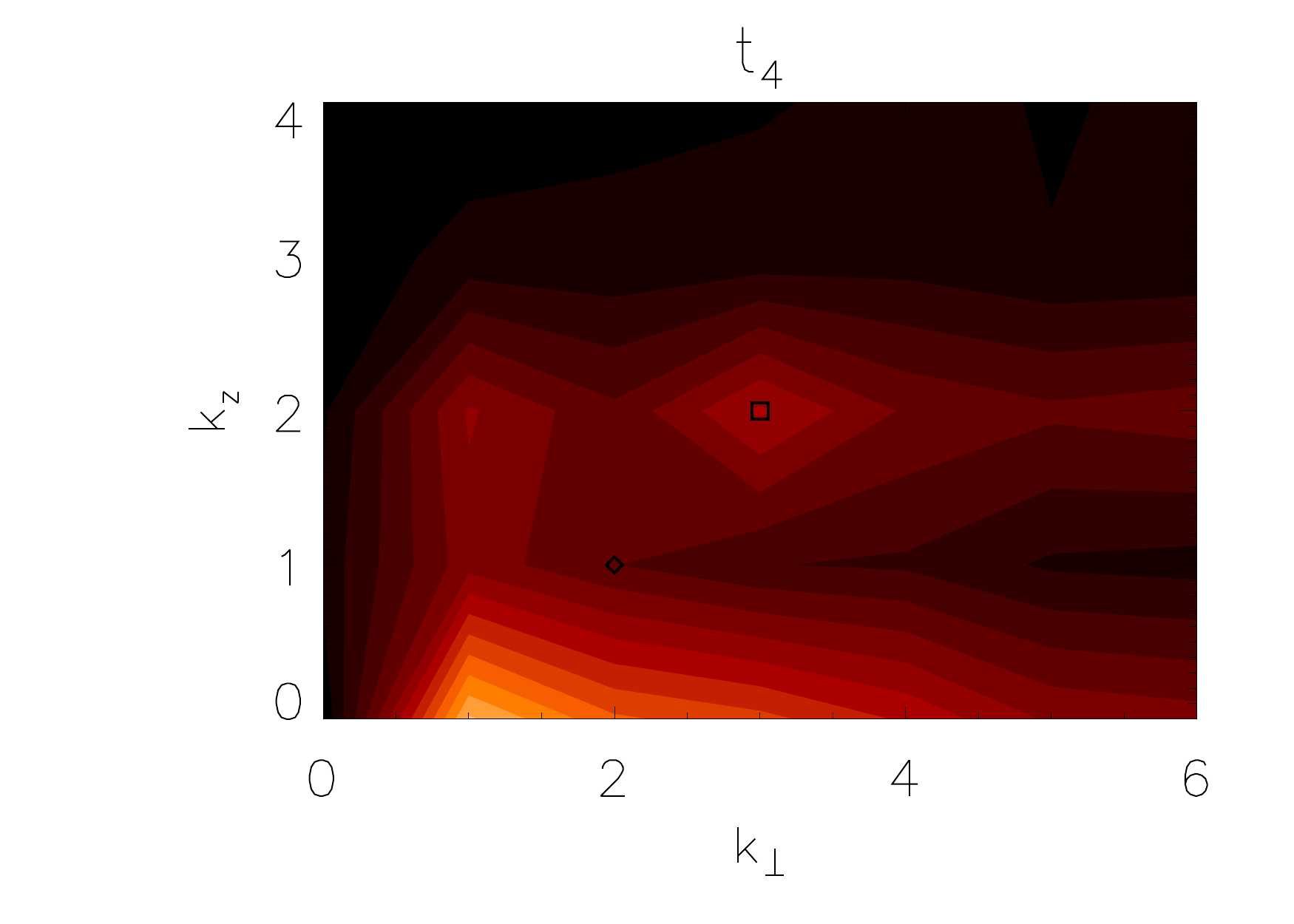}                             %%
}                                                                                     %%
\caption{Gray-scale (color online) images of the energy spectrum on the               %%
         $k_\perp,k_z$ plane. The four images correspond to the four times indicated  %%
         by the vertical dashed lines in the right panel of figure \ref{fig_6}. }     %%
\label{fig_7}                                                                         %%
\end{figure*}                                                                         %%
%%%%%%%%%%%%%%%%%%%%%%%%%%%%%%%%%%%%%%%%%%%%%%%%%%%%%%%%%%%%%%%%%%%%%%%%%%%%%%%%%%%%%%%%
More detail on the energy distribution during a burst can be seen in figure \ref{fig_7}
where a gray-scale (color online) image of the energy spectrum on the $k_\perp,k_z$ plane
is shown. The four images correspond to the four times indicated by the vertical dashed lines in 
the right panel of figure \ref{fig_6}. The forcing wave number on theses plots ($k_z=2,k_\perp=2\sqrt{2}\simeq3$)
is indicated by a square while the first unstable mode predicted by the linear theory ($k_z=1,k_x=2,k_y=0,\to k_\perp=2$, see eq.\ref{2mode})
is denoted by the diamond. The evolution of the burst then can be described  as follows:
during the decay phase most of the energy is in the $k_z=0$ plane (2D3C-flow) and a small part on the forcing wavenumber.
As the amplitude of the 2D3C flow decays the amplitude of the forcing mode increases approaching the solution \ref{v01}. 
As this solution is approached  a point is reached that the (2,0,1), (0,-2,-1) modes becomes unstable and start to grow.
When their amplitude becomes large enough (peak of the burst) the system becomes strongly nonlinear and allows for the violation
of the resonant conditions.
It thus couples to a large number of modes including modes with $k_z=0$. 
%The transfer of energy to the 2D3C flow is a sign that the first order expansion fails as it predicts no transfer to the 2D3C flow.
The energy of the unstable mode is transfered in all these modes and cascades to the dissipation scales.
This is true for all but the 2D3C modes that due to their quasi-two-dimensionality  do not cascade their energy to the 
small scales but in the large scales.
They thus form a large scale  condensate at $k_z=0,k_\perp=1$ that suppresses the instability. Afterwards it decays on the slow diffusive time scale since
there is no further injection of energy to sustain it. This process is then repeated.

%%%%%%%%%%%%%%%%%%%%%%%%%%%%%%%%%%%%%%%%%%%%%%%%%%%%%%%%%%%%%%%%%%%%%%%%%%%%%%%%%%%%%%%%
\begin{figure*}                                                                       %%
%\begin{center}                                                                       %%
\centerline{                                                                          %%
 \includegraphics[width=8cm]{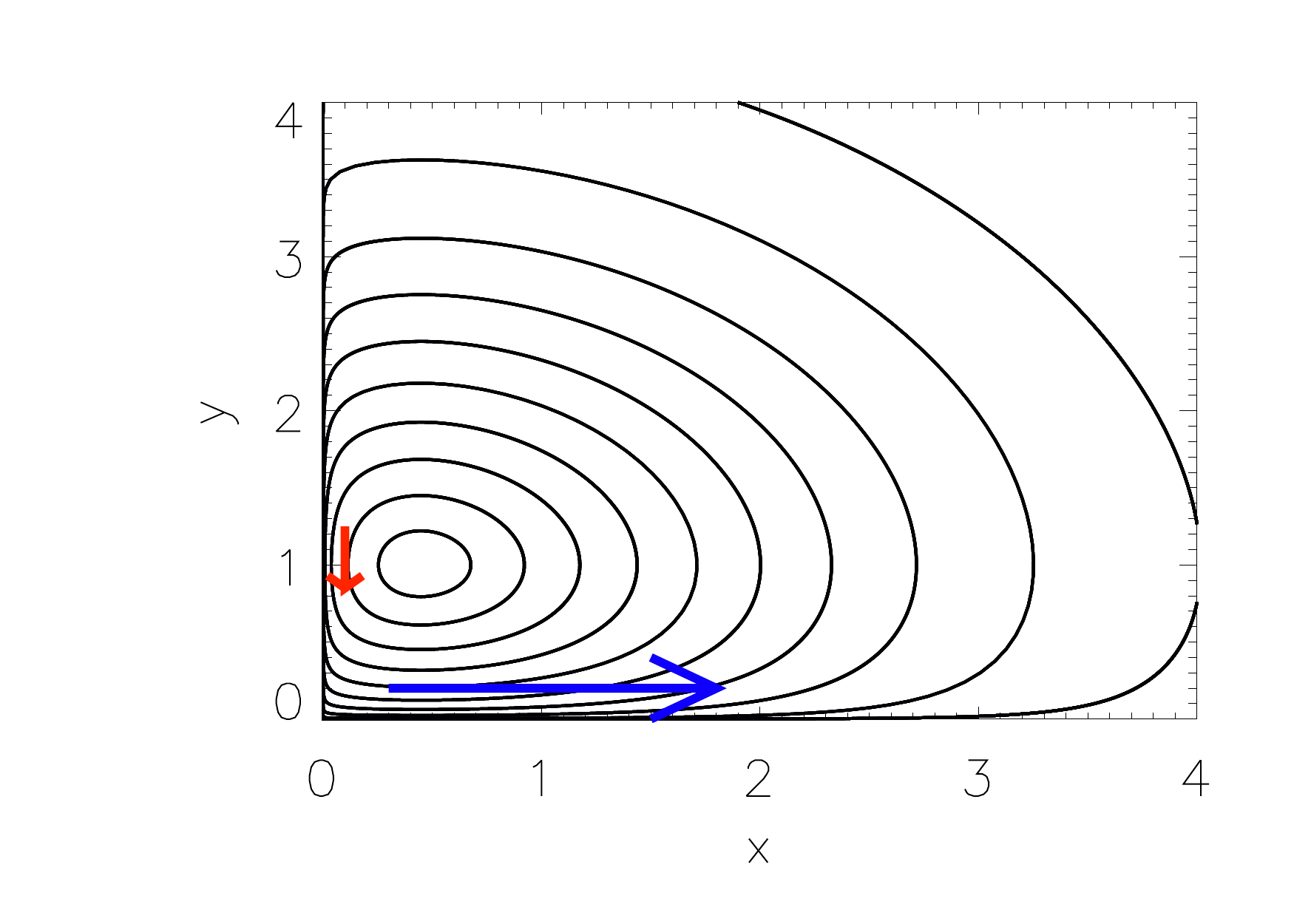}                                     %%
 \includegraphics[width=8cm]{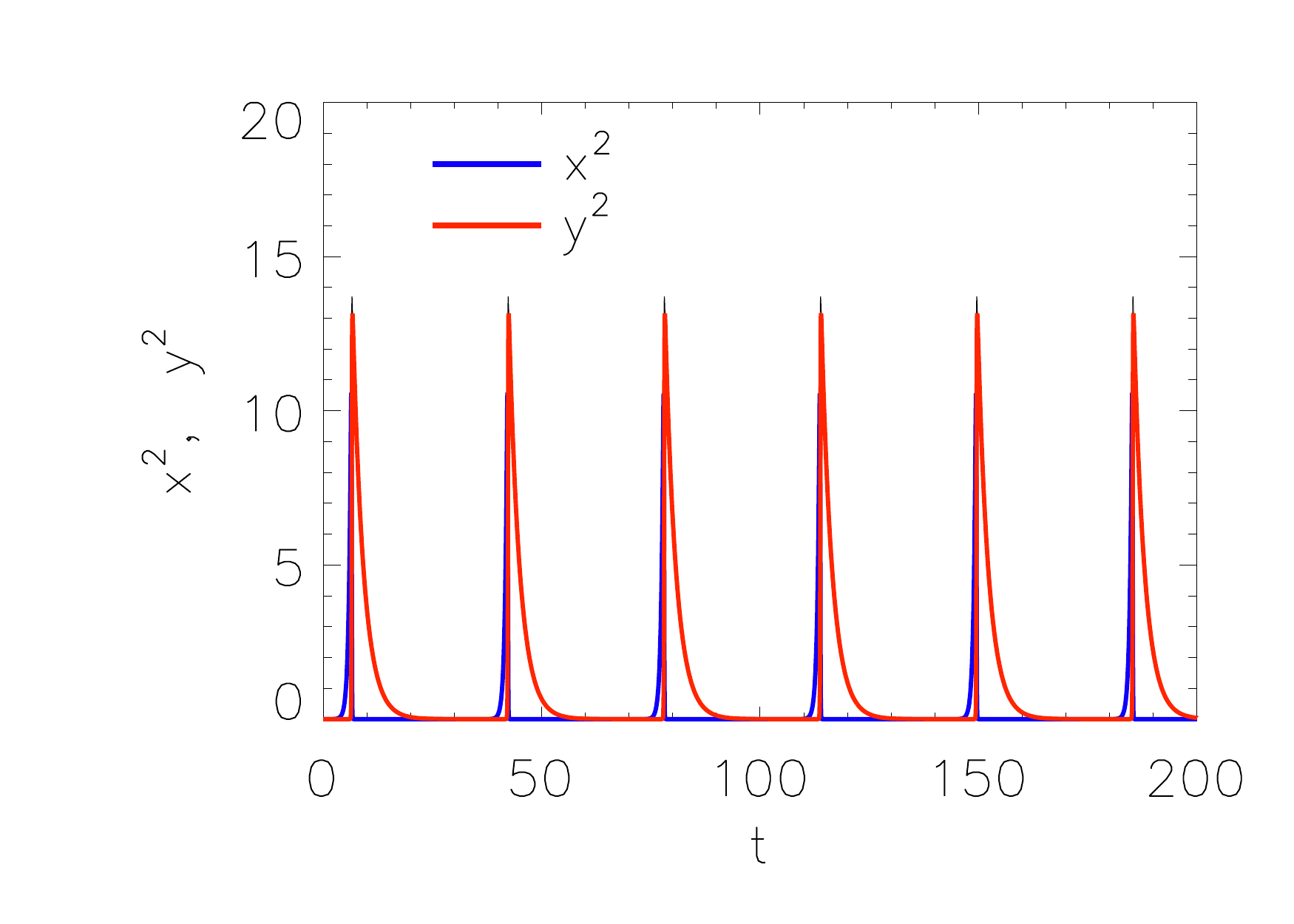}                                    %%
}                                                                                     %%
\caption{Phase-space trajectories of the model \ref{mdl1} (right panel), and time     %%
         evolution of a particular solution of the same model. }                      %%
\label{fig_8}                                                                        %%
\end{figure*}                                                                         %%
%%%%%%%%%%%%%%%%%%%%%%%%%%%%%%%%%%%%%%%%%%%%%%%%%%%%%%%%%%%%%%%%%%%%%%%%%%%%%%%%%%%%%%%%
This behavior of a fast increase followed by a slow decay is typical in dynamical systems that pass through a hyperbolic point for which the 
rate of attraction at the attracting manifold is small. The simplest version perhaps of such a model is written as:   
%%%%%%%%%%%%%%%%%%%%%%%%%%%%%%%%%%%%%%%%%%%%%%%%%%%%%%
\begin{equation}                                    %%
\begin{array}{ccr}                                  %% 
\dot{ x}  & =&            x - y^2x        \\        %% 
\dot{ y}  &= &  -\epsilon y + x^2y                  %%
\end{array}                                         %%
\label{mdl1}                                        %%
\end{equation}                                      %%
%%%%%%%%%%%%%%%%%%%%%%%%%%%%%%%%%%%%%%%%%%%%%%%%%%%%%%
where $\epsilon\ll 1$. This model has closed flow lines given by $C=\ln(yx^\epsilon)-\frac{1}{2}(x^2+y^2)$
shown in figure \ref{fig_8}, with $x=y=0$ being a hyperbolic unstable point and $x=\sqrt{\epsilon}$ and $y=1$ a neutrally stable point.
Trajectories passing near the unstable point (0,0) slowly approach it along the $y$-axis, until the $x$-mode becomes unstable leading 
a sudden increase of energy followed again by the slow decay. This procedure leads to the bursts shown in the right panel of \ref{fig_8}.

A more realistic model can perhaps be written taking in to account the dynamically most relevant modes in the system.
Based on the results of the linear study we consider two complex amplitudes $x,y$ that represent the two unstable modes
$(2,0,1)$ and $(0,-2,-1)$ respectively with frequency $\omega_x=\omega_y$
%=1/\sqrt{5}\epsilon$, 
one complex amplitude $z$ for the forcing mode $(2,2,2)$  with frequency $\omega_z$
%=1/\sqrt{3}\epsilon$ 
and one real amplitude $u$ representing the 2D3C field not affected by rotation.
%%%%%%%%%%%%%%%%%%%%%%%%%%%%%%%%%%%%%%%%%%%%%%%%%%%%%%%%%%%%%%%%%%%%%%%%%%%%%%%%%%%%%%%%%%%%%%%%%%%%%%%%%%%%%%%%%%%%%%%%%%%%
\begin{equation}                                                                                                          %%
\begin{array}{rcllrll}                                                                                                    %%
 \dot{x}   &=& \frac{i\omega_x}{\epsilon}x  &+& & z\,\, y    - \epsilon  |u|^{2} x  &-\lambda x    \\                     %%
 \dot{y}   &=& \frac{i\omega_y}{\epsilon}y  &+& & z^*   x    - \epsilon  |u|^{2} y  &-\lambda y    \\                     %%
 \dot{z}   &=& \frac{i\omega_z}{\epsilon}z  &-&2& y^*   x                           &-\lambda z + \frac{1}{\epsilon}   \\ %%
 \dot{u}   &=&                              &+&\epsilon&  u (|x|^2+|y^2|)                          &-\lambda u            %%
\end{array}     \label{model2}                                                                                            %%
\end{equation}                                                                                                            %%
%%%%%%%%%%%%%%%%%%%%%%%%%%%%%%%%%%%%%%%%%%%%%%%%%%%%%%%%%%%%%%%%%%%%%%%%%%%%%%%%%%%%%%%%%%%%%%%%%%%%%%%%%%%%%%%%%%%%%%%%%%%%
The coupling of the modes $x,y,z$ follows the coupling of the $(2,0,1)$,$(0,-2,-1)$,$(2,2,2)$ modes from the Navier-Stokes equation
while the coupling with 2D3C mode $u$ that is through strong multi mode interaction is modeled as a higher order nonlinearity.
The nonlinear term conserves the energy $|z|^2+|x|^2+|y|^2+|u|^2$. Just like the rotating Navier Stokes in the 
fast rotating limit $\epsilon\to0$  the non resonant modes decouple. 
%%%%%%%%%%%%%%%%%%%%%%%%%%%%%%%%%%%%%%%%%%%%%%%%%%%%%%%%%%%%%%%%%%%%%%%%%%%%%%%%%%%%%%%%%%%%
\begin{figure*}                                                                           %%
%\begin{center}                                                                           %%
\centerline{                                                                              %%
 \includegraphics[width=8cm]{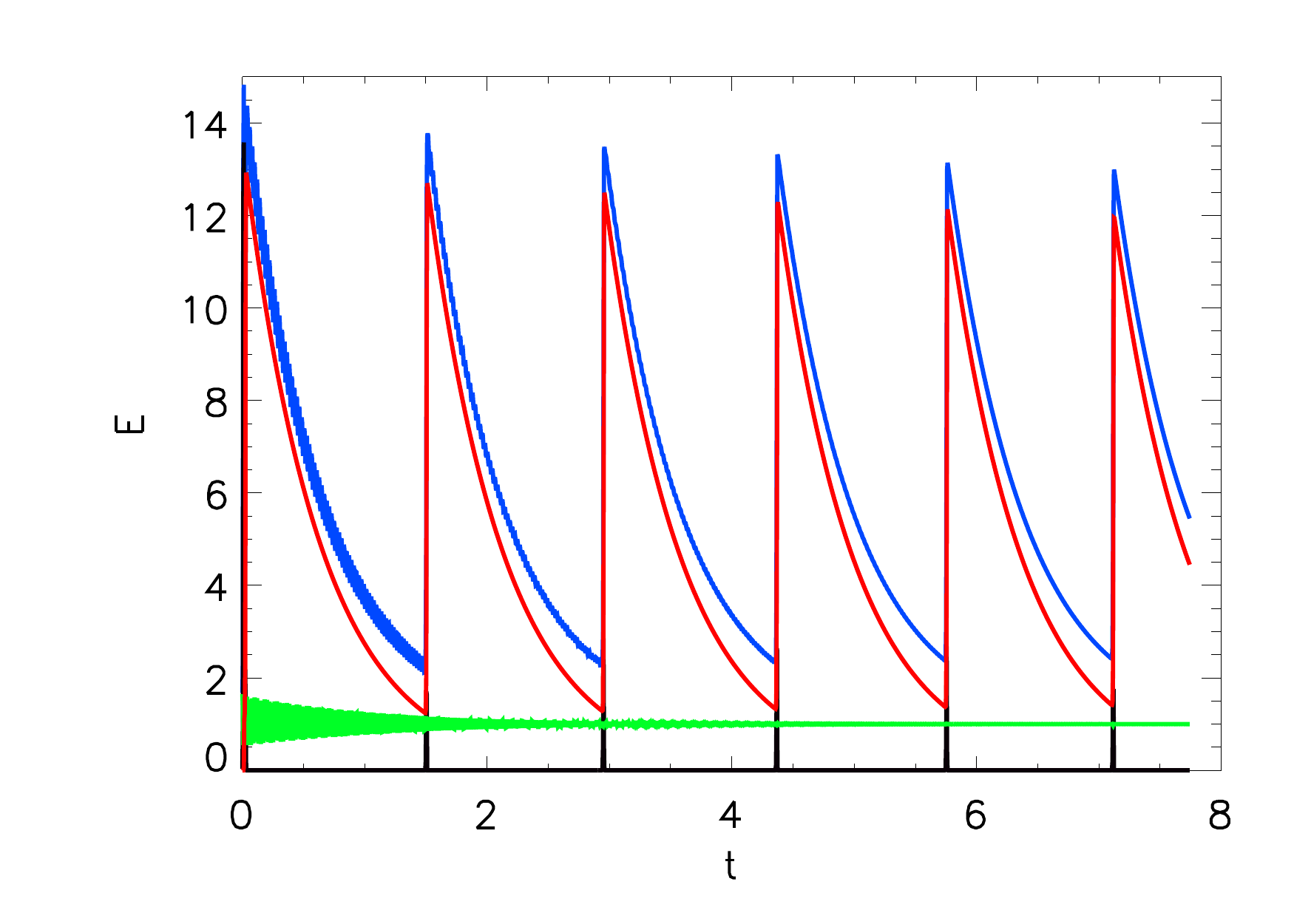}                                         %%
 \includegraphics[width=8cm]{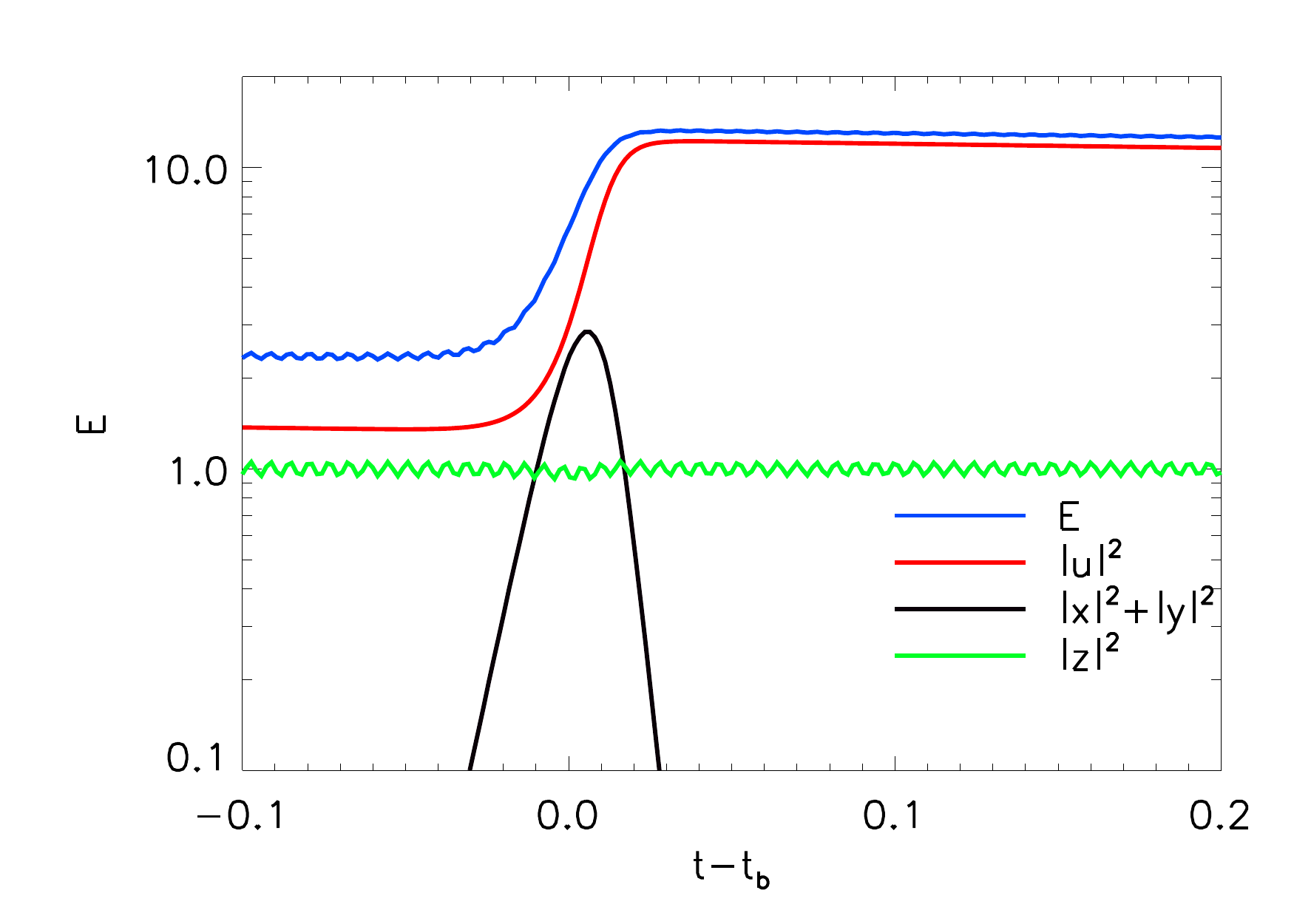}                                        %%
}                                                                                         %%
\caption{The evolution of different components of the energy of the model \ref{model2}.}  %%
\label{fig_9}                                                                             %%
\end{figure*}                                                                             %%
%%%%%%%%%%%%%%%%%%%%%%%%%%%%%%%%%%%%%%%%%%%%%%%%%%%%%%%%%%%%%%%%%%%%%%%%%%%%%%%%%%%%%%%%%%%%
In the absence of $U$ the unstable modes would saturate at large amplitude $\mathcal{O}(1/\sqrt{\epsilon})$.
However before this amplitude is reached $u$ becomes unstable and absorbs all energy. It then decays 
exponentially on a timescale $1/\lambda$.

A comparison of the results of the model with the direct numerical simulations can be seen 
by comparing figure \ref{fig_9} that shows the evolution of the energy of the model with
figure \ref{fig_6} that shows the results from direct numerical simulations. 
Although the full complexity of the DNS is not recovered due to the simplicity of the  
coupling to the 2D3C modes in the model, the basic features are reproduced.

%%%%%%%%%%%%%%%%%%%%%%%%%%%%%%%%%%%%%%%%%%%%%%%%%%%%%%%%%%%%%%%%%%%%%%%%%%%%%%%%%%%%%%%%%%%%%%%%%%%%%%%%%%%%%%%%%%%%%%%%%%%%%%%
%%%%%%%%%%%%%%%%%%%%%%%%%%%%%%%%%%%%%%%%%%%%%%%%%%%%%%%%%%%%%%%%%%%%%%%%%%%%%%%%%%%%%%%%%%%%%%%%%%%%%%%%%%%%%%%%%%%%%%%%%%%%%%%
\subsection{ 2D condensates and fully turbulent flows  }%%%%%%%%%%%%%%%%%%%%%%%%%%%%%%%%%%%%%%%%%%%%%%%%%%%%%%%%%%%%%%%%%%%%%%%
%%%%%%%%%%%%%%%%%%%%%%%%%%%%%%%%%%%%%%%%%%%%%%%%%%%%%%%%%%%%%%%%%%%%%%%%%%%%%%%%%%%%%%%%%%%%%%%%%%%%%%%%%%%%%%%%%%%%%%%%%%%%%%%
%%%%%%%%%%%%%%%%%%%%%%%%%%%%%%%%%%%%%%%%%%%%%%%%%%%%%%%%%%%%%%%%%%%%%%%%%%%%%%%%%%%%%%%%%%%%%%%%%%%%%%%%%%%%%%%%%%%%%%%%%%%%%%%

As discussed in Section \ref{sec:Param} in the parameter space for $\Ref>300$ and $1 \ge \Rof \ge \Ref^{-\alpha}$
the flow forms 2D3C-condensates. This class of flows describes the ultimate fate of rotating turbulence 
that is obtained in the limit $\Ref\to \infty$ for any fixed value of $\Rof$ with $\Rof<0.4$. 
The value of $\alpha$ depends at which order an injection of energy to the 2D3C flow first appears.
$\alpha$ is less than one (ie $n=1$) for in which case the bursts are observed.
If  second order terms couple inertial waves with the 2D3C flow and inject to it energy then $\alpha=1/3$ (ie $n=2$).
If not,  $\alpha=1/5$ (ie $n=3$) in which case the stationary flow alone injects energy to the 2D3C component.
 
%For any value of the exponent $\alpha$ in the range $1>\alpha\ge 1/5$ the asymptotic behavior of turbulence
%in the limit $\Ref\to\infty$ for any value of $\Rof<1$ falls in this class. 
%This class of flows then represent fully turbulent rotating flows.

\begin{figure*}                                                                       %%
%\begin{center}                                                                       %%
\centerline{                                                                          %%
 \includegraphics[width=8cm]{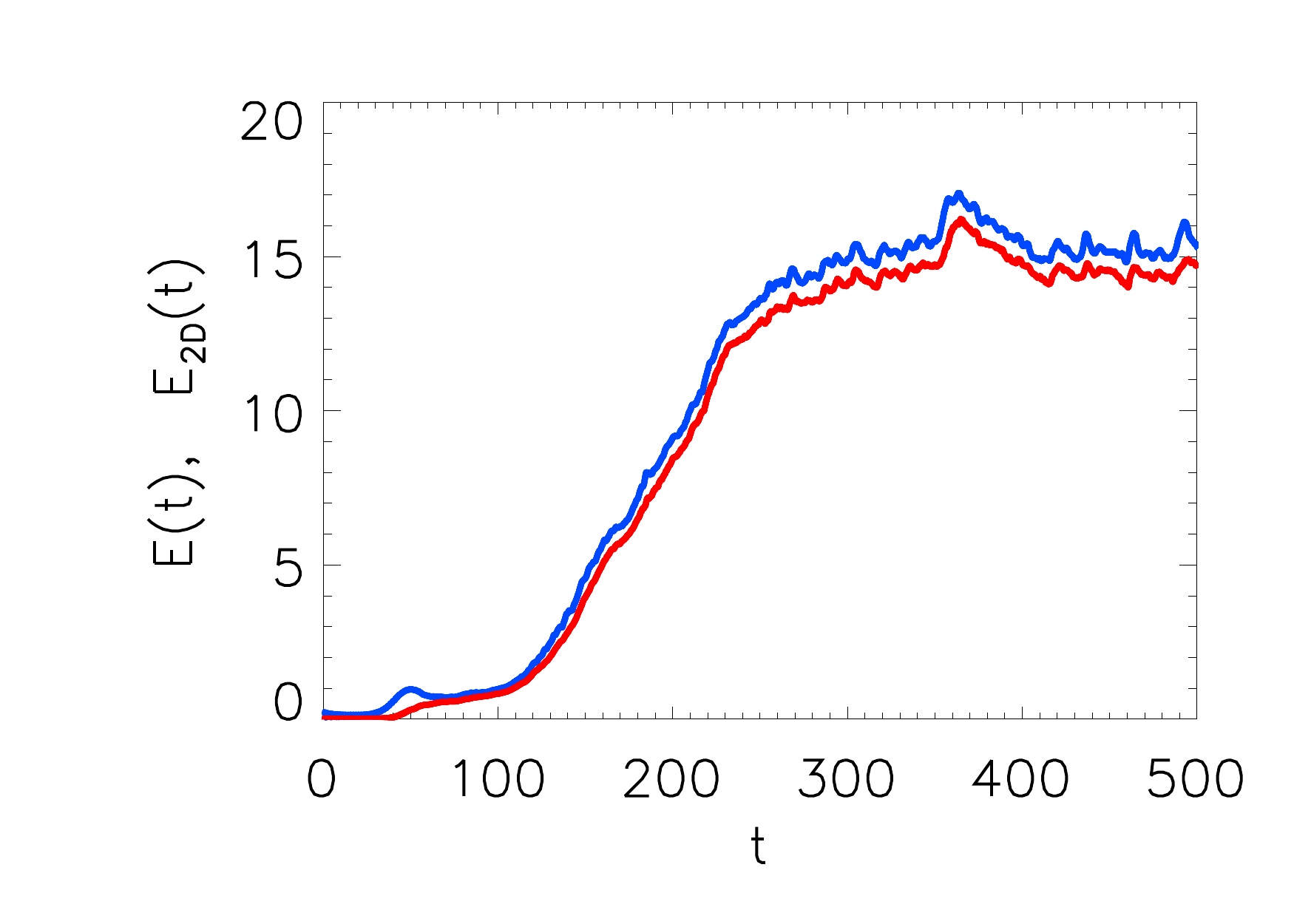}                                     %%
 \includegraphics[width=8cm]{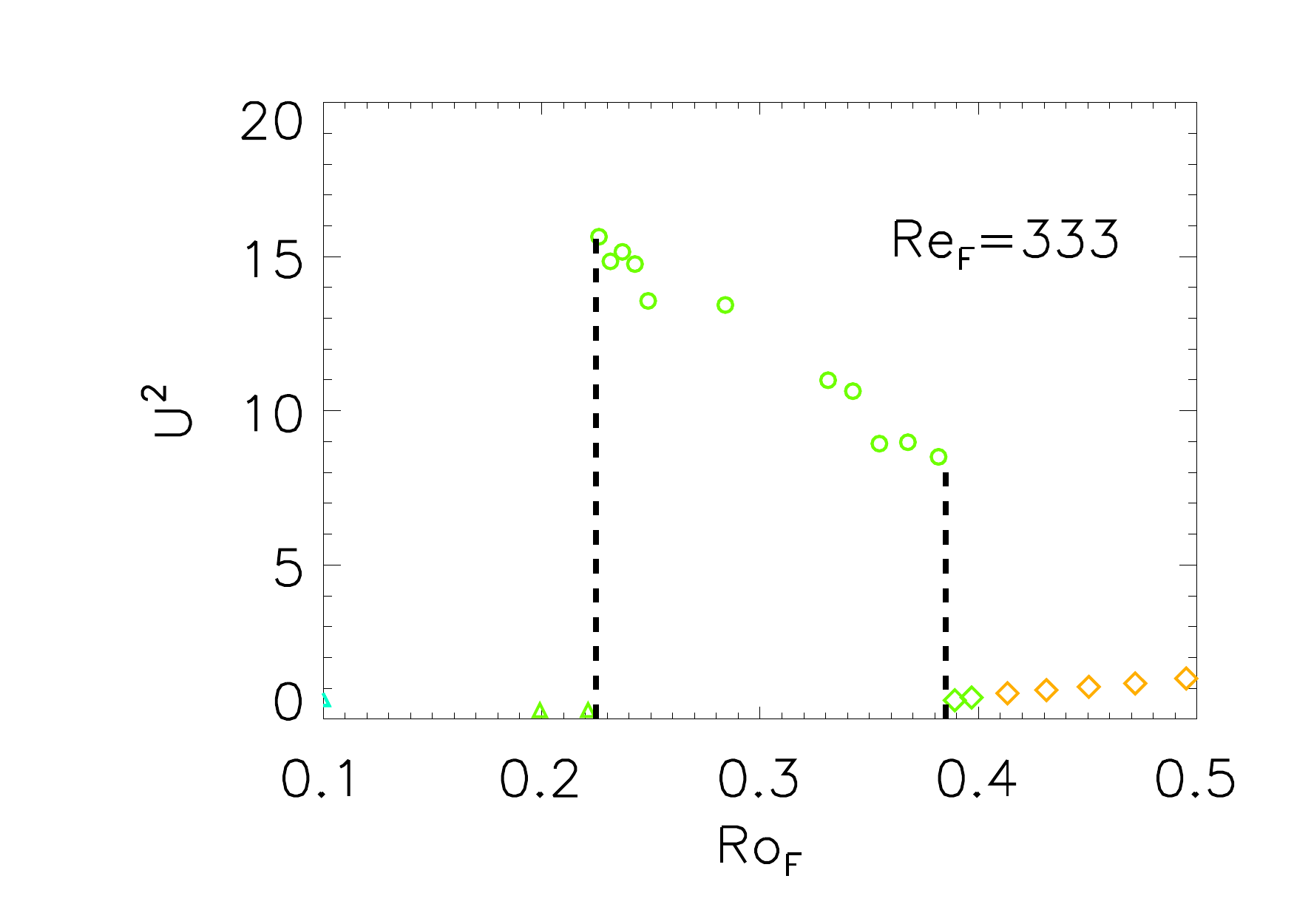}                                    %%
}                                                                                     %%
\caption{Left panel: The evolution of the total energy and of the 2D3C component of   %%
the energy for the run with $\Rof=0.25$ and $\Ref=500$. Right panel: The saturation   %%
level of the $U^2$ as a function of the Rossby $\Rof$                                 %%
for a fixed value of the Reynolds number $\Ref=333$.  }                               %%
\label{fig_10}                                                                        %%
\end{figure*}

Flows in this state are characterized by large values of energy that is concentrated
in the largest scales with $k_z=0$. The evolution of the total energy and of the 2D3C component of the energy 
for the run with $\Rof=0.25$ and $\Ref=500$ is shown in the left panel of figure \ref{fig_10}. After a short time ($t<100$) 
during which the energy appears to saturate at a value close to unity the 2D3C flow starts to grow linearly  until 
saturation is reached at very large values of energy. The non-2D3C part of the energy that consist of the energy 
of the inertial waves and the forced modes comprises only a small part of the energy. This behavior was observed 
for all runs that showed a quasi-2D condensate behavior. 

The saturation amplitude is much larger than
unity. The right panel of figure \ref{fig_10} shows the saturation level of the $U^2$ as a function of the Rossby $\Rof$
for a fixed value of the Reynolds number $\Ref=333$. $U^2$ varies discontinuously as $\Rof$
is increased, both at the critical value that it transitions from isotropic turbulence to the condensate state at $\Rof=0.39$
and at the critical value that it transitions from the condensate state to the intermittent bursts behavior at $\Rof=0.22$.
Thus the transition from the isotropic turbulent state to this condensate state is found to be sub-critical. 
%%%%%%%%%%%%%%%%%%%%%%%%%%%%%%%%%%%%%%%%%%%%%%%%%%%%%%%%%%%%%%%%%%%%%%%%%%%%%%%%%%%%%%%%
                                                                         %%
%%%%%%%%%%%%%%%%%%%%%%%%%%%%%%%%%%%%%%%%%%%%%%%%%%%%%%%%%%%%%%%%%%%%%%%%%%%%%%%%%%%%%%%%

\begin{figure*}                                                                       %%
\centerline{                                                                          %%
 \includegraphics[width=8cm]{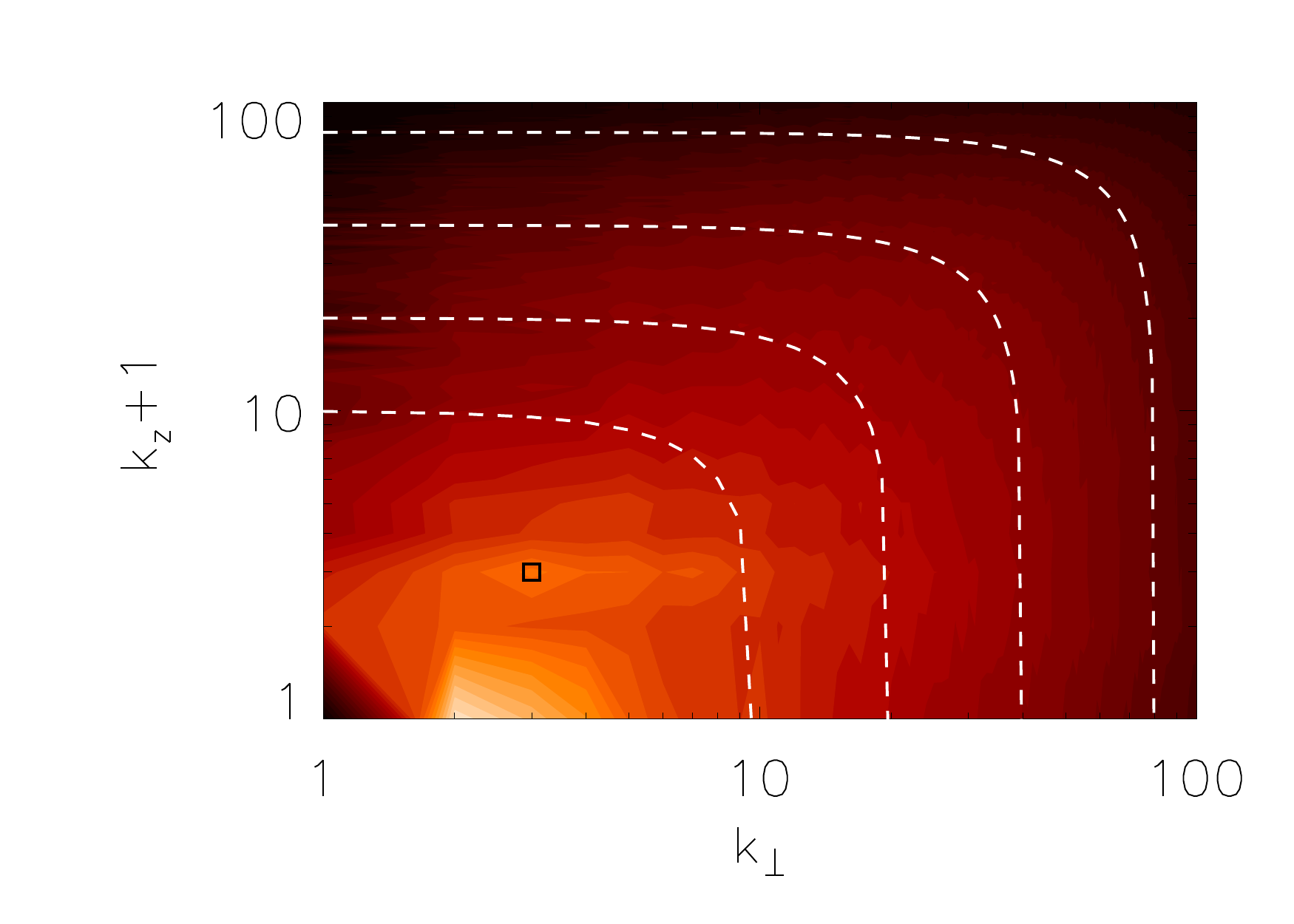}                                     %%
 \includegraphics[width=8cm]{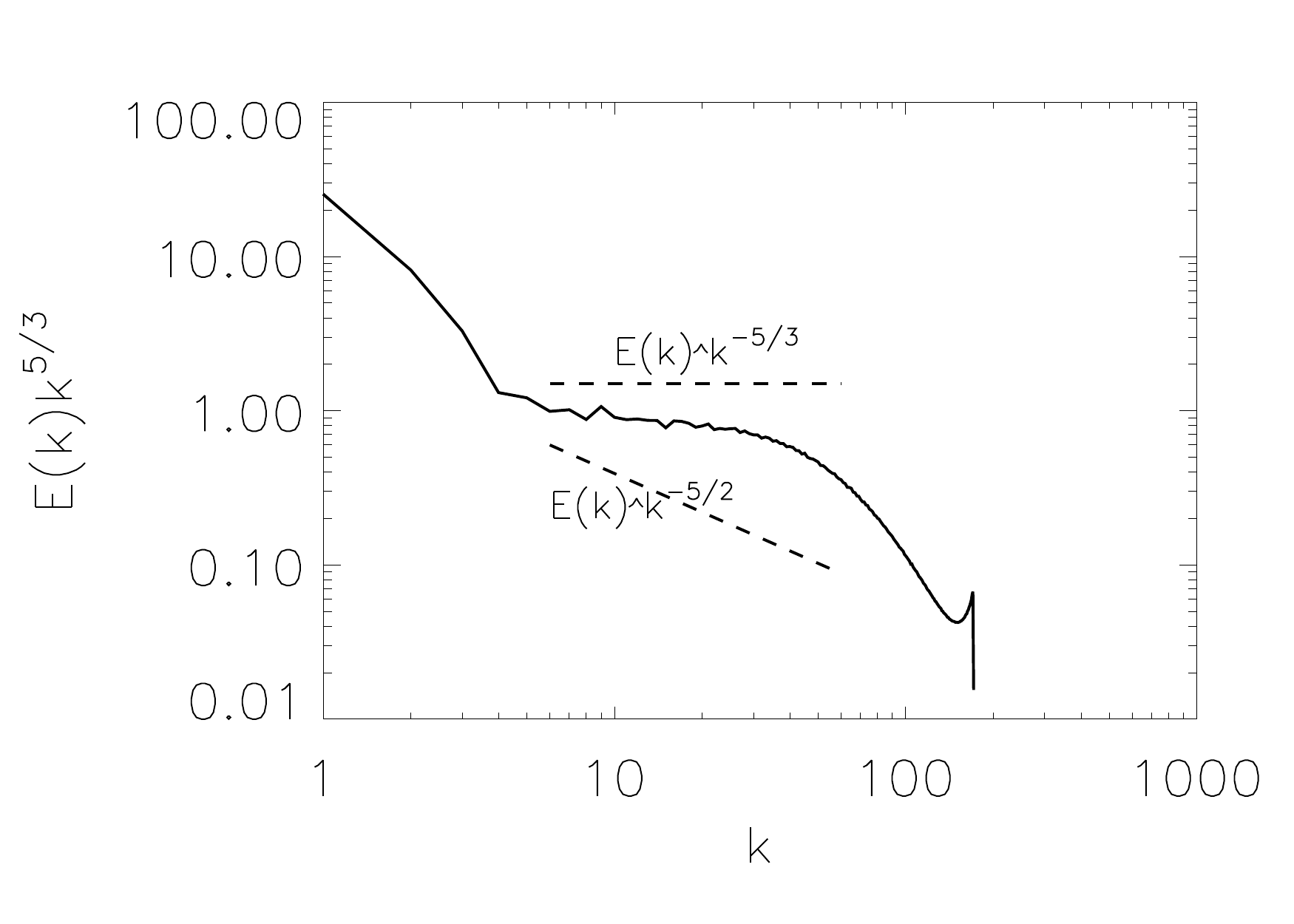}                                    %%
}                                                                                     %%
\caption{The two-dimensional energy spectrum $E_{_{2D}}(k_z,k_\perp)$                 %%
(left panel) and the 1D energy spectrum $E_{_{1D}}$ (right panel) }                   %%
\label{fig_11}                                                                        %%
\end{figure*}

The quasi-2D3C behavior of the flow can also be seen by looking at the energy spectra.
In figure \ref{fig_11} we show the two-dimensional energy spectrum $E_{_{2D}}(k_z,k_\perp)$ 
(left panel) and the 1D energy spectrum $E_{_{1D}}$ (right panel) compensated by $k^{5/3}$ defined as: 
\[
E_{_{2D}}(k_z,k_\perp)=\sum_{ \scriptsize
 \begin{array}{c} {k_z\le p_z< k_z+1} \\ {k_\perp \le p_\perp < k_\perp+1}\end{array} } |{\bf u_p}|^2,\qquad \quad
E_{_{1D}}(k_z,k_\perp)=\sum_{k \le |{\bf p} | < k+1 } |{\bf u_p}|^2
\]
of the run with $\Rof=0.20$ and $\Ref=2000$ that corresponds to the run with the largest $\Ref$
and smallest $\Rof$ that displayed the quasi-2D3C condensate behavior.
%%%%%%%%%%%%%%%%%%%%%%%%%%%%%%%%%%%%%%%%%%%%%%%%%%%%%%%%%%%%%%%%%%%%%%%%%%%%%%%%%%%%%%%%
                                                                         %%
%%%%%%%%%%%%%%%%%%%%%%%%%%%%%%%%%%%%%%%%%%%%%%%%%%%%%%%%%%%%%%%%%%%%%%%%%%%%%%%%%%%%%%%%
The square in the left panel indicates the location of the forcing. The dashed lines indicate 
the location of the circles $k_z^2+k_\perp^2=10,20,40,80$. An isotropic spectrum would be
constant along these lines. 
%%%%%%%%%%%%%%%%%
%Although the 2D spectrum does not appear isotropic, 
%the deviations from isotropy do not appear so pronounced.
In the large scales the $k_z=0$ modes have significantly larger 
amplitude and energy is concentrated in the in $k_\perp=\sqrt{2}$.
The small scales appear more isotropic with the exception of the 
$k_z\gg k_\perp$ modes that appear to be quenched and deviate stronger from isotropy. 
 The dominance of the large scale modes can be observed in the 
1D energy spectrum. A large amount of energy is concentrated at small
wave-numbers followed by steep drop. The small scales on the other hand display a power law behavior with an index slightly 
smaller than the Kolmogorov prediction $-5/3$, and much bigger than the wave turbulence 
prediction $-5/2$.   It is expected that at even higher values of $\Rof$ the isotropic Kolmogorov spectrum 
will be even better established in the small scales.

\begin{figure*}                                                                       %%
%\begin{center}                                                                       %%
\centerline{                                                                          %%
 \includegraphics[width=6cm]{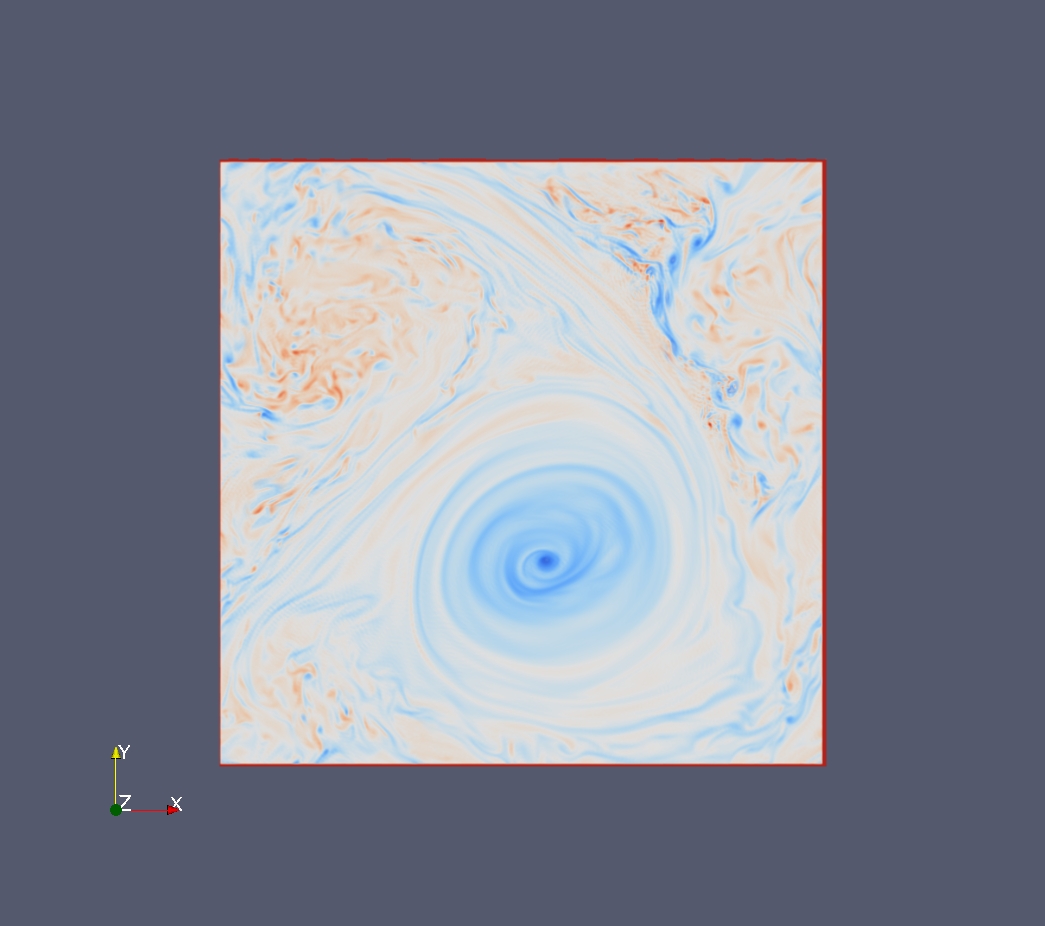}  \hspace{1cm }                      %%
 \includegraphics[width=6cm]{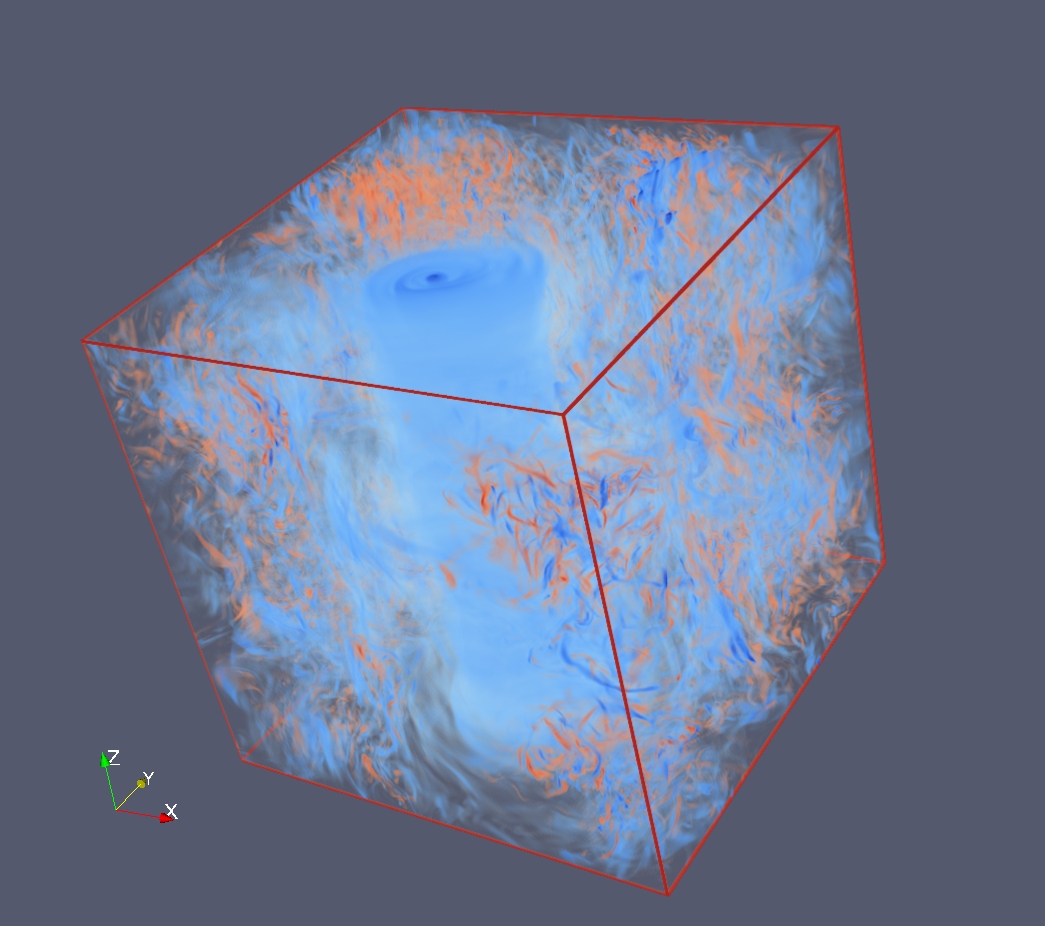}                                    %%
}                                                                                     %%
\caption{A visualization of the flow where the $z$-component of the vorticity         %% 
is shown.  The panel on the left shows the computational box viewed from the top,     %%
while the same flow is shown viewed from the side on the right panel where the        %%
opacity has been reduced so that the structures inside the                            %%
computational box can be seen.}                                                       %%
\label{fig_12}                                                                        %%
\end{figure*}

A visualization of the flow in the
condensate state is depicted in figure \ref{fig_12} where the $z$-component of the vorticity
is shown.  The panel on the left shows the computational box viewed from the top, while the same result 
is shown viewed from the side on the right panel where the opacity has been reduced so that the structures inside the 
computational box can be seen. The blue color indicates that vorticity is parallel to the rotation (the flow co-rotates)
while the red color indicates that the vorticity is anti-parallel to the rotation.  
Clearly both large and small scales coexist in the flow. 
A large scale, 2D co-rotating columnar structure can be seen. 
Opposite to this columnar structure a second columnar structure rotating in the opposite direction 
can be seen. This second structure (best seen in the right panel of figure \ref{fig_12}) is fluctuating strongly 
having more small scales. The persistence of co-rotating structures in rotating turbulence has been observed in 
experiments \cite{Hopfinger1982,Morize2005,Gallet2014} and has been discussed in various works \cite{Bartello1994,Hopfinger1993,Gence2001,Staplehurst2008,Sreenivasan2008,Gallet2014}. 
Perhaps it is not surprising that structures with vorticity anti-aligned to rotation are more responsive to 3D perturbations.
Following \citep{Bartello1994} the local Rossby number $\mathrm{Ro}_{loc}({\bf x})$
(defined by the local rotation rate $2\Omega_{local}={2\bf\Omega+w}({\bf x},t)$) will be increased and thus
are less likely to show a quasi-2D3C behavior.  
%%%%%%%%%%%%%%%%%%%%%%%%%%%%%%%%%%%%%%%%%%%%%%%%%%%%%%%%%%%%%%%%%%%%%%%%%%%%%%%%%%%%%%%%
                                                                         %%
%%%%%%%%%%%%%%%%%%%%%%%%%%%%%%%%%%%%%%%%%%%%%%%%%%%%%%%%%%%%%%%%%%%%%%%%%%%%%%%%%%%%%%%%

\begin{figure*}                                                                         %%
\centerline{                                                                            %%
 \includegraphics[width=8cm]{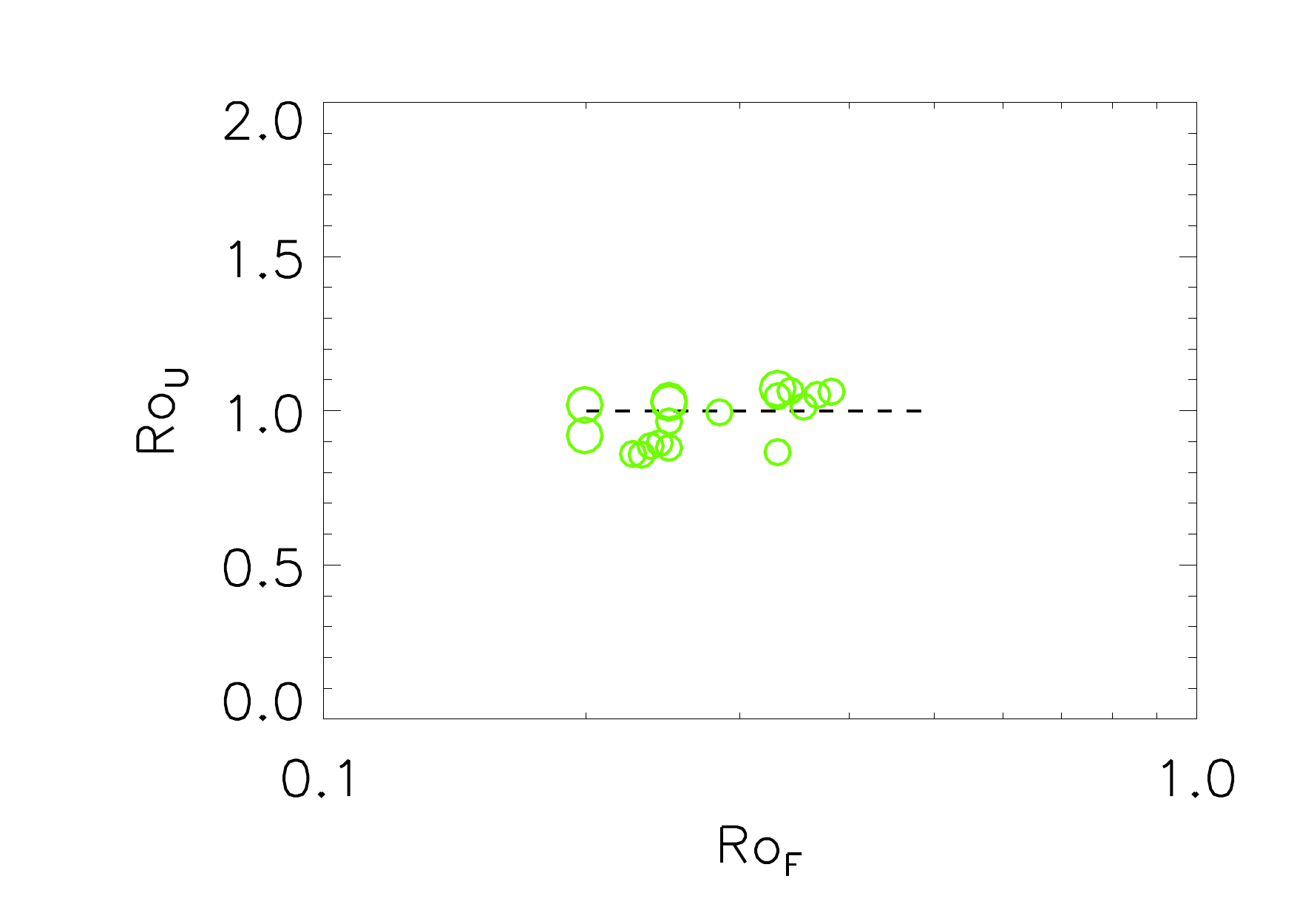}                                       %%
 \includegraphics[width=8cm]{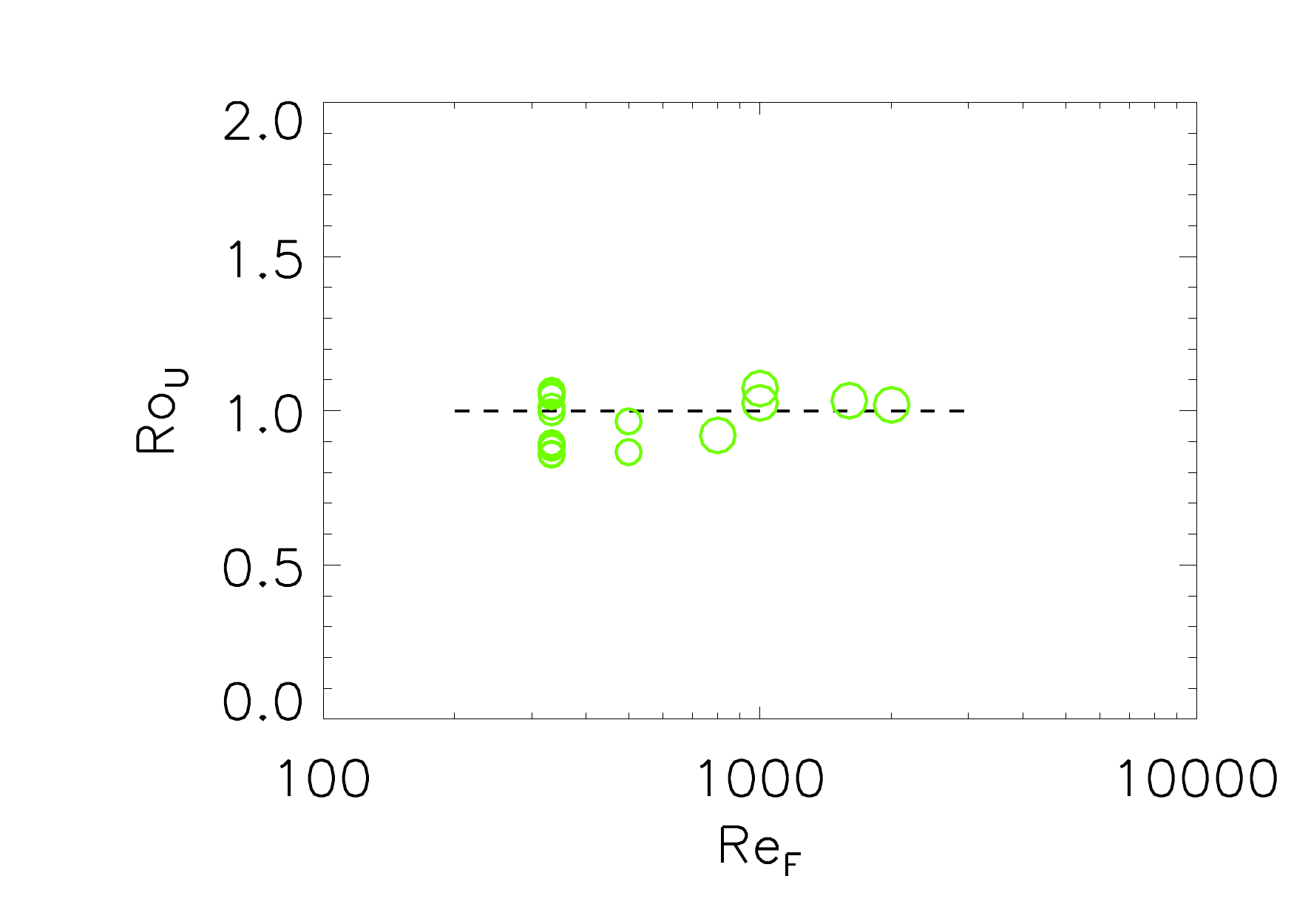}                                      %%
}                                                                                       %%
\caption{$\Rou$ as a function of $\Rof$ (left panel) and $\Ref$ (right panel) for all   %%
         the runs that lead to a quasi-2ED condensate. The $\Rou=1$ behavior shown here %%
         can also be seen in the left panel of figure \ref{fig_2}.}                     %%
\label{fig_13}                                                                          %%
\end{figure*}

With this observation one can easily understand how saturation is reached in the large scales.
The fast rotation leads to a quasi-two-dimensionalization of the system that leads to an inverse cascade
of energy. As energy of the 2D3C part of the flow increases it reaches the largest scale of the system 
where it forms two counter rotating vortexes. As the amplitude of these two vortexes is increased
a point is reached that the vertical vorticity of the counter rotating vortex will cancel 
the effect of rotation locally leading to an order one value of the local Rossby number.
Then the quasi-2D constrain that leads to the inverse cascade and the pile-up of energy to 
the large scales is broken and energy starts to flow back to the small scales.
Such a mechanism of course implies that saturation is reached when 
the eddy turn-over-time of the condensate is the same order with the rotation period or more simply  $\Rou\sim1$. 
Indeed in figure \ref{fig_13} we plot $\Rou$ from all the runs that showed
a condensate behavior as a function of $\Rof$ (left panel) and as a function of $\Ref$ (right panel)
%%%%%%%%%%%%%%%%%%%%%%%%%%%%%%%%%%%%%%%%%%%%%%%%%%%%%%%%%%%%%%%%%%%%%%%%%%%%%%%%%%%%%%%%%%
                                                                           %%
%%%%%%%%%%%%%%%%%%%%%%%%%%%%%%%%%%%%%%%%%%%%%%%%%%%%%%%%%%%%%%%%%%%%%%%%%%%%%%%%%%%%%%%%%%%
The value of $\Rou$ appears to be independent of both $\Rof$ and $\Ref$ thus the
scaling $\Rou\sim1$ is verified. This result implies that this state in unlikely to be described
by an $\Rou\to0$ expansion as the eddy turn-over time is the same order with the rotation period.
It also implies that condensates saturate by reaching a state of marginal inverse cascade \citep{Seshasayanan2014}.

%
%%%%%%%%%%%%%%%%%%%%%%%%%%%%%%%%%%%%%%%%%%%%%%%%%%%%%%%%%%%%%%%%%%%%%%%%%%%%%%%%%%%%%%%%%%%%%%%%%%%%%%%%%%%%%%%%%%%%%%%%%%%%%%%
%%%%%%%%%%%%%%%%%%%%%%%%%%%%%%%%%%%%%%%%%%%%%%%%%%%%%%%%%%%%%%%%%%%%%%%%%%%%%%%%%%%%%%%%%%%%%%%%%%%%%%%%%%%%%%%%%%%%%%%%%%%%%%% 
%%%%%%%%%%%%%%%%%%%%%%%%%%%%%%%%%%%%%%%%%%%%%%%%%%%%%%%%%%%%%%%%%%%%%%%%%%%%%%%%%%%%%%%%%%%%%%%%%%%%%%%%%%%%%%%%%%%%%%%%%%%%%%% 
%%%%%%%%%%%%%%%%%%%%%%%%%%%%%%%%%%%%%%%%%%%%%%%%%%%%%%%%%%%%%%%%%%%%%%%%%%%%%%%%%%%%%%%%%%%%%%%%%%%%%%%%%%%%%%%%%%%%%%%%%%%%%%% 
\section{Large $\Ref$, small $\Rof$ asymptotic behavior of turbulence }   %%%%%%%%%%%%%%%%%%%%%%%%%%%%%%%%%%%%%%%%%%%%%%%%%%%%%
%%%%%%%%%%%%%%%%%%%%%%%%%%%%%%%%%%%%%%%%%%%%%%%%%%%%%%%%%%%%%%%%%%%%%%%%%%%%%%%%%%%%%%%%%%%%%%%%%%%%%%%%%%%%%%%%%%%%%%%%%%%%%%%
%%%%%%%%%%%%%%%%%%%%%%%%%%%%%%%%%%%%%%%%%%%%%%%%%%%%%%%%%%%%%%%%%%%%%%%%%%%%%%%%%%%%%%%%%%%%%%%%%%%%%%%%%%%%%%%%%%%%%%%%%%%%%%% 
%%%%%%%%%%%%%%%%%%%%%%%%%%%%%%%%%%%%%%%%%%%%%%%%%%%%%%%%%%%%%%%%%%%%%%%%%%%%%%%%%%%%%%%%%%%%%%%%%%%%%%%%%%%%%%%%%%%%%%%%%%%%%%% 
%%%%%%%%%%%%%%%%%%%%%%%%%%%%%%%%%%%%%%%%%%%%%%%%%%%%%%%%%%%%%%%%%%%%%%%%%%%%%%%%%%%%%%%%%%%%%%%%%%%%%%%%%%%%%%%%%%%%%%%%%%%%%%%

The results so far have shown that different behaviors are present for large $\Ref$ and small $\Rof$ depending on the ordering
of these parameters. It is thus also expected that the basic energy balance relations ) that link the  
the forcing amplitude and the velocity amplitude with the energy injection/dissipation rate) will differ 
for the different states of the system.
Knowledge of the relation of these quantities allows to determine the map between the control parameter 
pairs $(\Rof,\Ref)$, $(\Rou,\Reu)$ and $(\Rod,\Red)$.

In non-rotating turbulence in the high $\Ref$ limit it is expected that the role of viscosity is unimportant in the large scales. 
With this assumption the relation between the forcing amplitude, the velocity fluctuations amplitude $U$ and the energy 
dissipation rate can be derived by dimensional analysis. It results in the following two relations:
\beq
U \propto C_{_U}^{^T},   \qquad \epsilon \propto C_{_D}^{^T}   U^3 q,
\label{turb_rel}
\eeq
where %$\tilde{U}$ is the dimensional root mean square velocity ({\it ie} $\tilde{U}=U\sqrt{F_0L}$) and
the proportionality coefficients $C_{_U}^{^T},C_{_D}^{^T}$ are order one numbers. 
(Here it is reminded to the reader that $U$ stands for the root mean square velocity non-dimentionalized by $\sqrt{F_0L}$.
Thus the first part of equation \ref{turb_rel} simply implies the dimensional velocity amplitude is proportional to $\sqrt{F_0L}$).
In the other limit for which $\Ref$ is very small then the nonlinearity can be neglected and obtain the laminar scaling
\beq
U \propto C_{_U}^{^L} \Ref q^{-2},   \qquad \epsilon \propto  C_{_U}^{^L} \Reu^{-1} {U}^3q \propto \Ref^{-1} q^{-2}  
\label{lam_rel} 
\eeq
by balancing the forcing with the viscous term. These scalings are expected to be valid for weakly rotating flows
$\Rof\gg1$ where the effect of rotation can be neglected.

%In rotating turbulence another parameter is introduced, 
For $\Rof\lesssim 1$  however the rotation can not be neglected and this makes the energy balance relations 
for the high $\Ref$ limit less straight forward. The increase in complexity stems from the fact that:
first even if $\Ref$ is large we cannot conclude independence on the viscosity with out also specifying the value of $\Rof$. 
This limitation exists because in general rotation diminishes the energy cascade thus for any value of $\Reu$ 
there would exist a value of $\Rou$ small enough so that  energy cascade flux is comparable to the dissipation.
Second even if viscosity is neglected in the large $\Ref$ limit we are still left with on non-dimensional parameter the Rossby number,
making the derivation of scaling laws by simple dimensional analysis to require further physically motivated arguments.

For the Taylor-Green flow we have already showed that the flow is guaranteed to be laminar when the relations
\ref{enstb1},\ref{enstb2} hold. Thus in this range 
the desired relations can be directly be obtained from the laminar solutions that for $\Rof \ll 1 \,\, \& \,\, \Ref \propto \Rof^{-1}$ become:
\beq
U \simeq  \sqrt{3} \Rof,   \qquad \epsilon \simeq 3q^2U^3\Reu^{-1} \simeq 9 q^2 \Rof^2/\Ref  %\qquad \mathrm{for} \quad \Rof \le c_1 \Ref 
\label{rlam_rel}
\eeq

For large $\Ref$ and fast rotation rate wave-turbulence arguments based on three wave-interactions 
(see chapter 3 of \cite{Nazarenko_book2011}) suggest that 
energy cascade rate to the small scales will be decreased by a factor $\Rou$. This reasoning leads to the relations
\beq
U \propto C_{_U}^{^{WT}} \Rof,  \qquad \epsilon \propto  C_{_D}^{^{WT}} \Rou U^3 q 
\label{lam_rel} 
\eeq
will hold.
The first relation comes from a balance of the Coriolis term in the Navier Stokes equation with the forcing 
while the second relation is a weak turbulence estimate that is derived assuming an ensemble of random
traveling waves whose fast de-correlation time leads to the reduction in the energy cascade rate by a factor proportional
to their inverse speed (here $\Rof$). These arguments however assume uniform and isotropic forcing and do not take into account 
the formation of condensates. Thus are not necessarily expected to hold for a structured forcing like the Taylor-Green.

%%%%%%%%%%%%%%%%%%%%%%%%%%%%%%%%%%%%%%%%%%%%%%%%%%%%%%%%%%%%%%%%%%%%%%%%%%%%%%%%%%%%%%%%
\begin{figure*}                                                                       %%
%\begin{center}                                                                       %%
\centerline{                                                                          %%
 \includegraphics[width=8cm]{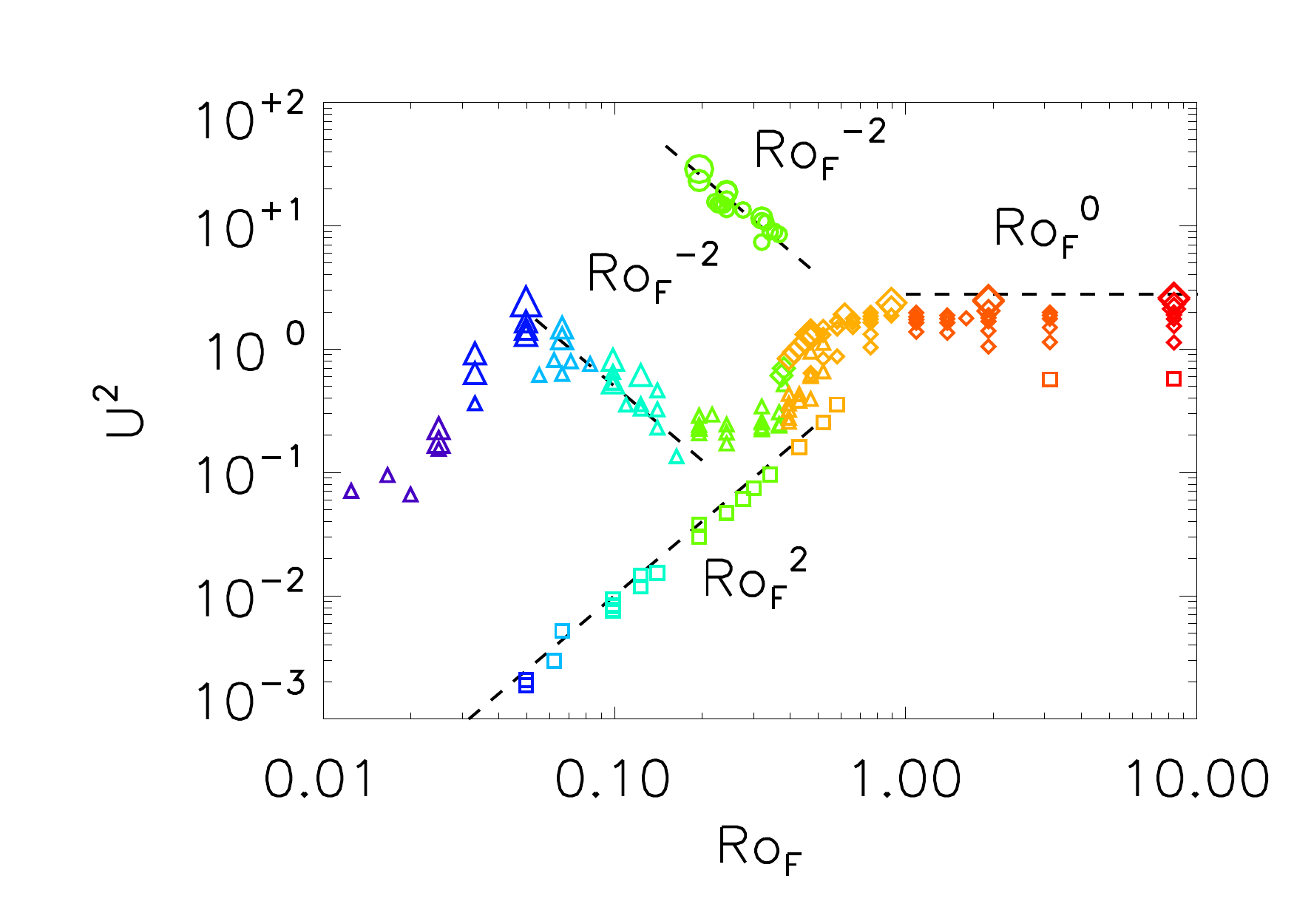}                                     %%
 \includegraphics[width=8cm]{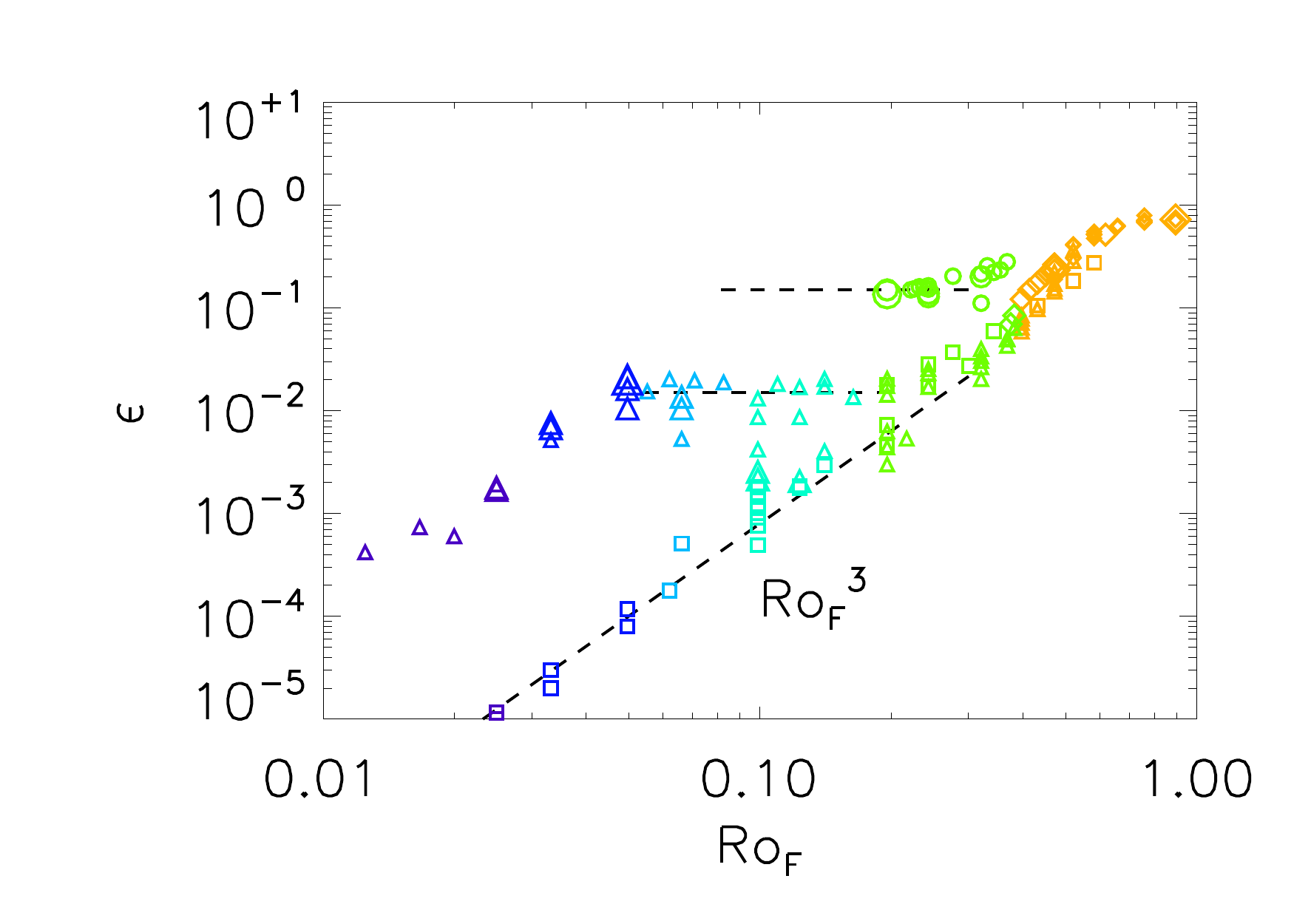}                                    %%
}                                                                                     %%
\caption{$U^2$ (left panel) and $\epsilon$ (right panel) as a function of $\Rof$      %%
         for all runs. Same symbols,colors,sizes are used as in figure \ref{fig_1}. } %%
\label{fig_14}                                                                        %%
\end{figure*}                                                                         %%
%%%%%%%%%%%%%%%%%%%%%%%%%%%%%%%%%%%%%%%%%%%%%%%%%%%%%%%%%%%%%%%%%%%%%%%%%%%%%%%%%%%%%%%%

In figure \ref{fig_14} we plot the basic quantities $U^2$ (left panel) and $\epsilon$ (right panel). 
as a function of $\Rof$. We remind the reader that the same symbols were used as in figure \ref{fig_1}
thus large symbols imply larger $\Ref$ while dark (violet online) symbols imply small $\Rof$.
Clearly different phases in the flow follow different scaling laws. 
For large $\Rof$ the effect of rotation is not felt and thus the turbulent scaling \ref{turb_rel} is recovered
with both $U^2$ and $\epsilon$ being independent from $\Rof\ge1$ and independent from $\Ref$ for sufficiently
large $\Ref$.
As $\Rof$ is decreased different behaviors are observed.  
The laminar runs reproduce the relation \ref{rlam_rel} with a clear scaling $U^2\propto \Rof^2$. 
Since most of the laminar states examined are close to the stability boundary $\Rof\propto \Ref^{-1}$, 
the $\epsilon \propto \Rof^2/\Ref$ scaling appears as $\epsilon \propto \Rof^3$
\footnote{Note that $\epsilon \propto \Rof^3$ is not a true scaling for the laminar flows. 
It originates from a bias in the choice of runs. Such biases are commonly met in numerical 
simulations where computational costs put strong restrictions, and sometimes are misinterpreted as physical scaling laws.}
This however provides only an upper limit for the dissipation of the laminar states.
The quasi-2D condensate states result in the scaling 
$U^2\propto \Rof^{-2}$ in accordance with the results shown in figure \ref{fig_13}. Note also the
discontinuous change in $U^2$ that is a direct result of the sub-critical behavior of the condensate modes.
The energy dissipation rate for these runs decreases initially as $\Rof$ is decreased but then it seems to saturate
at the smallest values of $\Rof$ attained. This however will need to be verified at smaller $\Rof$.
A similar behavior is also observed for the runs that displayed a bursting behavior.
The scaling  $U^2\propto \Rof^{-2}$ and $\epsilon\propto \Rof^0$ is also observed for these modes
for some range of $\Rof$. This behavior however transitions to a laminar behavior at even smaller $\Rof$ propably
because not large enough $\Ref$ has been reached for these runs.

%%%%%%%%%%%%%%%%%%%%%%%%%%%%%%%%%%%%%%%%%%%%%%%%%%%%%%%%%%%%%%%%%%%%%%%%%%%%%%%%%%%%%%%%%%%%%
\begin{figure*}                                                                            %%
\centerline{                                                                               %%
 \includegraphics[width=8cm]{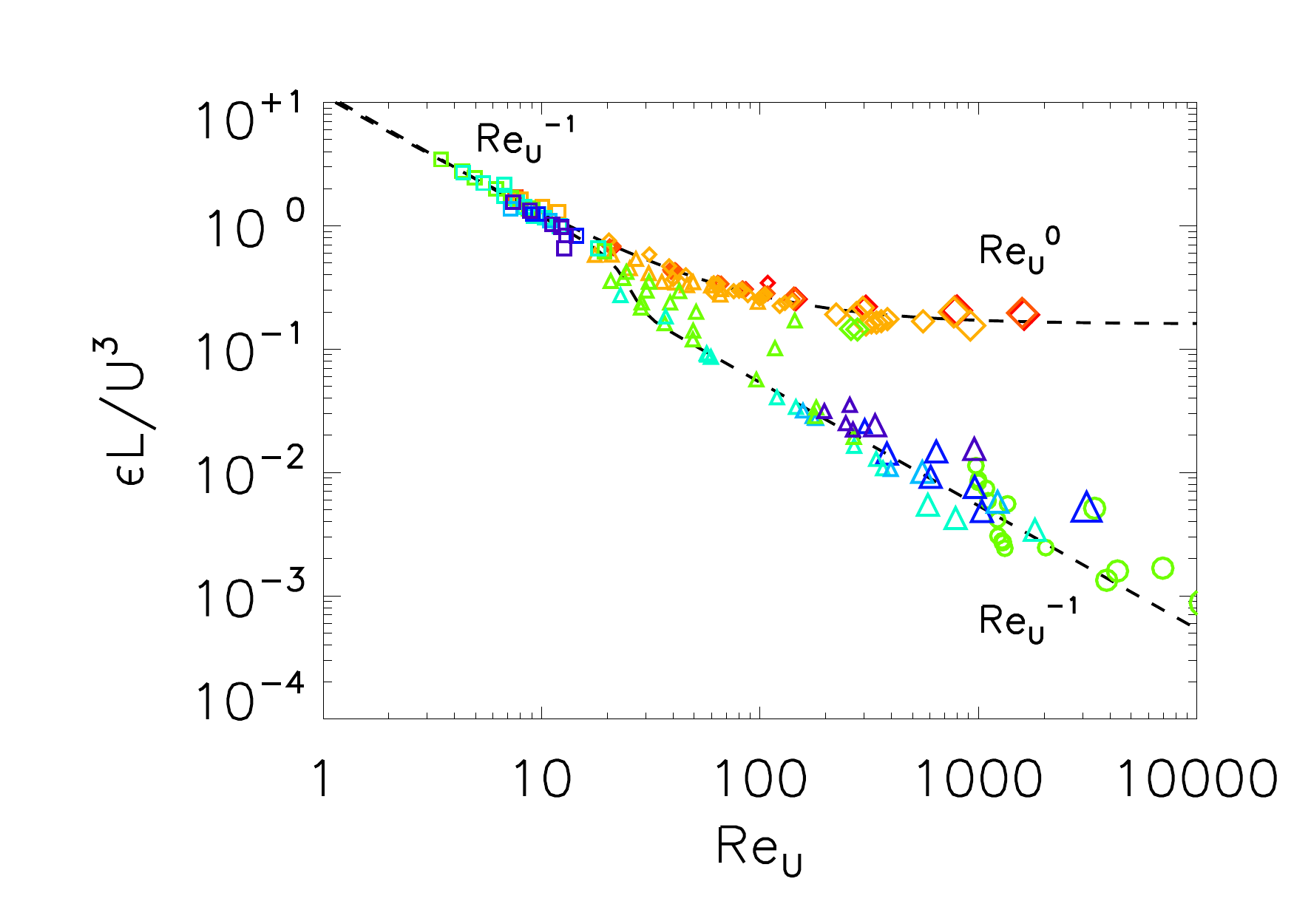}                                          %%
 \includegraphics[width=8cm]{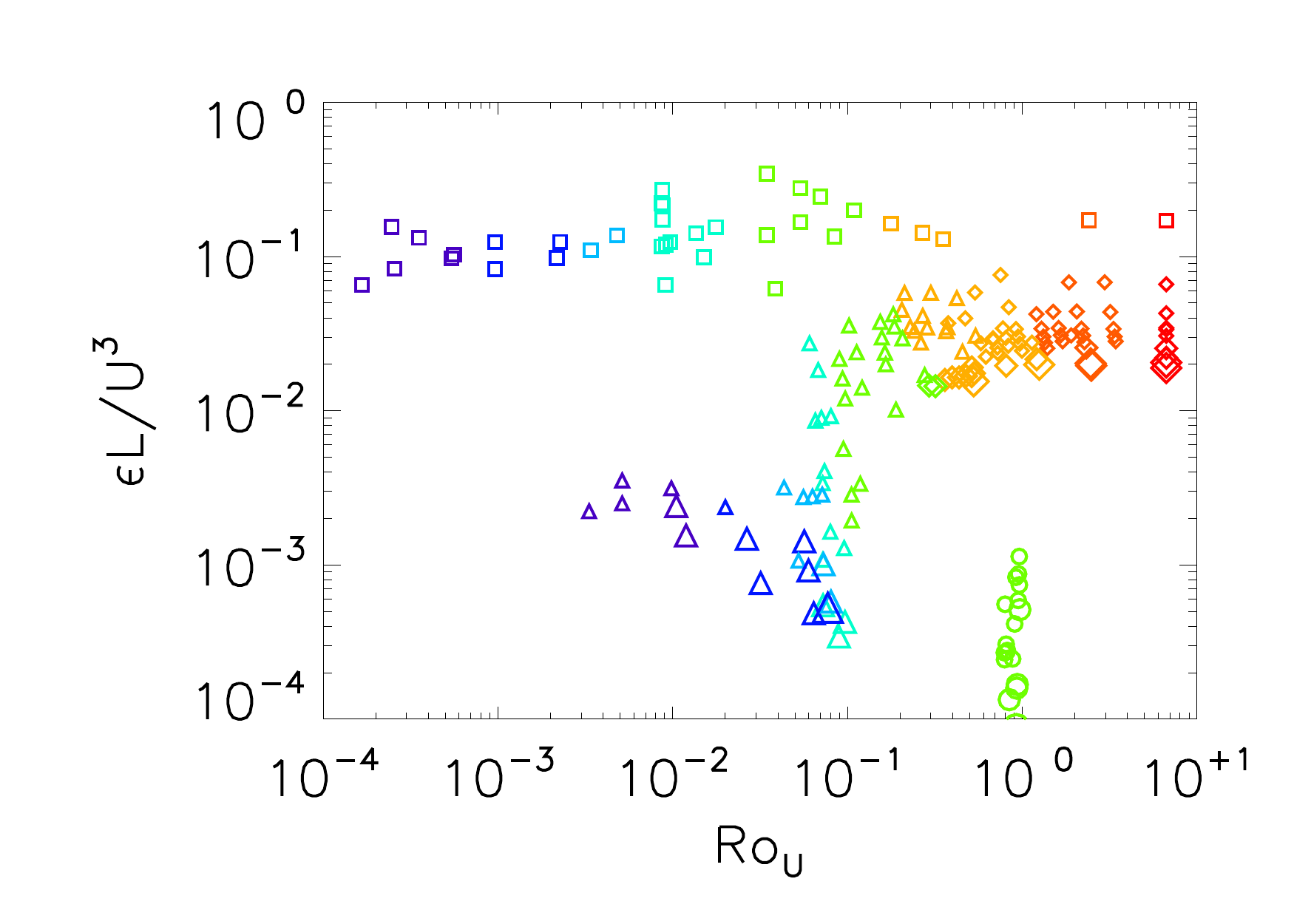}                                         %%
}                                                                                          %%
\caption{$\epsilon L/U^3$ as a function of $\Reu$ (left panel) and $\Rou$ (right panel)    %%
         for all runs. Same symbols,colors,sizes are used as in figure \ref{fig_1}. }      %%
\label{fig_15}                                                                             %%
\end{figure*}                                                                              %%
%%%%%%%%%%%%%%%%%%%%%%%%%%%%%%%%%%%%%%%%%%%%%%%%%%%%%%%%%%%%%%%%%%%%%%%%%%%%%%%%%%%%%%%%%%%%%

This section is concluded by
discussing the properties of the energy dissipation rate as a function of the more commonly used control 
parameters $\Reu$ and $\Rou$. Figure \ref{fig_15} displays $\epsilon L/U^3$ as a function of $\Reu$
(left panel) and $\Rou$ (right panel). 
The right panel serves mostly to demonstrate that although some of the data points are grouped together in the left panel
they correspond to different processes as can be realised by the different values of $\Rou$ they occupy in the parameter space.
%%%
The ratio $ \epsilon L/U^3$ for non-rotating turbulence  based on eq.\ref{turb_rel} is
expected to scale like $\Reu^{-1}$ for small values of $\Reu$ while an asymptotic value (independent of $\Reu$)
is expected to be reached at large enough $\Reu$. 
Reaching this asymptotic value indicates that the systems has reached a
turbulent state for which large scale properties do not depend on viscosity and energy injection is balanced only
by the energy flux to the small scales due to the nonlinearity \citep{Kaneda2003}. 
Indeed such a state is reached for our weakly rotating runs for $\Reu\gtrsim 800$ and $\Rof \ge 0.4$ (ie diammond runs). 
For the fast rotating runs however (including both bursts and condensates) such a state is not reached and the ratio  $\epsilon L/U^3$
continues to decrease as $\Reu^{-1}$ even at $\Reu\sim 10^4$.  
For these states most of the energy (and vorticity) is concentrated in a few large scales modes
and thus they have a laminar scaling. 
However, since the saturation amplitude at the large scale is viscosity independent but depends only on $\Omega$, it is expected that
at even larger $\Reu$ the vorticity at small scales will grow enough
so that despite the large energy concentration in the large scale modes, vorticity will be dominated
by the small scales and not the large and a viscosity free scaling will be obtained. 
This can be explained best by looking at the 1D energy spectrum in the right panel of figure \ref{fig_11}.
As $\Ref$ is increased the amplitude of energy in the large scales will remain fixed
(so that $\Rou=1$) but the large wavenumber viscous cut-off $k_c$ will extend to larger wavenumbers. 
Since the spectrum is sufficiently flat (less flat than $k^{-3}$) at large $\Ref$ the dissipation
($\propto \Ref^{-1}\int k^2E_{_{1D}}dk$) will be dominated by the small scales provided $k_c$
is sufficiently large. 
Thus a viscosity independent scaling is expected to be obtained in the $\Ref\to \infty $ limit.  This however would require resolutions not
attainable in the present study.

%%%%%%%%%%%%%%%%%%%%%%%%%%%%%%%%%%%%%%%%%%%%%%%%%%%%%%%%%%%%%%%%%%%%%%%%%%%%%%%%%%%%%%%%%%%%%%%%%%%%%%%%%%%%%%%%%%%%%%%%%%%%%%%
%%%%%%%%%%%%%%%%%%%%%%%%%%%%%%%%%%%%%%%%%%%%%%%%%%%%%%%%%%%%%%%%%%%%%%%%%%%%%%%%%%%%%%%%%%%%%%%%%%%%%%%%%%%%%%%%%%%%%%%%%%%%%%% 
%%%%%%%%%%%%%%%%%%%%%%%%%%%%%%%%%%%%%%%%%%%%%%%%%%%%%%%%%%%%%%%%%%%%%%%%%%%%%%%%%%%%%%%%%%%%%%%%%%%%%%%%%%%%%%%%%%%%%%%%%%%%%%% 
%%%%%%%%%%%%%%%%%%%%%%%%%%%%%%%%%%%%%%%%%%%%%%%%%%%%%%%%%%%%%%%%%%%%%%%%%%%%%%%%%%%%%%%%%%%%%%%%%%%%%%%%%%%%%%%%%%%%%%%%%%%%%%% 
\section{Summary and conclusions  }   %%%%%%%%%%%%%%%%%%%%%%%%%%%%%%%%%%%%%%%%%%%%%%%%%%%%%%%%%%%%%%%%%%%%%%%%%%%%%%%%%%%%%%%%%
%%%%%%%%%%%%%%%%%%%%%%%%%%%%%%%%%%%%%%%%%%%%%%%%%%%%%%%%%%%%%%%%%%%%%%%%%%%%%%%%%%%%%%%%%%%%%%%%%%%%%%%%%%%%%%%%%%%%%%%%%%%%%%%
%%%%%%%%%%%%%%%%%%%%%%%%%%%%%%%%%%%%%%%%%%%%%%%%%%%%%%%%%%%%%%%%%%%%%%%%%%%%%%%%%%%%%%%%%%%%%%%%%%%%%%%%%%%%%%%%%%%%%%%%%%%%%%% 
%%%%%%%%%%%%%%%%%%%%%%%%%%%%%%%%%%%%%%%%%%%%%%%%%%%%%%%%%%%%%%%%%%%%%%%%%%%%%%%%%%%%%%%%%%%%%%%%%%%%%%%%%%%%%%%%%%%%%%%%%%%%%%% 
%%%%%%%%%%%%%%%%%%%%%%%%%%%%%%%%%%%%%%%%%%%%%%%%%%%%%%%%%%%%%%%%%%%%%%%%%%%%%%%%%%%%%%%%%%%%%%%%%%%%%%%%%%%%%%%%%%%%%%%%%%%%%%%

Perhaps the most intriguing result of the present work is the demonstration that a simple two parameter system 
like the one under study can display such richness and complexity of behavior. Depending on the location in 
the parameter space the system can display {\it laminar behavior, intermittent bursts, quasi-2D condensate states, 
and weakly rotating turbulence}. All these behaviors can be obtained in the $\Ref\to\infty$ limit provided 
the appropriate scaling of $\Rof$ with $\Ref$ is considered.

For high rotation rates laminar solution can be found in terms
of an asymptotic expansion. Up to the ordering $\Rof\propto\Ref^{-1/3}$ the laminar solution has no zeroth order
projection to the 2D3C-flows. For  $\Rof\propto\Ref^{-1/5}$  the flow has an order one projection to 2D3C-flows
and for even larger values of $\Ref$ no laminar solution that can be captured by the expansion exists.
The realizability of the laminar flows is determined by their stability properties.
%For fast rotation rates stable laminar flows, are restricted in the parameter space $\Ref \le c \Rof^{-1}$. 
%In terms of the parameters ($\Reu,\Rou$) (using the relations \ref{RefRouMap1}) the stability relation reads $\Reu\le c$.  

When $\Ref$ is increased and the conditions for stability are violated the system transitions sub-critically to 
a time dependent flow that exhibits intermittent bursts. The unstable modes involved can be predicted by 
an asymptotic theory, that takes in to account exact resonances and is valid in the limit $\Rof\to 0$.
The existence of exact resonances at the first order of the expansion that can drive the system unstable 
implies that the linear instability boundary is along the $\Rof \propto \Ref^{-1}$ line. 
These modes drive the system on the dynamical time scale to high levels of energy where the asymptotic expansion fails
and the resonance conditions are violated. After the burst the remaining energy is concentrated in a 2D3C flow
that condensates in the largest available scale and decays on the viscous time scale. 
These bursts can be described  of a low dimensional dynamical system, and thus do not describe a truly turbulent state.
 
As $\Ref$ is increased with respect to $\Rof^{-1}$ and a relation $\Rof \le c \Ref^{-\alpha}$ is satisfied the system transitions again subcritically 
to a quasi-2D condensate. This state represents the rotating turbulence regime for the Taylor-Green flow as 
it the one obtained in the limit $\Ref\to \infty$ for any value of $\Rof\le 0.4$. The flow at this state is composed of a 
quasi2D condensate of vertical vorticity at the large scales and only weakly anisotropic small scale turbulence. 
Saturation of the large scale condensate comes from reaching values of $\Rou=1$ at which point the 
counter-rotating vortex from the pair of vortexes that formed becomes unstable and cascades the energy to the small scales. 
The value of the exponent $\alpha$ is either $1/3$ or $1/5$. 
%For $\Ref\propto\Rof^{-1}$ the asymptotic expansion leads to
%a complete decoupling of the 2D3C flow that would decay freely without any energy source.
If energy injection is achieved by the coupling of inertial waves at second order then $\alpha=1/3$. 
If not then at third order it has already been shown that the forcing alone is capable at injecting energy to the flow
and $\alpha=1/5$. The numerical simulations give more support to the scaling $\Rof^{-1}\propto \Ref^{1/3}$ without 
however being decisive. 
Finally, the energy dissipation rate has not reached the viscosity independent scaling in this regime
due to the large amplitude of the condensates, but it is expected to be reached at higher values of $\Ref$.

%Further increase of $\Ref$ does not lead to an other state and thus this state is found in the $\Ref\to \infty$ limit.
%However if $\Rof$ is increased the system returns (again subcritically) to a weakly rotating state for which only
%small deviations from the non-rotating case were noticed. At this state the flow reproduces all the results 
%of non-rotating turbulence such as the asymptotic  viscous independent scaling for large $\nu$. 

From these results there are a few points that worth being pointed out.
First of all  the first order expansion fails to describe 
the evolution of the 2D3C part of the flow. The small $\Rof$ expansion that predicts
independence of the 2D3C part of the flow can not predict the increase of the 2D3C flow
that was observed both during the intermittent bursts stage and during to the quasi-2D
state. Thus to describe the flow higher terms in the expansion need to be kept.

It was also found that for any value of the parameters examined ($\Ref,\Rof$) a phase that
could be described as weak wave turbulence was not met. Fast rotating flows were either
dominated by condensates or intermittent bursts.
Possibly this is the case because only 
the value $q=2$ was used. For larger values of $q$ both the number of exact resonances
and quasi-resonances would increase and the system would come closer  to a continuous 
Fourier space where the assumptions of the weak-wave-turbulence theory better hold.
However, it is noted that fully turbulent flows 
displayed the formation of quasi-2D condensates that reduced all flows to $\Rou=1$.
Thus in the steady state there is no guaranty that the prerequisite condition for weak turbulence $\Rou\ll1$
will be met at the steady state for this choice of forcing and for $\Rof\ll1$ even at larger $q$.

This last point also raises an interesting aspect of systems with inverse cascade.
In the absence of a large scale dissipative mechanisms,
the need to saturate the inverse cascade drove
the system to marginality of the inverse cascade by reaching in the present case $\Rou=1$.
At this state there is a very weak inverse energy transfer just sufficient to sustain the large scale flow against viscosity.
Aspects of such marginal states of inverse cascades have been recently investigated in more 
simplified models \citep{Seshasayanan2014}.

A further point worth pointed out by the present work is that referring to the 
{\it large}-Reynolds-{\it small}-Rossby limit without specifically prescribing the 
limiting procedure is meaningless, since different behaviors can be obtained
for different scaling of the Reynolds number with the Rossby number.
Different ordering of $\Ref$ with $\Rof$ leads to different behaviors and scalings. Also 
the velocity amplitude used in the definition of the Reynolds and Rossby number 
is more than just a conventional formality in rotating turbulence. Thus
one needs to be precise on the definitions used and the limiting procedure considered. 
From the present results the $\Rod,\Red$ were shown to best describe the level of turbulence with respect
to the rotation rate and the viscosity.
 
Finally, it is pointed out the necessity of numerical studies to cover systematically the parameter space 
in order to draw any general conclusions.

\acknowledgements

I would like to thank Martin Schrinner for motivating this work that its initial steps were 
thoroughly discussed with him. I would also like to thank the members of the Nonlinear physics 
group at LPS/ENS for their very useful comments. The present work benefited from the computational support 
of the HPC resources of GENCI-TGCC-CURIE \& GENCI-CINES-JADE (Project No. x2014056421, No. x2013056421 \& No. 2012026421).
and MesoPSL financed by the Region Ile de France and the project EquipMeso (reference ANR-10-EQPX-29-01) where the
present numerical simulations have been performed in the last 3 years.

%%%%%%%%%%%%%%%%%%%%%%%%%%%%%%%%%%%%%%%%%%%%%%%%%%%%%%%%%%%%%%%%%%%%%%%%%%%%%%%%%%%%%%%%%%%%%%%%%%%%%%%%%%%%%%%%%%%%%%%%%%%%%%%
%%%%%%%%%%%%%%%%%%%%%%%%%%%%%%%%%%%%%%%%%%%%%%%%%%%%%%%%%%%%%%%%%%%%%%%%%%%%%%%%%%%%%%%%%%%%%%%%%%%%%%%%%%%%%%%%%%%%%%%%%%%%%%%
%%%%%%%%%%%%%%%%%%%%%%%%%%%%%%%%%%%%%%%%%%%%%%%%%%%%%%%%%%%%%%%%%%%%%%%%%%%%%%%%%%%%%%%%%%%%%%%%%%%%%%%%%%%%%%%%%%%%%%%%%%%%%%%
%%%%%%%%%%%%%%%%%%%%%%%%%%%%%%%%%%%%%%%%%%%%%%%%%%%%%%%%%%%%%%%%%%%%%%%%%%%%%%%%%%%%%%%%%%%%%%%%%%%%%%%%%%%%%%%%%%%%%%%%%%%%%%%

\bibliographystyle{jfm2}

\bibliography{MHDturb}

\begin{thebibliography}{56}
\expandafter\ifx\csname natexlab\endcsname\relax\def\natexlab#1{#1}\fi

\bibitem[{Babin} {\em et~al.\/}(1969){Babin}, {Mahalov} \&
  {Nicolaenko}]{Babin1996}
{\sc {Babin}, A., {Mahalov}, A. \& {Nicolaenko}, B.} 1969 {Global splitting,
  integrability and regularity of three-dimensional Euler and Navier–Stokes
  equations for uniformly rotating fluids}. {\em Eur. J. Mech.B/Fluids\/} {\bf
  15}, 291--300.

\bibitem[{Bardina} {\em et~al.\/}(1985){Bardina}, {Ferziger} \&
  {Rogallo}]{Bardina1985}
{\sc {Bardina}, J., {Ferziger}, J.~H. \& {Rogallo}, R.~S.} 1985 {Effect of
  rotation on isotropic turbulence - Computation and modelling}. {\em Journal
  of Fluid Mechanics\/} {\bf 154}, 321--336.

\bibitem[{Bartello} {\em et~al.\/}(1994){Bartello}, {Metais} \&
  {Lesieur}]{Bartello1994}
{\sc {Bartello}, P., {Metais}, O. \& {Lesieur}, M.} 1994 {Coherent structures
  in rotating three-dimensional turbulence}. {\em Journal of Fluid Mechanics\/}
  {\bf 273}, 1--29.

\bibitem[{Bewley} {\em et~al.\/}(2007){Bewley}, {Lathrop}, {Maas} \&
  {Sreenivasan}]{Sreenivasan2007}
{\sc {Bewley}, G.~P., {Lathrop}, D.~P., {Maas}, L.~R.~M. \& {Sreenivasan},
  K.~R.} 2007 {Inertial waves in rotating grid turbulence}. {\em Physics of
  Fluids\/} {\bf 19}~(7), 071701.

\bibitem[{Boubnov} \& {Golitsyn}(1986)]{Boubnov1986}
{\sc {Boubnov}, B.~M. \& {Golitsyn}, G.~S.} 1986 {Experimental study of
  convective structures in rotating fluids}. {\em Journal of Fluid Mechanics\/}
  {\bf 167}, 503--531.

\bibitem[{Brachet} {\em et~al.\/}(1992){Brachet}, {Meneguzzi}, {Vincent},
  {Politano} \& {Sulem}]{Brachet1992}
{\sc {Brachet}, M.~E., {Meneguzzi}, M., {Vincent}, A., {Politano}, H. \&
  {Sulem}, P.~L.} 1992 {Numerical evidence of smooth self-similar dynamics and
  possibility of subsequent collapse for three-dimensional ideal flows}. {\em
  Physics of Fluids\/} {\bf 4}, 2845--2854.

\bibitem[{Bustamante} \& {Hayat}(2013)]{Bustamante2013}
{\sc {Bustamante}, M.~D. \& {Hayat}, U.} 2013 {Complete classification of
  discrete resonant Rossby/drift wave triads on periodic domains}. {\em
  Communications in Nonlinear Science and Numerical Simulations\/} {\bf 18},
  2402--2419.

\bibitem[{Cambon} \& {Jacquin}(1989)]{Cambon1989}
{\sc {Cambon}, C. \& {Jacquin}, L.} 1989 {Spectral approach to non-isotropic
  turbulence subjected to rotation}. {\em Journal of Fluid Mechanics\/} {\bf
  202}, 295--317.

\bibitem[{Chen} {\em et~al.\/}(2005){Chen}, {Chen}, {Eyink} \&
  {Holm}]{Chen2005}
{\sc {Chen}, Q., {Chen}, S., {Eyink}, G.~L. \& {Holm}, D.~D.} 2005 {Resonant
  interactions in rotating homogeneous three-dimensional turbulence}. {\em
  Journal of Fluid Mechanics\/} {\bf 542}, 139--164.

\bibitem[{Cortet} {\em et~al.\/}(2010){Cortet}, {Chiffaudel}, {Daviaud} \&
  {Dubrulle}]{Cortet2010}
{\sc {Cortet}, P.-P., {Chiffaudel}, A., {Daviaud}, F. \& {Dubrulle}, B.} 2010
  {Experimental Evidence of a Phase Transition in a Closed Turbulent Flow}.
  {\em Physical Review Letters\/} {\bf 105}~(21), 214501.

\bibitem[{Davidson} {\em et~al.\/}(2006){Davidson}, {Staplehurst} \&
  {Dalziel}]{Davidson2006}
{\sc {Davidson}, P.~A., {Staplehurst}, P.~J. \& {Dalziel}, S.~B.} 2006 {On the
  evolution of eddies in a rapidly rotating system}. {\em Journal of Fluid
  Mechanics\/} {\bf 557}, 135--144.

\bibitem[{Gallet} {\em et~al.\/}(2014){Gallet}, {Campagne}, {Cortet} \&
  {Moisy}]{Gallet2014}
{\sc {Gallet}, B., {Campagne}, A., {Cortet}, P.-P. \& {Moisy}, F.} 2014
  {Scale-dependent cyclone-anticyclone asymmetry in a forced rotating
  turbulence experiment}. {\em Physics of Fluids\/} {\bf 26}~(3), 035108.

\bibitem[{Galtier}(2003)]{Galtier2003}
{\sc {Galtier}, S.} 2003 {Weak inertial-wave turbulence theory}. {\em Phys.
  Rev. E\/} {\bf 68}~(1), 015301.

\bibitem[{Gence} \& {Frick}(2001)]{Gence2001}
{\sc {Gence}, J.-N. \& {Frick}, C.} 2001 {Naissance des corr{\'e}lations
  triples de vorticit{\'e} dans une turbulence statistiquement homog{\`e}ne
  soumise {\`a} une rotation}. {\em Academie des Sciences Paris Comptes Rendus
  Serie B Sciences Physiques\/} {\bf 329}, 351--356.

\bibitem[{Godeferd} \& {Lollini}(1999)]{Godeferd1999}
{\sc {Godeferd}, F.~S. \& {Lollini}, L.} 1999 {Direct numerical simulations of
  turbulence with confinement and rotation}. {\em Journal of Fluid Mechanics\/}
  {\bf 393}, 257--308.

\bibitem[{G{\'o}mez} {\em et~al.\/}(2005){G{\'o}mez}, {Mininni} \&
  {Dmitruk}]{Minini_code2}
{\sc {G{\'o}mez}, D.~O., {Mininni}, P.~D. \& {Dmitruk}, P.} 2005 {Parallel
  Simulations in Turbulent MHD}. {\em Physica Scripta Volume T\/} {\bf 116},
  123--127.

\bibitem[Greenspan(1968)]{Greenspan1968}
{\sc Greenspan, H.} 1968 {\em The Theory of Rotating Fluids\/}. Cambridge
  University Press.

\bibitem[{Hopfinger} {\em et~al.\/}(1982){Hopfinger}, {Gagne} \&
  {Browand}]{Hopfinger1982}
{\sc {Hopfinger}, E.~J., {Gagne}, Y. \& {Browand}, F.~K.} 1982 {Turbulence and
  waves in a rotating tank}. {\em Journal of Fluid Mechanics\/} {\bf 125},
  505--534.

\bibitem[{Hopfinger} \& {van Heijst}(1993)]{Hopfinger1993}
{\sc {Hopfinger}, E.~J. \& {van Heijst}, G.~J.~F.} 1993 {Vortices in rotating
  fluids}. {\em Annual Review of Fluid Mechanics\/} {\bf 25}, 241--289.

\bibitem[{Hossain}(1994)]{Hossain1994}
{\sc {Hossain}, M.} 1994 {Reduction in the dimensionality of turbulence due to
  a strong rotation}. {\em Physics of Fluids\/} {\bf 6}, 1077--1080.

\bibitem[{Kaneda} {\em et~al.\/}(2003){Kaneda}, {Ishihara}, {Yokokawa},
  {Itakura} \& {Uno}]{Kaneda2003}
{\sc {Kaneda}, Y., {Ishihara}, T., {Yokokawa}, M., {Itakura}, K. \& {Uno}, A.}
  2003 {Energy dissipation rate and energy spectrum in high resolution direct
  numerical simulations of turbulence in a periodic box}. {\em Physics of
  Fluids\/} {\bf 15}, L21--L24.

\bibitem[{Kolvin} {\em et~al.\/}(2009){Kolvin}, {Cohen}, {Vardi} \&
  {Sharon}]{Sharon2009}
{\sc {Kolvin}, I., {Cohen}, K., {Vardi}, Y. \& {Sharon}, E.} 2009 {Energy
  Transfer by Inertial Waves during the Buildup of Turbulence in a Rotating
  System}. {\em Physical Review Letters\/} {\bf 102}~(1), 014503.

\bibitem[{Lamriben} {\em et~al.\/}(2011){Lamriben}, {Cortet} \&
  {Moisy}]{Lamriben2011}
{\sc {Lamriben}, C., {Cortet}, P.-P. \& {Moisy}, F.} 2011 {Anisotropic energy
  transfers in rotating turbulence}. {\em Journal of Physics Conference
  Series\/} {\bf 318}~(4), 042005.

\bibitem[{Mansour} {\em et~al.\/}(1992){Mansour}, {Gambon} \&
  {Speziale}]{Mansour1992}
{\sc {Mansour}, N.~N., {Gambon}, C. \& {Speziale}, C.~G.} 1992 {\em
  {Theoretical and computational study of rotating isotropic turbulence}\/},
  pp. 59--75. Springer-Verlag.

\bibitem[{Maurer} \& {Tabeling}(1998)]{Mauer1998}
{\sc {Maurer}, J. \& {Tabeling}, P.} 1998 {Local investigation of superfluid
  turbulence}. {\em EPL (Europhysics Letters)\/} {\bf 43}, 29--34.

\bibitem[{Mininni} {\em et~al.\/}(2006){Mininni}, {Alexakis} \&
  {Pouquet}]{Mininni2006}
{\sc {Mininni}, P.~D., {Alexakis}, A. \& {Pouquet}, A.} 2006 {Large-scale flow
  effects, energy transfer, and self-similarity on turbulence}. {\em Phys. Rev.
  E\/} {\bf 74}~(1), 016303.

\bibitem[{Mininni} {\em et~al.\/}(2009){Mininni}, {Alexakis} \&
  {Pouquet}]{Mininni2009}
{\sc {Mininni}, P.~D., {Alexakis}, A. \& {Pouquet}, A.} 2009 {Scale
  interactions and scaling laws in rotating flows at moderate Rossby numbers
  and large Reynolds numbers}. {\em Physics of Fluids\/} {\bf 21}~(1), 015108.

\bibitem[{Mininni} \& {Pouquet}(2009)]{Mininni2009a}
{\sc {Mininni}, P.~D. \& {Pouquet}, A.} 2009 {Helicity cascades in rotating
  turbulence}. {\em Phys. Rev. E\/} {\bf 79}~(2), 026304.

\bibitem[Mininni \& Pouquet(2010)]{Mininni2010}
{\sc Mininni, P.~D. \& Pouquet, A.} 2010 Rotating helical turbulence. i. global
  evolution and spectral behavior. {\em Phys. Fluids\/} {\bf 22}, 035105.

\bibitem[{Mininni} {\em et~al.\/}(2012){Mininni}, {Rosenberg} \&
  {Pouquet}]{Mininni2012}
{\sc {Mininni}, P.~D., {Rosenberg}, D. \& {Pouquet}, A.} 2012 {Isotropization
  at small scales of rotating helically driven turbulence}. {\em Journal of
  Fluid Mechanics\/} {\bf 699}, 263--279.

\bibitem[{Monchaux} {\em et~al.\/}(2009){Monchaux}, {Berhanu}, {Auma{\^i}tre},
  {Chiffaudel}, {Daviaud}, {Dubrulle}, {Ravelet}, {Fauve}, {Mordant},
  {P{\'e}tr{\'e}lis}, {Bourgoin}, {Odier}, {Pinton}, {Plihon} \&
  {Volk}]{Monchaux2009}
{\sc {Monchaux}, R., {Berhanu}, M., {Auma{\^i}tre}, S., {Chiffaudel}, A.,
  {Daviaud}, F., {Dubrulle}, B., {Ravelet}, F., {Fauve}, S., {Mordant}, N.,
  {P{\'e}tr{\'e}lis}, F., {Bourgoin}, M., {Odier}, P., {Pinton}, J.-F.,
  {Plihon}, N. \& {Volk}, R.} 2009 {The von K{\'a}rm{\'a}n Sodium experiment:
  Turbulent dynamical dynamos}. {\em Physics of Fluids\/} {\bf 21}~(3), 035108.

\bibitem[{Monchaux} {\em et~al.\/}(2007){Monchaux}, {Berhanu}, {Bourgoin},
  {Moulin}, {Odier}, {Pinton}, {Volk}, {Fauve}, {Mordant}, {P{\'e}tr{\'e}lis},
  {Chiffaudel}, {Daviaud}, {Dubrulle}, {Gasquet}, {Mari{\'e}} \&
  {Ravelet}]{Monchaux2007}
{\sc {Monchaux}, R., {Berhanu}, M., {Bourgoin}, M., {Moulin}, M., {Odier}, P.,
  {Pinton}, J.-F., {Volk}, R., {Fauve}, S., {Mordant}, N., {P{\'e}tr{\'e}lis},
  F., {Chiffaudel}, A., {Daviaud}, F., {Dubrulle}, B., {Gasquet}, C.,
  {Mari{\'e}}, L. \& {Ravelet}, F.} 2007 {Generation of a Magnetic Field by
  Dynamo Action in a Turbulent Flow of Liquid Sodium}. {\em Physical Review
  Letters\/} {\bf 98}~(4), 044502.

\bibitem[{Morinishi} {\em et~al.\/}(2001){Morinishi}, {Nakabayashi} \&
  {Ren}]{Morinishi2001}
{\sc {Morinishi}, Y., {Nakabayashi}, K. \& {Ren}, S.~Q.} 2001 {Dynamics of
  anisotropy on decaying homogeneous turbulence subjected to system rotation}.
  {\em Physics of Fluids\/} {\bf 13}, 2912--2922.

\bibitem[{Morize} \& {Moisy}(2006)]{Morize2006}
{\sc {Morize}, C. \& {Moisy}, F.} 2006 {Energy decay of rotating turbulence
  with confinement effects}. {\em Physics of Fluids\/} {\bf 18}~(6), 065107.

\bibitem[{Morize} {\em et~al.\/}(2005){Morize}, {Moisy} \&
  {Rabaud}]{Morize2005}
{\sc {Morize}, C., {Moisy}, F. \& {Rabaud}, M.} 2005 {Decaying grid-generated
  turbulence in a rotating tank}. {\em Physics of Fluids\/} {\bf 17}~(9),
  095105.

\bibitem[{M{\"u}ller} \& {Thiele}(2007)]{Muller2007}
{\sc {M{\"u}ller}, W.-C. \& {Thiele}, M.} 2007 {Scaling and energy transfer in
  rotating turbulence}. {\em EPL (Europhysics Letters)\/} {\bf 77}, 34003.

\bibitem[Nazarenko(2011)]{Nazarenko_book2011}
{\sc Nazarenko, S.~V.} 2011 {\em Wave Turbulence\/}. Springer-Verlag.

\bibitem[{Newell}(1969)]{Newell1969}
{\sc {Newell}, A.~C.} 1969 {Rossby wave packet interactions}. {\em Journal of
  Fluid Mechanics\/} {\bf 35}, 255--271.

\bibitem[{Ponty} {\em et~al.\/}(2008){Ponty}, {Mininni}, {Laval}, {Alexakis},
  {Baerenzung}, {Daviaud}, {Dubrulle}, {Pinton}, {Politano} \&
  {Pouquet}]{Ponty2008}
{\sc {Ponty}, Y., {Mininni}, P.~D., {Laval}, J.-P., {Alexakis}, A.,
  {Baerenzung}, J., {Daviaud}, F., {Dubrulle}, B., {Pinton}, J.-F., {Politano},
  H. \& {Pouquet}, A.} 2008 {Linear and non-linear features of the Taylor Green
  dynamo}. {\em Comptes Rendus Physique\/} {\bf 9}, 749--756.

\bibitem[{Ruppert-Felsot} {\em et~al.\/}(2005){Ruppert-Felsot}, {Praud},
  {Sharon} \& {Swinney}]{Swinney2005}
{\sc {Ruppert-Felsot}, J.~E., {Praud}, O., {Sharon}, E. \& {Swinney}, H.~L.}
  2005 {Extraction of coherent structures in a rotating turbulent flow
  experiment}. {\em Phys. Rev. E\/} {\bf 72}~(1), 016311.

\bibitem[{Salort} {\em et~al.\/}(2010){Salort}, {Baudet}, {Castaing},
  {Chabaud}, {Daviaud}, {Didelot}, {Diribarne}, {Dubrulle}, {Gagne},
  {Gauthier}, {Girard}, {H{\'e}bral}, {Rousset}, {Thibault} \&
  {Roche}]{Salort2010}
{\sc {Salort}, J., {Baudet}, C., {Castaing}, B., {Chabaud}, B., {Daviaud}, F.,
  {Didelot}, T., {Diribarne}, P., {Dubrulle}, B., {Gagne}, Y., {Gauthier}, F.,
  {Girard}, A., {H{\'e}bral}, B., {Rousset}, B., {Thibault}, P. \& {Roche},
  P.-E.} 2010 {Turbulent velocity spectra in superfluid flows}. {\em Physics of
  Fluids\/} {\bf 22}~(12), 125102.

\bibitem[{Seshasayanan} {\em et~al.\/}(2014){Seshasayanan}, {Benavides} \&
  {Alexakis}]{Seshasayanan2014}
{\sc {Seshasayanan}, K., {Benavides}, S.~J. \& {Alexakis}, A.} 2014 {On the
  edge of an inverse cascade}. {\em ArXiv e-prints\/} .

\bibitem[{Smith} \& {Waleffe}(1999)]{Smith1999}
{\sc {Smith}, L.~M. \& {Waleffe}, F.} 1999 {Transfer of energy to
  two-dimensional large scales in forced, rotating three-dimensional
  turbulence}. {\em Physics of Fluids\/} {\bf 11}, 1608--1622.

\bibitem[{Squires} {\em et~al.\/}(1994){Squires}, {Chasnov}, {Mansour} \&
  {Cambon}]{Squires1994}
{\sc {Squires}, K.~D., {Chasnov}, J.~R., {Mansour}, N.~N. \& {Cambon}, C.}
  (ed.) 1994 {\em {The asymptotic state of rotating homogeneous turbulence at
  high Reynolds numbers}\/}.

\bibitem[{Sreenivasan} \& {Davidson}(2008)]{Sreenivasan2008}
{\sc {Sreenivasan}, B. \& {Davidson}, P.~A.} 2008 {On the formation of cyclones
  and anticyclones in a rotating fluid}. {\em Physics of Fluids\/} {\bf
  20}~(8), 085104.

\bibitem[{Staplehurst} {\em et~al.\/}(2008){Staplehurst}, {Davidson} \&
  {Dalziel}]{Staplehurst2008}
{\sc {Staplehurst}, P.~J., {Davidson}, P.~A. \& {Dalziel}, S.~B.} 2008
  {Structure formation in homogeneous freely decaying rotating turbulence}.
  {\em Journal of Fluid Mechanics\/} {\bf 598}, 81--105.

\bibitem[{Sugihara} {\em et~al.\/}(2005){Sugihara}, {Migita} \&
  {Honji}]{Sugihara2005}
{\sc {Sugihara}, Y., {Migita}, M. \& {Honji}, H.} 2005 {Orderly flow structures
  in grid-generated turbulence with background rotation}. {\em Fluid Dynamics
  Research\/} {\bf 36}, 23--34.

\bibitem[{Teitelbaum} \& {Mininni}(2009)]{Teitelbaum2009}
{\sc {Teitelbaum}, T. \& {Mininni}, P.~D.} 2009 {Effect of Helicity and
  Rotation on the Free Decay of Turbulent Flows}. {\em Physical Review
  Letters\/} {\bf 103}~(1), 014501.

\bibitem[{Teitelbaum} \& {Mininni}(2010)]{Teitelbaum2010}
{\sc {Teitelbaum}, T. \& {Mininni}, P.~D.} 2010 {Large-scale effects on the
  decay of rotating helical and non-helical turbulence}. {\em Physica Scripta
  Volume T\/} {\bf 142}~(1), 014003.

\bibitem[{Thiele} \& {M{\"u}ller}(2009)]{Muller2009}
{\sc {Thiele}, M. \& {M{\"u}ller}, W.-C.} 2009 {Structure and decay of rotating
  homogeneous turbulence}. {\em Journal of Fluid Mechanics\/} {\bf 637}, 425.

\bibitem[{van Bokhoven} {\em et~al.\/}(2009){van Bokhoven}, {Clercx}, {van
  Heijst} \& {Trieling}]{Bokhoven2009}
{\sc {van Bokhoven}, L.~J.~A., {Clercx}, H.~J.~H., {van Heijst}, G.~J.~F. \&
  {Trieling}, R.~R.} 2009 {Experiments on rapidly rotating turbulent flows}.
  {\em Physics of Fluids\/} {\bf 21}~(9), 096601.

\bibitem[{Waleffe}(1992)]{Waleffe1992}
{\sc {Waleffe}, F.} 1992 {The nature of triad interactions in homogeneous
  turbulence}. {\em Physics of Fluids\/} {\bf 4}, 350--363.

\bibitem[{Waleffe}(1993)]{Waleffe1993}
{\sc {Waleffe}, F.} 1993 {Inertial transfers in the helical decomposition}.
  {\em Physics of Fluids\/} {\bf 5}, 677--685.

\bibitem[{Yarom} {\em et~al.\/}(2013){Yarom}, {Vardi} \& {Sharon}]{Sharon2013}
{\sc {Yarom}, E., {Vardi}, Y. \& {Sharon}, E.} 2013 {Experimental
  quantification of inverse energy cascade in deep rotating turbulence}. {\em
  Physics of Fluids\/} {\bf 25}~(8), 085105.

\bibitem[{Yeung} \& {Zhou}(1998)]{Yeung1998}
{\sc {Yeung}, P.~K. \& {Zhou}, Y.} 1998 {Numerical study of rotating turbulence
  with external forcing}. {\em Physics of Fluids\/} {\bf 10}, 2895--2909.

\bibitem[{Yoshimatsu} {\em et~al.\/}(2011){Yoshimatsu}, {Midorikawa} \&
  {Kaneda}]{Yoshimatsu2011}
{\sc {Yoshimatsu}, K., {Midorikawa}, M. \& {Kaneda}, Y.} 2011 {Columnar eddy
  formation in freely decaying homogeneous rotating turbulence}. {\em Journal
  of Fluid Mechanics\/} {\bf 677}, 154--178.

\end{thebibliography}

\end{document}